\documentclass[aip,apl,amsmath,amssymb,reprint]{revtex4-2}

\usepackage{dcolumn,amsmath,bm,epsfig,graphicx}
\usepackage[colorlinks,linkcolor=blue,citecolor=blue]{hyperref}
\usepackage[T1]{fontenc}
\setcitestyle{super}
\usepackage{epsfig,graphicx,times}
\usepackage{amstext}
\usepackage{amssymb}             
\usepackage{latexsym}
\usepackage{bm}
\usepackage[dvipsnames]{xcolor}
\usepackage[normalem]{ulem}
\usepackage{afterpage} 

\usepackage[utf8]{inputenc}
\usepackage{mathptmx}
\usepackage{etoolbox}
\usepackage[utf8]{inputenc}
\usepackage[framemethod=tikz]{mdframed}

\usepackage[most]{tcolorbox}
\usepackage{enumitem,lipsum}
\tcbuselibrary{breakable}

\makeatletter
\def\@email#1#2{
 \endgroup
 \patchcmd{\titleblock@produce}
  {\frontmatter@RRAPformat}
  {\frontmatter@RRAPformat{\produce@RRAP{*#1\href{mailto:#2}{#2}}}\frontmatter@RRAPformat}
  {}{}
}
\makeatother

\newcommand*{\citen}[1]{
  \begingroup
    \romannumeral-`\x 
    \setcitestyle{numbers}
    \cite{#1}
  \endgroup   
}

\begin{document}

\title{Gate-defined Josephson weak-links in monolayer WTe\textsubscript{2}}

\author{Michael D. Randle}
\affiliation{Advanced Device Laboratory, RIKEN, 2-1 Hirosawa, Wako, Saitama 351-0198, Japan}

\author{Masayuki Hosoda}
\affiliation{Fujitsu Research, Fujitsu Ltd., Atsugi, Kanagawa 243-0197, Japan}

\author{Russell S. Deacon}
\affiliation{RIKEN Center for Emergent Matter Science (CEMS), 2-1 Hirosawa, Wako, Saitama 351-0198, Japan}
\affiliation{Advanced Device Laboratory, RIKEN, 2-1 Hirosawa, Wako, Saitama 351-0198, Japan}
\email[Corresponding author: ]{russell@riken.jp}

\author{Manabu Ohtomo}
\affiliation{Fujitsu Research, Fujitsu Ltd., Atsugi, Kanagawa 243-0197, Japan}

\author{Patrick Zellekens}
\affiliation{RIKEN Center for Emergent Matter Science (CEMS), 2-1 Hirosawa, Wako, Saitama 351-0198, Japan}

\author{Kenji Watanabe}
\affiliation{Research Center for Electronic and Optical Materials, National Institute for Materials Science, 1-1 Namiki, Tsukuba, Ibaraki, 305-0044, Japan}

\author{Takashi Taniguchi}
\affiliation{Research Center for Materials Nanoarchitectonics, National Institute for Materials Science, 1-1 Namiki, Tsukuba, Ibaraki, 305-0044, Japan}

\author{Shota Okazaki}
\affiliation{Laboratory for Materials and Structures, Tokyo Institute of Technology, 4259 Nagatsuta, Midori-ku, Yokohama 226-8503, Japan}

\author{Takao Sasagawa}
\affiliation{Laboratory for Materials and Structures, Tokyo Institute of Technology, 4259 Nagatsuta, Midori-ku, Yokohama 226-8503, Japan}

\author{Kenichi Kawaguchi}
\affiliation{Fujitsu Research, Fujitsu Ltd., Atsugi, Kanagawa 243-0197, Japan}

\author{Shintaro Sato}
\affiliation{Fujitsu Research, Fujitsu Ltd., Atsugi, Kanagawa 243-0197, Japan}

\author{Koji Ishibashi}
\affiliation{RIKEN Center for Emergent Matter Science (CEMS), 2-1 Hirosawa, Wako, Saitama 351-0198, Japan}
\affiliation{Advanced Device Laboratory, RIKEN, 2-1 Hirosawa, Wako, Saitama 351-0198, Japan}
\begin{abstract}
Systems combining superconductors with topological insulators offer a platform for the study of Majorana bound states and a possible route to realize fault tolerant topological quantum computation. Among the systems being considered in this field, monolayers of tungsten ditelluride (WTe\textsubscript{2}) have a rare combination of properties. Notably, it has been demonstrated to be a Quantum Spin Hall Insulator (QSHI) and can easily be gated into a superconducting state. We report measurements on gate-defined Josephson weak-link devices fabricated using monolayer WTe\textsubscript{2}. It is found that consideration of the two dimensional superconducting leads are critical in the interpretation of magnetic interference in the resulting junctions. The reported fabrication procedures suggest a facile way to produce further devices from this technically challenging material and the results mark the first step toward realizing versatile all-in-one topological Josephson weak-links using monolayer WTe\textsubscript{2}.
\end{abstract}

\maketitle 
\pagenumbering{arabic}

\color{black}

\section{Introduction}

A two-dimensional (2D) topological insulator or Quantum Spin Hall Insulator (QSHI) exhibits a bulk energy gap that is created due to a particular crossing of bands near the Fermi level (band inversion) which can then interact through spin-orbit coupling. The arrangement of bands in this scenario can be represented by the nontrivial $Z_{2}$ topology, an implication of which are the existence of one-dimensional, conducting edge states that accompany the typical bulk spectrum\cite{Hasan2010,Ando2013,Kane2005_1,Kane2005_2}. These edge states are helical (exhibit spin-momentum locking) and thus are robust to elastic scattering, implying ballistic transport and the observation of a quantized conductance ($ne^2/h$), where $n$ is the number of edge channels. The original proposal for a QSHI concerned the opening of a spin-gap in the 2D material graphene due to the spin-orbit interaction\cite{Kane2005_1,Kane2005_2}. However, graphene's intrinsic spin-orbit interaction was found to be too weak to enable observation of the QSHI state experimentally. Recently graphene/$\mathrm{WS_{2}}$ and graphene/$\mathrm{WSe_{2}}$ heterostructures have been studied as a route toward enhancing the spin-orbit interaction\cite{AvsarNatCommun2014}, increasing the effective spin gap to make the QSHI state experimentally accessible. One class of 2D materials, the transition metal dichalcogenides (TMDs), present another promising path toward identifying QSHI materials. They possess a number of polytypic crystal phases (2H, 1T, and 1T') and are predicted to exhibit a QSHI state if the structure is in the monoclinic 1T'-phase\cite{Qian2014}. Among TMDs, the monolayer tungsten ditelluride ($\mathrm{WTe_{2}}$) is predicted to stably exist within the 1T'-phase, making it a suitable system for the study of a 2D crystalline QSHI. A number of experimental studies have identified strong evidence for the QSHI state in monolayer $\mathrm{WTe_{2}}$ through conductance quantization\cite{Fei2017,Wu2018,ZhaoNatMaterials2020,SunArXiv2021,ZhaoPhysRevX2021}, scanning tunnelling microscopy measurements\cite{Tang2017,Jia2017,Peng2017}, and direct observation of edge states through scanning microwave impedance spectroscopy\cite{Shi2019}. Strikingly, it has also been demonstrated that monolayer WTe\textsubscript{2} can host an intrinsic superconducting phase by means of gate-induced electrostatic tuning\cite{Fatemi2018,Sajadi2018}. Such behavior is also reported in twisted bilayer graphene\cite{CaoNature2018}, where surface gates have been used to form Josephson weak-links and tune the system through a complex phase space\cite{Folkert2021,Rodan-Legrain2021,DiezMeridNatComm2023}. The coexistence of superconductivity and intrinsic topological states in WTe\textsubscript{2} makes it an ideal platform to study Majorana zero modes in a single material system by gate-definition of a Josephson weak-link. 

\begin{figure*}[t!]
\includegraphics[width=15 cm,angle=0]{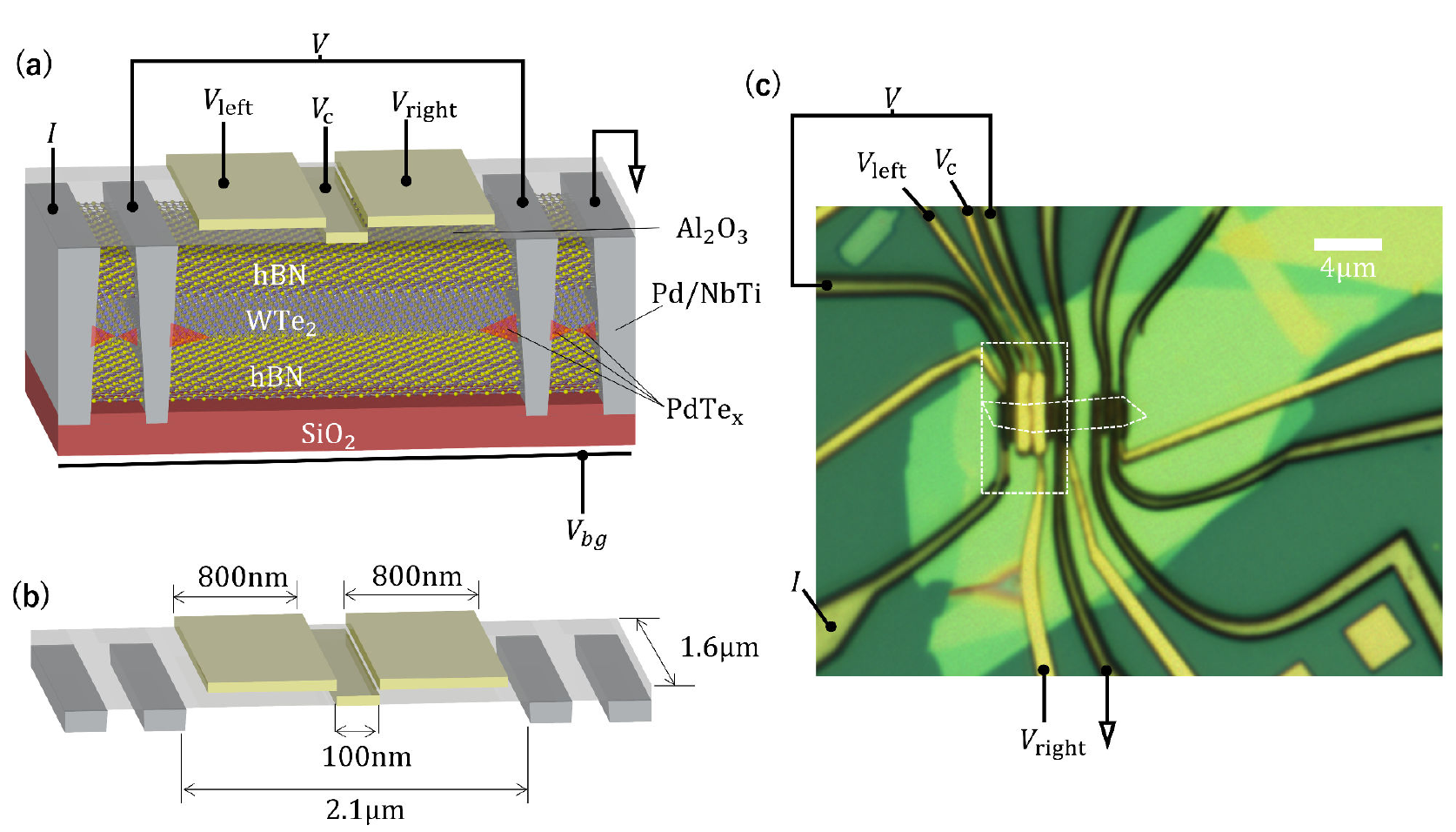}
\caption{\textbf{Device schematic}. (a) Cross-sectional schematic of the device design showing an hBN encapsulated $\mathrm{WTe_{2}}$ monolayer contacted with sputtered $\mathrm{Pd/NbTi}$ "edge" contacts. The device is gated using both a global back-gate through $280\,$nm of thermal grown $\mathrm{SiO_{2}}$ on the device substrate, and with three surface-gates. The two wide surface-gates ($V_{\mathrm{left}}$ and $V_{\mathrm{right}}$) are isolated from the center surface-gate ($V_{\mathrm{c}}$) with a $20\,$nm layer of atomic layer deposition grown alumina. Regions in which palladium diffusion is expected to occur resulting in formation of $\mathrm{PdTe_{x}}$ are indicated in red. (b) Nominal fabricated dimensions of the device channel gates and flake. (c) Optical image of the device with the section being measured indicated.}
\label{Fig1}
\end{figure*}

Josephson weak-links created by interfacing a QSHI with an \textit{s}-wave superconductor are predicted to host Majorana modes, as proposed in the seminal work of Fu and Kane\cite{Fu2008,Fu2009}. In combination with device designs to break time-reversal symmetry in the edge states (via local gating or ferromagnetic materials) such a system can be a platform to study localized Majorana Fermions, which have applications in topological quantum computing\cite{Kitaev2001,Kitaev2003,Nayak2008,Qi2011}, with realistic proposals for the design for topological qubits\cite{ShuoPhysRevB2013}. A number of QSHI systems have been studied in combination with superconductors to probe the presence of topological superconducting states with efforts in the $\mathrm{HgTe/CdTe}$ based system\cite{Hart2014,Deacon2017b,Deacon2017} and the $\mathrm{InAs/GaSb}$ system\cite{PribiagNatNano2015,NicheleNewJPhys2016}. Reports of planar weak-links using 2D materials are surprisingly uncommon and they are almost exclusively fabricated using graphene\cite{LeeRepProgPhys2018,HeercheNature2007,Calado2015,Amet2016,BenShalom2016,Folkert2021,Rodan-Legrain2021,DiezMeridNatComm2023}, with other recent reports using black phosphorus monolayers\cite{TelesioACSNano2022} and efforts using $\mathrm{MoS_{2}}$\cite{RamezaniACSnanolett2021,SeredinskiAIPAdvanced}. To our knowledge, there are no reports of such devices made with a 2D crystal QSHI. The ability of monolayer WTe\textsubscript{2} to be tuned from the QSHI to superconducting state suggests a clear path toward realizing a planar Josephson weak-link in a single material where the superconducting and QSHI elements are explicitly defined by gates\cite{Fatemi2018,Sajadi2018}. Despite this and the not-so-recent prediction of that WTe\textsubscript{2} may host topological states\cite{Qian2014,Soluyanov2015}, no Josephson weak-link devices have been reported with the monolayer material to date. The primary reason is that few- and monolayer $\mathrm{WTe_{2}}$ is challenging to process into devices due to the rapid oxidization of its surface under ambient conditions\cite{YeSmall2016,YangPhysChemLett2018,Lee2015,Kim2016,Naylor2DMaterials2017} and even in a high purity $\mathrm{N_{2}}$ environment\cite{Hou2020}. Oxidation effectively destroys the monolayer material and severely impedes the transport properties of multilayers, for example, by greatly suppressing its magnetoresistance\cite{WoodsACS2017,YangPhysChemLett2018}. Therefore, fabrication must be performed quickly and adapted to inert gas (e.g. argon) environments.

In this report, we present the fabrication and measurement of gate-defined Josephson weak-link devices with $\mathrm{hBN}$ encapsulated monolayer $\mathrm{WTe_{2}}$. Contacts to the encapsulated $\mathrm{WTe_{2}}$ are formed using an edge contacting technique, utilizing in-situ reactive ion etching of the $\mathrm{hBN}/\mathrm{WTe_{2}}/\mathrm{hBN}$ stack followed by dc-sputter deposition of superconducting contacts, which provides superconducting connections to the external device circuit. Device contacts are formed using a thin contacting layer of sputtered Pd, motivated by recent reports\cite{Kononov2021,Ohtomo2022,EndresPhysRevMaterials2022} of induced superconductivity in multilayer $\mathrm{WTe_{2}}$ through diffusion of $\mathrm{Pd}$ forming regions of $\mathrm{PdTe_{x}}$. The clean Pd/WTe\textsubscript{2} interface is found to produce good superconducting contacts upon the annealing of the system in later processing steps (see Supplementary Note 1). The channel of the device is defined by electrostatic gates which can locally drive sections of the monolayer superconducting, creating a Josephson weak-link, in which the layer forming the weak-link is monolayer $\mathrm{WTe_{2}}$. We demonstrate that electrostatic gating can induce a superconducting state in the system (sections 2.1 and 2.2) and then confirm the presence of a Josephson weak-link through measurements of the magnetic response and the Shapiro response (inverse ac-Josephson effect). We report measurements on a single device in the main text alongside complementary measurements of a second device in supplementary Note 2a.

A schematic of the device design and optical image is shown in \textbf{figure \ref{Fig1}}. The device studied consists of a channel with three surface-gates. Namely, a narrow center gate ($V_{\mathrm{c}}$) on the capping hBN and a pair of wider gates ($V_{\mathrm{left}}$ and $V_{\mathrm{right}}$), isolated from $V_{\mathrm{c}}$ using a layer of $\mathrm{Al_{2}O_{3}}$ grown by atomic layer deposition. These terminals facilitate gating of the underlying $\mathrm{WTe_{2}}$ into the superconducting state. The edge contacts span the width of the $\mathrm{WTe_{2}}$ flake. Measurements taken between the contact pairs used for source or drain terminals reveal supercurrent and only a weak gate response, so we conclude that the processing has resulted in diffusion of $\mathrm{Pd}$ forming $\mathrm{PdTe_{x}}$ and effectively shorting the contacts (Supplementary Note 1). In contrast, the device channel resides far beyond the distance anticipated for the $\mathrm{Pd}$ diffusion and reveals the strong gate response indicative of the monolayer $\mathrm{WTe_{2}}$. Fabrication details are given in methods section 4.1. 

\begin{figure*}[t!]
\centering
\includegraphics[width=15 cm,angle=0]{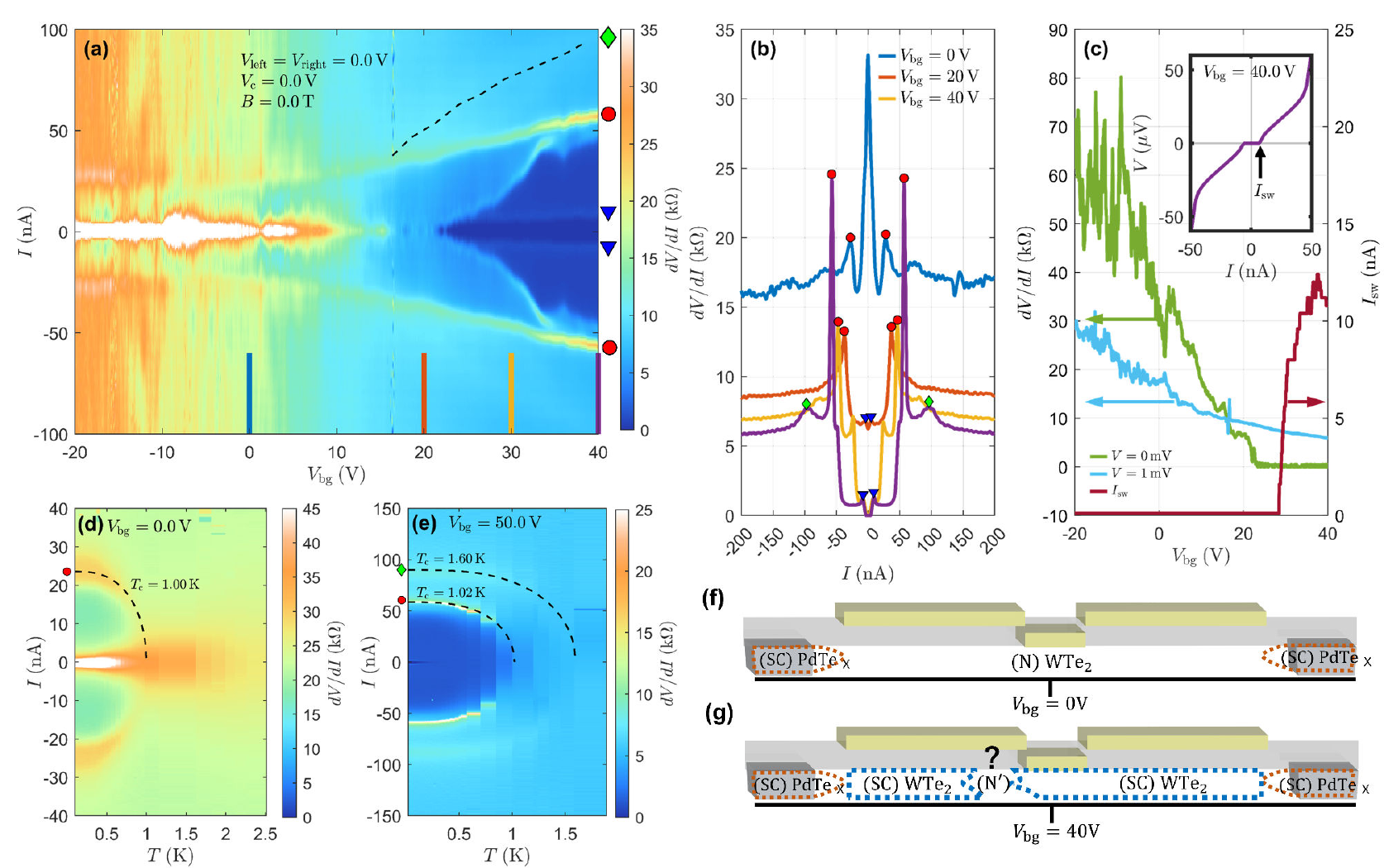}
\caption{\textbf{Influence of the global back-gate}. (a) Plot of differential resistance ($dV/dI$) as a function of device current (I) and global back-gate voltage ($V_{\mathrm{bg}}$). All other gate voltages are set to zero. Colored lines indicate the positions of traces shown in (b). The dashed black line indicates the weak transport feature discussed in the main text. (b) Differential resistance as a function of current for three values of $V_{\mathrm{bg}}$, indicated by solid lines in (a). The colored symbols indicate the transport features discussed in the main text. (c) Plot of differential resistance as a function of back-gate at $V=0\,$V and $V=1\,$mV, alongside extracted switching current ($I_{\mathrm{sw}}$). (inset) $I(V)$ trace at $V_{\mathrm{bg}}=40\,\mathrm{V}$ showing the switching current $I_{\mathrm{sw}}$ at transition from non-dissipative to resistive branches of the trace. (d) and (e) Plots of $dV/dI$ as a function of temperature and applied current for $V_{\mathrm{bg}}=0\,$V and $V_{\mathrm{bg}}=50\,$V respectively. In (d) and (e) all other gate voltages are set to $0\,$V. Dashed lines indicate fits of transport features to the interpolated BCS gap relationship with indicated critical temperatures $T_{\mathrm{c}}$ with standard error of $\pm 0.01\,\mathrm{K}$. (f) shows illustrations of the device at $T=0$ for $V_{\mathrm{bg}}=0\,$V where no weak-link has formed. (g) illustration at $V_{\mathrm{bg}}=40\,$V where an S-N'-S junction has formed at some position. The exact position of the weak-link is unknown but is speculated to be close to the $V_{\mathrm{c}}$ gate region. Here N' may indicate a normal region due to gate response or a micro-constriction that is normal due to transport current.}
\label{Fig2}
\end{figure*}

\section{Results and discussion}

\subsection{Josephson weak-link in monolayer WTe\textsubscript{2} formed by a global back-gate}

\begin{figure*}[t!]
\centering
\includegraphics[width=15 cm,angle=0]{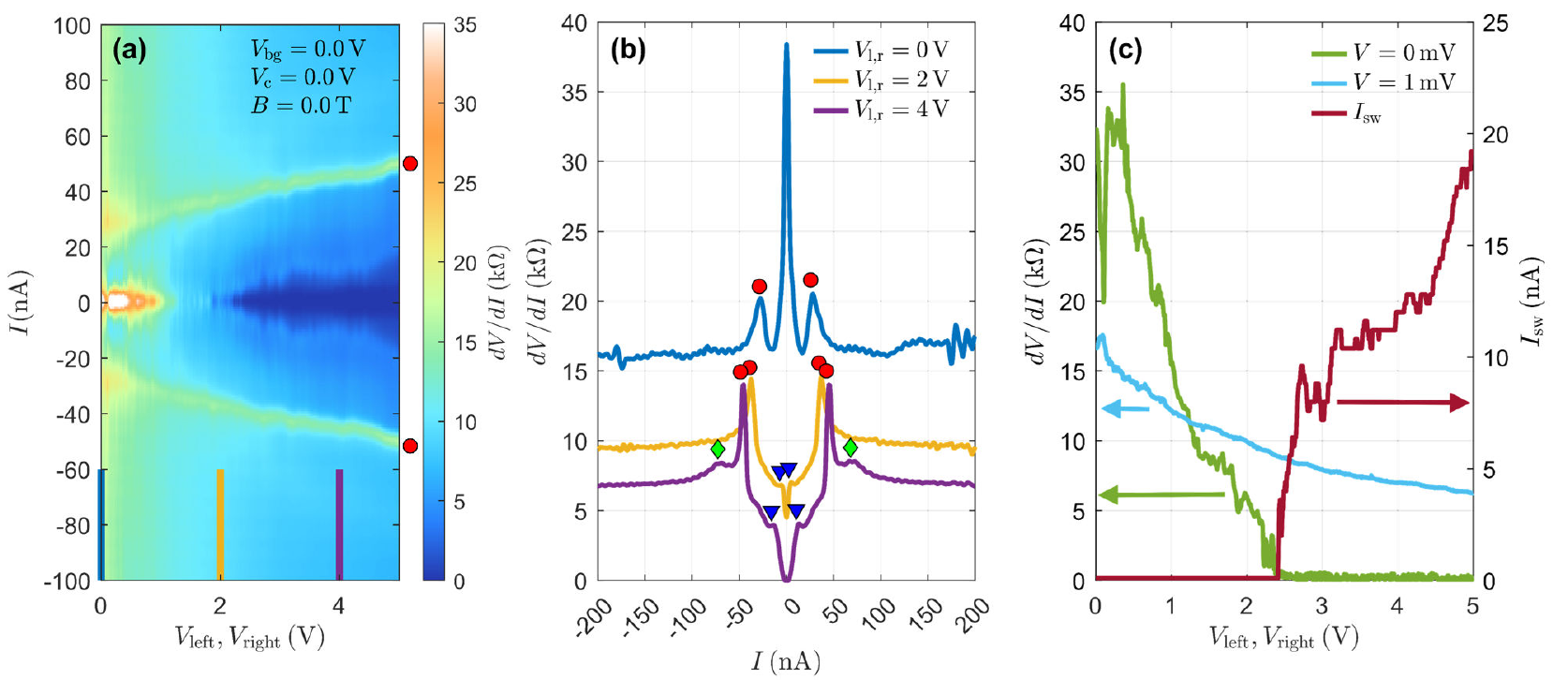}
\caption{\textbf{Influence of the top-gates $\bm{V_{\mathrm{left}}}$ and $\bm{V_{\mathrm{right}}}$}. (a) Plot of differential resistance ($dV/dI$) as a function of device current and a voltage applied to both $V_{\mathrm{left}}$ and $V_{\mathrm{right}}$. All other gate voltages are set to zero. (b) Differential resistance as a function of current for three values of $V_{\mathrm{left}}=V_{\mathrm{right}}$, indicated by solid lines in (a). Colored symbols indicate equivalent features to those discussed in figure \ref{Fig2} (b). (c) Plot of differential resistance as a function of voltage applied on $V_{\mathrm{left}}$ and $V_{\mathrm{right}}$ at $V=0\,$V and $V=1\,$mV.}
\label{Fig3}
\end{figure*}

Before discussing the situation in which the channel is explicitly defined by local gates, we initially demonstrate that the device can exhibit superconducting transport when a positive bias is applied to the global back-gate ($V_{\mathrm{bg}}$) which acts through $280\,$nm of thermally grown $\mathrm{SiO_{2}}$ on the silicon substrate. In \textbf{figure \ref{Fig2}} (a), we present a color plot of the differential resistance ($dV/dI$) as a function of both applied current ($I$) and back-gate voltage ($V_{\mathrm{bg}}$). Figure \ref{Fig2} (b) then shows cross-sections of this plot at selected values of $V_{\mathrm{bg}}$. For $V_{\mathrm{bg}}=0\,$V, the chemical potential of the monolayer is sitting near or within the gap such that its contribution to the transport is nearly insulating. Conversely, the $\mathrm{PdTe_{x}}$ contacts are far below their $T_{\mathrm{c}}$ and superconducting. In this case, we effectively produce an S-N-S junction (figure \ref{Fig2} (f)). In the corresponding trace of $dV/dI$ as a function of $I$ in figure \ref{Fig2} (b), we observe a sharp resistance peak at zero current flanked by a pair of smaller satellite peaks at $I=\pm 20\,$nA (red circles). The zero current feature can be qualitatively understood through the scattering dynamics of normal electrons and quasiparticles at a low transparency S-N-S interface. In the context of BTK theory\cite{Blonder1982,Klapwijk1982,Octavio1983}, one expects for a very low transparency junction a suppressed current within the bias window of the superconducting gap due to the greater influence of normal reflections as opposed to Andreev reflections. Consistent with this scenario, the satellite peaks arise from the superconducting $\mathrm{PdTe_{x}}$ regions transitioning to the normal state at a critical current of $I=\pm 20\,$nA. Indeed, temperature dependent measurements of these peaks reveal a BCS-like dependence (figure \ref{Fig2} (d)) with an associated $T_{\mathrm{c}}\approx 1\,$K. At higher currents, all elements of the system are normal and the net resistance stabilizes to a constant value near $16\,\mathrm{k\Omega}$.

For $V_{\mathrm{bg}}=40\,$V, the system is significantly doped such that the monolayer $\mathrm{WTe_{2}}$ undergoes a transition to a superconducting state. Consequently, several new features appear in the transport. The first of which is a narrow supercurrent branch about zero bias with associated resistance peaks at $I=\pm 6.5\,$nA (blue triangles). We attribute these features to the formation of a Josephson weak-link somewhere within the device channel (figure \ref{Fig2} (g)). In this scenario, the peaks (blue triangles) correspond to the switching current ($I_{\mathrm{sw}}$) of the junction, which we define as the maximum supercurrent that the junction can support before transitioning or switching to the dissipative transport branch (as shown in the inset of figure \ref{Fig2} (c)). The formation of a weak-link is corroborated by observation of interference patterns in the magnetic response (presented in section 2.4) and observation of a Shapiro response (section 2.3). The origin of the weak-link may be a region of normal $\mathrm{WTe_{2}}$ or a micro-constriction within the monolayer which is driven normal by transport current. Accordingly, we denote the weak-link as S-N'-S. Possible origins will be discussed in further detail in section 2.5. At higher currents, we observe that the resistance peaks associated with the critical current of the superconducting $\mathrm{PdTe_{x}}$ (red circles) shift to higher currents ($\sim 48\,\mathrm{nA}$ at $V_{\mathrm{bg}}=40\,$V) and then finally a new pair of weak resistance maxima (green diamonds) appear. We associate these features with the superconducting gap of the $\mathrm{WTe_{2}}$ monolayer, as they appear approximately constant in bias (Supplementary Note 3). This interpretation is consistent with temperature dependent measurements (figure \ref{Fig2} (e)), which again reveal a BCS-like dependence of each feature, with the $\mathrm{PdTe_{x}}$ having ($T_{\mathrm{c}}\approx 1\,$K) and the monolayer having ($T_{\mathrm{c}}\approx 1.6\,$K). We note that these critical temperatures differ from reported values in the literature\cite{Kononov2021,Ohtomo2022,EndresPhysRevMaterials2022}, but the results are consistent with a secondary monolayer $\mathrm{WTe_{2}}$ device in which $T_{\mathrm{c}}$ was estimated in a different manner (Supplementary Note 2b). Finally, we conjecture that the $V_{\mathrm{bg}}=20\,$V data represents a transitive state in which a Josephson weak-link has formed, but cannot yet support a supercurrent due to poor transparency. The explicit effect of $V_{\mathrm{bg}}$ on device resistance and the observation of supercurrent is shown in figure \ref{Fig2} (c). For example, the orange trace indicates that a switching current can be observed at $V_{\mathrm{bg}}\approx30\,$V. The small critical current associated with the $\mathrm{PdTe_{x}}$ (indicated with red circles in all plots) complicates the analysis of features in the dissipative transport beyond the supercurrent branch because when it becomes normal, the junction bias is now dropped across not only the weak-link, but also the contacting regions. See Supplementary Note 3 for plots of transport features as a function of voltage. In addition to those already discussed we note an additional transport feature that emerges between the supercurrent branch (blue triangles) and the transition associated with the $\mathrm{PdTe_{x}}$ (red circles), appearing around $V_{\mathrm{bg}}=20\,\mathrm{V}$ and merging with the $\mathrm{PdTe_{x}}$ feature near $V_{\mathrm{bg}}=40\,\mathrm{V}$. The origin of this feature is inconclusive but its scaling in gate suggests that it originates in the gated superconducting $\mathrm{WTe_{2}}$ regime and so could be a sub-gap transport feature associated with the weak-link such as resonances due to multi-Andreev reflections\cite{BardasPhysRevLett1995} or possibly a region of the gated $\mathrm{WTe_{2}}$ exhibiting a lower critical current.

\subsection{Tunability of a gate-defined Josephson weak-link in monolayer $\mathrm{WTe_{2}}$}

\begin{figure*}[t!]
\centering
\includegraphics[width=15 cm,angle=0]{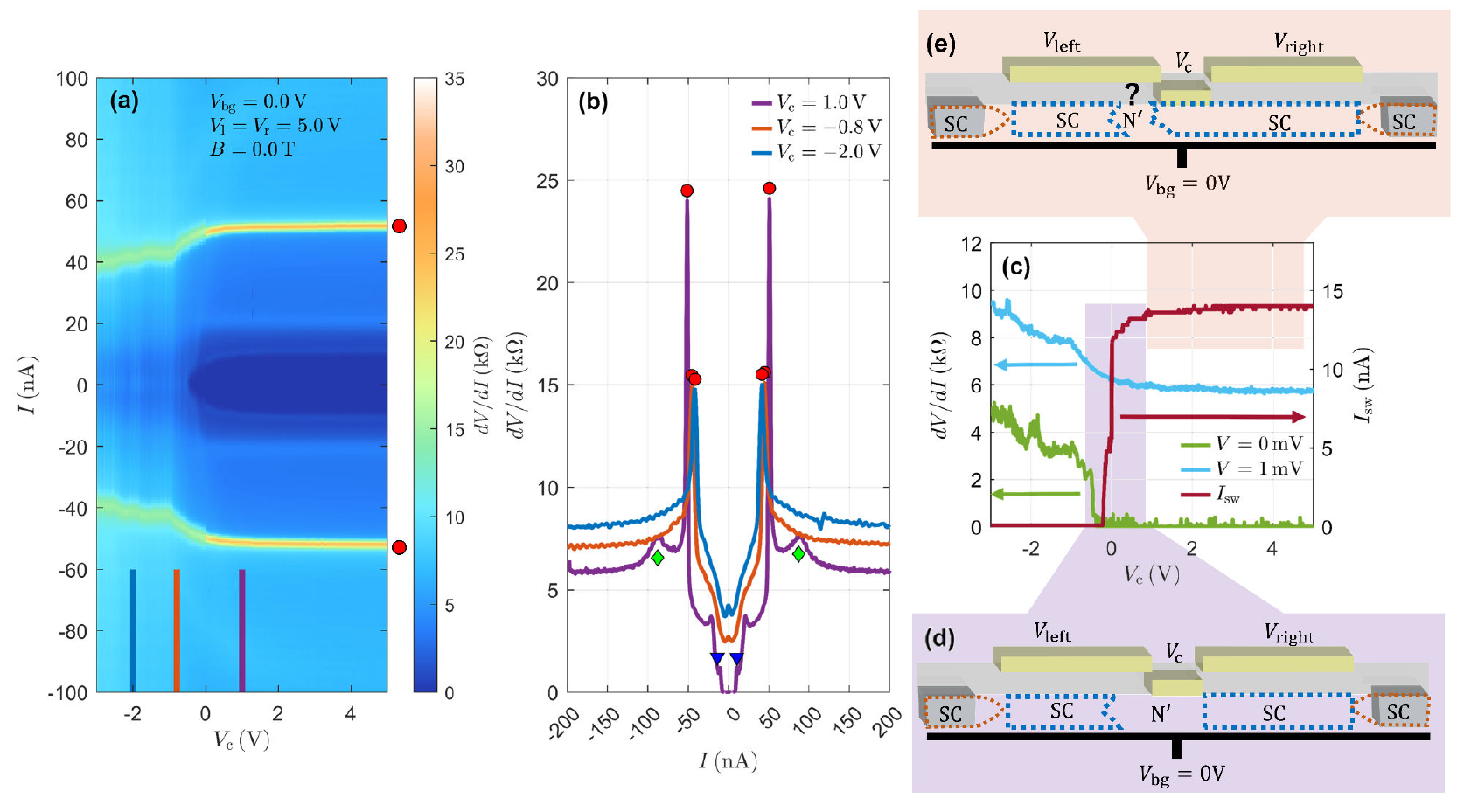}
\caption{\textbf{Control of channel supercurrent using gate $\bm{V_{\mathrm{c}}}$.} (a) Plot of differential resistance ($dV/dI$) as a function of device current and a voltage $V_{\mathrm{c}}$. Gates $V_{\mathrm{left}}$ and $V_{\mathrm{right}}$ are set at $5.0\,$V with $V_{\mathrm{bg}}=0.0\,$V. (b) Differential resistance as a function of current for four values of $V_{\mathrm{c}}$, indicated by solid lines in (a). Colored symbols indicate equivalent features to those discussed in figure \ref{Fig2} (b). (c) Plot of differential resistance as a function of voltage applied on $V_{\mathrm{c}}$ at $V=0\,$V and $V=1\,$mV, plotted alongside the extracted Josephson weak-link switching current ($I_\mathrm{sw}$). (d) Schematic of the gate controlled S-N'-S weak-link within the range of $V_{\mathrm{c}}$ for which the junction properties can be tuned (shown by the shaded region).(e) illustration of the S-N'-S weak-link formed when the center gate region is gated superconducting (shown by the shaded region).}
\label{Fig4}
\end{figure*}

We now demonstrate that the as-patterned top-gates ($V_{\mathrm{left}}$, $V_{\mathrm{right}}$) can perform a similar function as the global back-gate. In the present discussion, the formed Josephson weak-link is defined as a channel between the gated regions in a more deterministic manner, which is in contrast to the previous discussion where the weak-link's position is not explicitly known. Furthermore, we will demonstrate that a patterned center gate ($V_{\mathrm{c}}$) can turn on and off the supercurrent branch of the junction. The $dV/dI$ color plot of \textbf{figure \ref{Fig3}} (a) and its corresponding cross-sections (figure \ref{Fig3} (b)) should be compared to the equivalent panels of figure \ref{Fig2}. We find that for increasing ($V_{\mathrm{left}}$, $V_{\mathrm{right}}$) the device's transport features (identified by colored symbols in figure \ref{Fig3} (b)) qualitatively mimic the progression discussed previously for the back-gated junction. This indicates that through local electrostatic gating, we are able to locally drive sections of the monolayer $\mathrm{WTe_{2}}$ superconducting and create a Josephson weak-link in the central region (\textbf{figure \ref{Fig4}} (d)). Figure \ref{Fig3} (c) indicates that supercurrent can be observed at ($V_{\mathrm{left}}$, $V_{\mathrm{right}}>2.5\,$V). Here again we find properties of a Josephson S-N'-S weak-link, where we believe that the $N'$ region is in close proximity to the center gate. We corroborate this point by demonstrating the tunability of this junction by $V_{\mathrm{c}}$. Figure \ref{Fig4} (a-c) illustrates the behavior of the device when ($V_{\mathrm{left}}$, $V_{\mathrm{right}}=5\,$V), a value high enough to induce superconductivity in the monolayer, and $V_{\mathrm{c}}$ is changed to progressively more negative values. In figure \ref{Fig4} (a-b) it is observed that the channel supercurrent can be suppressed and for the highest negative $V_{\mathrm{c}}$, a resistance maximum near zero bias begins to form, confirming the expected behavior of an S-N-S junction without supercurrent. The proposed system behavior when a weak-link is formed is illustrated in figure \ref{Fig4} (d) and (e). For high values of $V_{\mathrm{c}}$ the center gate region is superconducting but still some region of the device forms a S-N'-S weak-link (figure \ref{Fig4} (e)) as was seen in the back-gate induced superconducting case in figure \ref{Fig2}. Here the influence of $V_{\mathrm{c}}$ is saturated as the region of the device forming the junction is elsewhere. For low $V_{\mathrm{c}}$ the center gate region is driven normal and the properties of the weak-link are tuned, figure \ref{Fig4} (e).

\begin{figure*}[t!]
\centering
\includegraphics[width=15 cm,angle=0]{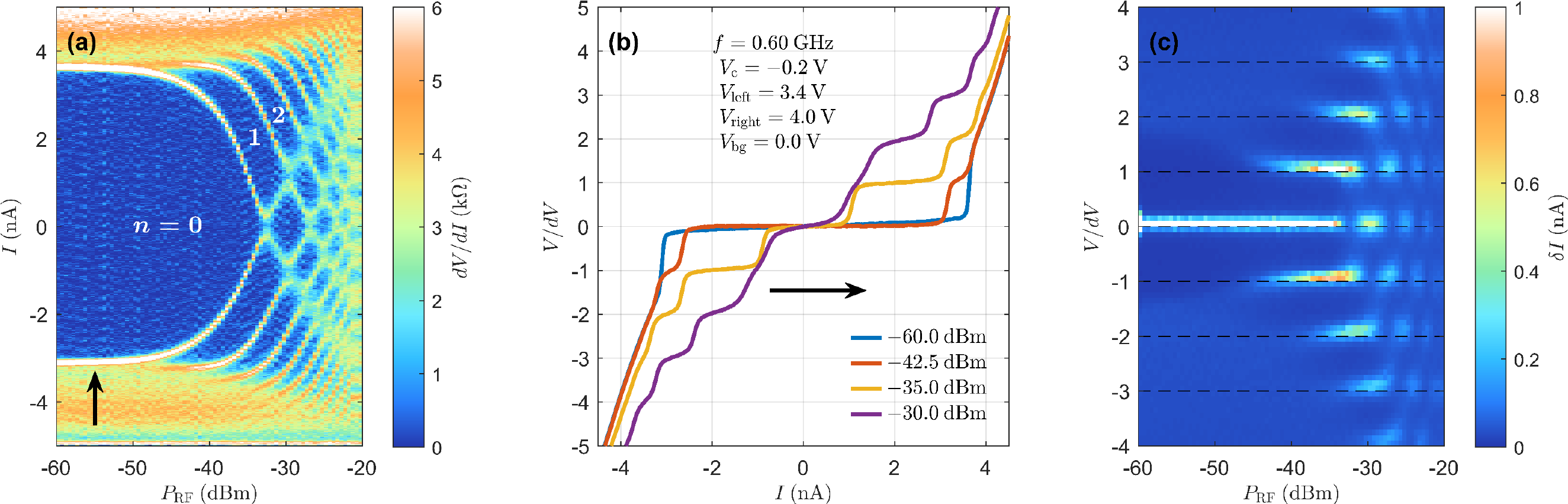}
\caption{\textbf{Shapiro response with external drive at $\bm{f=0.6\,}$GHz}.(a) Differential resistance ($dV/dI$) as a function of device current ($I$) and microwave power ($P_\mathrm{RF}$). The indicated microwave power is at the output from the signal generator and is not corrected for attenuation of lines on the fridge. Gate settings are $V_{\mathrm{c}}=-0.2\,$V, $V_{\mathrm{left}}=3.4\,$V, $V_{\mathrm{right}}=4.0\,$V and $V_{\mathrm{bg}}=0.0\,$V. The black arrow indicates the direction of the current sweep in the measurement. (b) Example $I(V)$ traces for the rf powers indicated. Parameter $dV=hf/2e=1.2424\,\mathrm{\mu V}$. (c) Voltage histograms of $I(V)$ traces revealing the quantization of the Shapiro steps. Histogram bins have a width of $\delta V=0.1dV$.}
\label{Fig5}
\end{figure*}

\subsection{Shapiro response of the gate-defined weak-link}

We demonstrate the properties of the gate-defined Josephson weak-link by measuring the response of the device to a microwave drive (\textbf{figure \ref{Fig5}}). In a conventional Josephson weak-link, with sinusoidal current phase relationship $I_{c}=I_{0}\sin(\varphi)$, steps appear quantized in voltage at multiples of $hf/2e$. With increasing ac-current amplitude the step widths in current exhibit an oscillating Bessel function like pattern\cite{Russer}. The oscillatory pattern and good quantization of steps at voltages $V_{n}=nhf/2e$, $n \in \mathbb{Z}$ are easily visualized by plotting histograms of binned measurement voltage data as shown in figure \ref{Fig5} (b). We observe conventional patterns with no evidence of sub-harmonic steps to indicate a non-sinusoidal current phase relationship\cite{RaesPRB2020,UedaPRR2020}. We also see no evidence of doubling of step voltages due to the fractional ac-Josephson effect associated with topological modes\cite{Deacon2016,Deacon2017,Deacon2017b}. Additional Shapiro response data for a collection of alternative gating conditions can be found in Supplementary Note 7. Broadly, we observe that the visibility of the Shapiro steps scales with the weak-link $I_{\mathrm{sw}}$.

\subsection{Magnetic field response and asymmetric current distribution}

\begin{figure*}[t!]
\centering
\includegraphics[width=15 cm,angle=0]{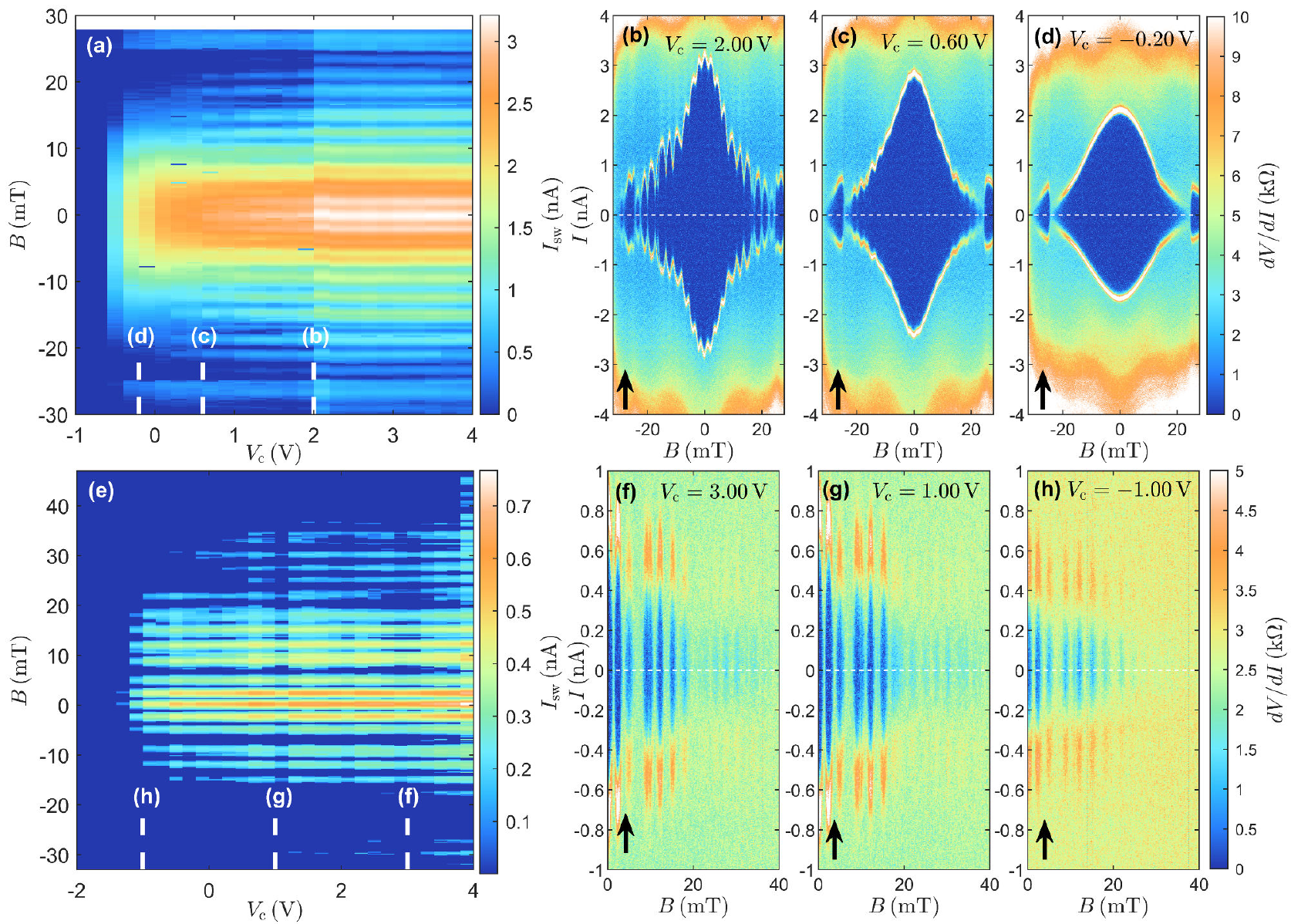}
\caption{\textbf{Influence of magnetic field on the junction transport}. (a) Plot of evaluated switching currents $I_{\mathrm{sw}}$ as a function of gate $V_{\mathrm{c}}$ with $V_{\mathrm{bg}}=0\,$V and $V_{\mathrm{left}}=V_{\mathrm{right}}=5\,$V. The SQUID-like oscillation period is $\delta B\sim 2.2\,\mathrm{mT}$. (b-d) Plots of $dV/dI$ as a function of $I$ and $B$ at values of $V_{\mathrm{c}}$ indicated in (a). Black arrows indicate the direction of the current sweep in the measurement. In all data, a correction for a $\mathrm{2.1\,}$mT  remnant field has been subtracted to center the observed patterns. (e) Plot of evaluated switching currents $I_{\mathrm{sw}}$ as a function of gate $V_{\mathrm{c}}$ with $V_{\mathrm{bg}}=50\,$V and $V_{\mathrm{left}}=V_{\mathrm{right}}=0\,$V. The SQUID-like oscillation period is $\delta B\sim 2.4\,\mathrm{mT}$. (f-h) Plots of $dV/dI$ as a function of $I$ and $B$ at values of $V_{\mathrm{c}}$ indicated in (a). Black arrows indicate the direction of the current sweep in the measurement. Some asymmetry present in the magnetic field response is discussed in Supplementary Note 4c. In all data, a correction for a $\mathrm{5.5\,}$mT remnant field has been subtracted to center the observed patterns.}
\label{Fig6}
\end{figure*} 

Additional confirmation of a weak link can be seen in the response of the junction to an out-of-plane magnetic field ($B$). \textbf{Figure \ref{Fig6}} shows evaluated junction switching currents ($I_{\mathrm{sw}}$) as a function of $B$ and $V_{\mathrm{c}}$ for the two different gating conditions. Explicitly, figure \ref{Fig6} (a-d) show the response for the superconducting state induced using $V_{\mathrm{left}}$ and $V_{\mathrm{right}}$, while figure \ref{Fig6} (e-h) show the response with superconductivity induced using $V_{\mathrm{bg}}$. Figure \ref{Fig6} (b-d) show differential resistance plotted as a function of $B$ for a range of different values of $V_{\mathrm{c}}$. For high $V_{\mathrm{c}}$ (figure \ref{Fig6} (b)) we observe a Fraunhofer-like background decorated with small SQUID-like oscillations that are periodic in flux with a short period ($\delta B\sim 2.2\,$mT). As $V_{\mathrm{c}}$ is reduced, the visibility of the SQUID-like oscillations is diminished until only the background remains (figure \ref{Fig6} (d)). Figure \ref{Fig6} (f-h) show a similar set of traces for the back-gated case (without bias on $V_{\mathrm{left}}$ and $V_{\mathrm{right}}$). Note that here the values of $I_{\mathrm{sw}}$ are suppressed as compared to the measurement data presented in section 2.1 as a consequence of the measurement being performed at a later date after application of high fields to the sample and thermal cycling. Similar SQUID-like oscillations with a slightly longer period ($\delta B\sim 2.4\,$mT) are observed and $V_{\mathrm{c}}$ can be used to tune and eventually fully suppress the non-dissipative current (for additional data see Supplementary Note 4a). The period of the SQUID-like oscillations under both gate conditions is small ($\delta B\sim 2.2$ - 2.4 mT), which is significantly below the field required to apply a flux quantum to the junction area defined by the $V_{\mathrm{c}}$ gate of nominally $\sim 0.16 \mu\mathrm{m}^{2}$. The short period of oscillations in $B$ is a consequence of the 2D nature of the gated superconducting $\mathrm{WTe_{2}}$ regions that define the junction and can be understood in terms of the Pearl theory. In thin superconductors with film thickness ($d$) below the London penetration depth ($\lambda$) the magnetic field screening is characterised by the Pearl length\cite{Pearl1964} ($\lambda_{\mathrm{Pearl}}=2\lambda^{2}/d$). For thin 2D materials, the Pearl length may exceed the size of the sample leading to uniform penetration of magnetic flux throughout the unscreened device (\textbf{figure \ref{SimExample2}} (a)). The uniform magnetic flux in this unscreened case causes modifications to the magnetic response of the junction as compared to more conventional systems with bulk leads in which flux only penetrates within the London penetration length around the lead perimeters due to screening. A Josephson junction with 2D superconducting leads thus has a magnetic interference pattern that is heavily influenced by the geometry of the leads themselves instead of the junction geometry itself. This can be visualized by the simulated response comparison plots of figure \ref{SimExample2} (c). The effect of unscreened 2D superconductors has been observed in other 2D systems such as twisted bilayer graphene\cite{Rodan-Legrain2021,Folkert2021} and thin NbSe\textsubscript{2} flakes\cite{SinkoPRM2021}. SQUID-like oscillations with lifted or non-zero minima on a Fraunhofer-like background can have a range of origins, the simplest being an asymmetry in the current density profile\cite{DynesPRB1971} across the junction giving rise to an asymmetric SQUID. We disregard any influence from self-screening currents and transparency (see Supplementary Notes 4d and 6a) and conclude that the observed magnetic response in the case of top-gate induced superconducting state is most likely due to such an asymmetry.

\subsection{Simulation of the magnetic field response}

\begin{figure*}[t!]
\centering
\includegraphics[width=15 cm,angle=0]{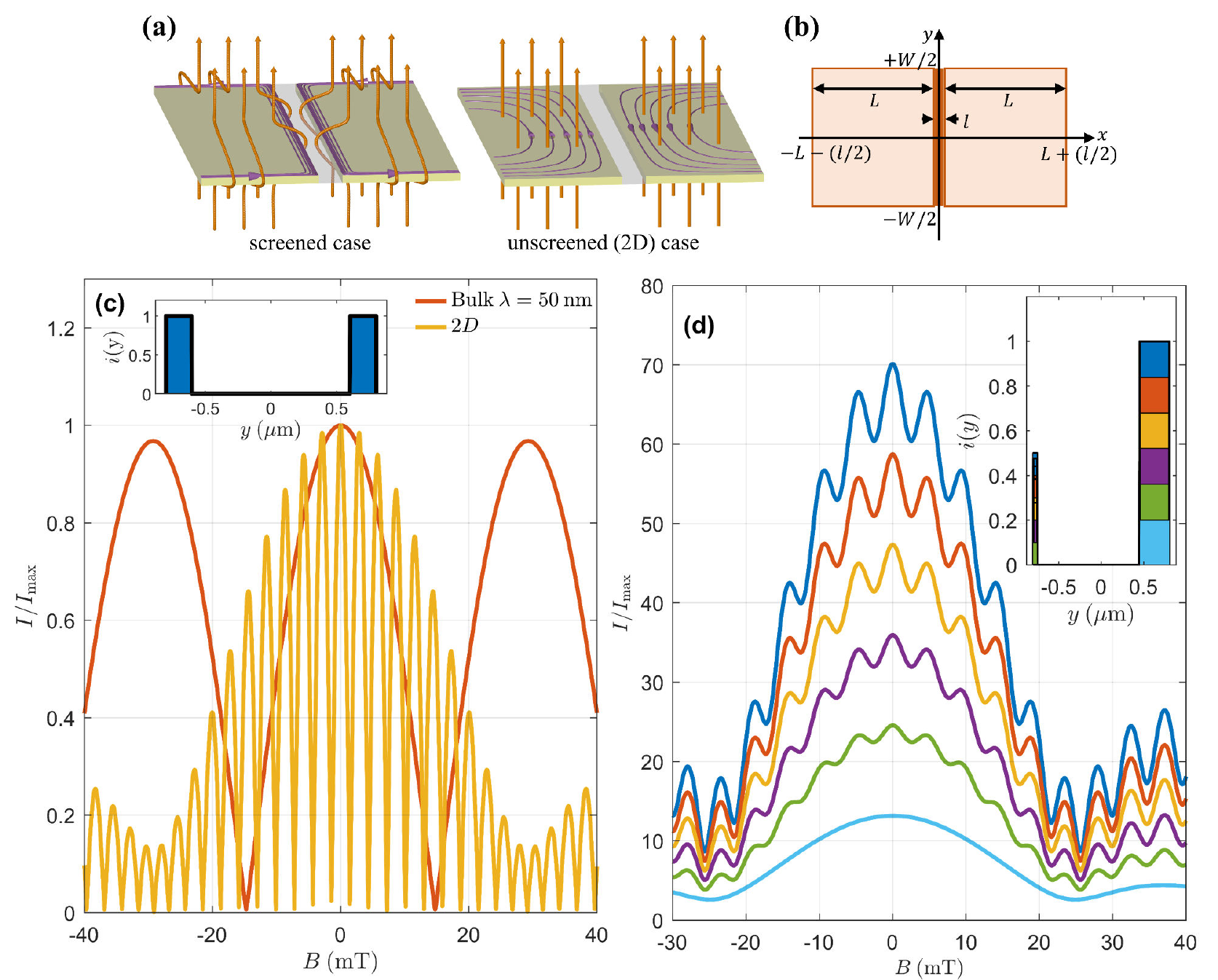}
\caption{\textbf{Simulations of junction magnetic response}. (a) Illustrations of screened and unscreened superconducting leads showing magnetic field vectors and in-plane screening currents. In a screened case current circulates within the London penetration depth ($\lambda$) about the perimeter of the lead and magnetic flux is expelled. (b) Schematic of device geometry used in simulations. All simulation parameters include $W=1.6\,\mu$m, $L=0.9\,\mu$m and $l=100\,$nm. (c) Comparison of results for screened and unscreened geometries with a SQUID-like current distribution $i(y)$ across the junction width as shown in the inset. In the unscreened case a London penetration depth of $\lambda=50\,$nm is used. (d) Plot of simulation results for the unscreened two-dimensional superconductor leads using the inset $i(y)$ current profiles. A slow reduction of currents in the current profile is shown to mimic experimental observations. The modification of the asymmetric SQUID profile reduces the SQUID-like oscillations, all while maintaining the Fraunhofer-like envelope.}
\label{SimExample2}
\end{figure*}

To qualitatively study the magnetic field response under both gating conditions, we follow the work of Clem in \citen{Clem2010} which describes a Josephson junction with rectangular leads which are smaller in dimensions than the Pearl length (see methods section 4.5). A rectangular geometry is well suited to our device layout assuming that the flake has an approximately uniform width. In figure \ref{SimExample2} we show simulation results using parameters estimated for the device with $W=1.6\,\mu$m, $L=0.9\,\mu$m and $l=100\,$nm (figure \ref{SimExample2} (b)). In the simulations, we select current profiles which can qualitatively reproduce the magnetic response of the junctions formed using the different gating schemes. The behavior seen in the back-gated junction is similar to that seen for a symmetric current distribution concentrated about the edges of the sample (figure \ref{SimExample2} (c)). Here, the influence of the SQUID-like features is strong and the minima have clearly defined zeros. As one tunes the asymmetry of the current distribution (figure \ref{SimExample2} (d)), the coexistence of the SQUID-like and Fraunhofer-like features is still seen, but the relative strength of the SQUID-like features are softened as found in the magnetic response of the top-gated junction. To further evaluate a suitable current distribution describing the behavior of our device, we have utilized a minimization fitting approach following the work of Hui et al. in \citen{HuiPRB2014} (see Supplementary Note 5). Both approaches support the conclusion that inducing superconductivity with the back-gate results in a relatively symmetric current distribution across the junction width, with peaks in current near the channel edges. Conversely, when superconductivity is induced using the top-gates the current is predominantly carried in a peak located near one edge.

Under all gate conditions that we have studied we find that the system displays the properties of a weak-link formed somewhere within the device (as indicated with the magnetic and Shapiro response). We find that for some parameter range we are able to tune the system efficiently using the narrow center gate $V_{\mathrm{c}}$ indicating that the weak-link is formed in proximity to the gate. The difference in current distributions for each gating condition could have several possible origins. The device as fabricated may possesses imperfections such as small bubbles after the stamping process which may lead to inhomogeneous gating and to more complex effective geometries. We speculate that such imperfections may be limiting the ability of one of the top-gates to tune part of the monolayer, leading to an effectively longer junction at one side of the flake. This would result in smaller critical currents along one edge and the observed asymmetric oscillations with a relatively longer $\delta B$. In contrast, the back-gate would be unaffected by this imperfection, which lies above the monolayer material. Furthermore the monolayer flake itself may possess a nontrivial geometry due to cracks either present before or produced following the stamping process, as have been observed in recent scanning microwave impedance measurements\cite{Shi2019,ZhaoPhysRevX2021}. Future devices may be improved through better pre-selection of materials for stacking and the use of graphite top and back-gates which have been demonstrated to improve gate disorder in graphene devices\cite{ZibrovNature2017}. As we find no evidence to indicate series junctions being formed we believe that the center gate allows the tuning of the geometry of the N' junction region. We note however that the geometry of the junction is challenging to evaluate from the magnetic response due to the unscreened nature of the 2D superconducting regions which causes phase to be accumulated both inside the weak-link and within the lead regions in a magnetic field (Supplementary Note 4b).

\section{Conclusions}

We have fabricated and measured for the first time Josephson weak-links made from monolayer $\mathrm{WTe_{2}}$. Our approach involves electrostatic gating to define the various regions of the Josephson weak-link by utilizing the material's recently reported ability to be tuned into the superconducting state. We confirm the Josephson effect through observation of a conventional Shapiro response and interference patterns in the weak-link's magnetic response. In both back-gated and top-gated configurations, we observe SQUID-like oscillations in the magnetic field response, whose relative strength is tuned by application of the gate on the weak-link. Our analysis indicates that the majority of current is carried along states situated at the edges of the junction in the back-gated case, while current is carried predominantly along one edge in the top-gated case. Whether these observations are related to the expected topological edge states or simply imperfections in the channel formed during fabrication cannot be discerned. Our results highlight the experimental difficulty of realizing and measuring the predicted topological character of monolayer $\mathrm{WTe_{2}}$, but nonetheless we have successfully fabricated a Josephson weak-link. With improved device fabrication techniques, it may be possible to access the predicted topological states of this system and separate them from other effects that can introduce asymmetry into the supercurrent distribution. Improved processing combined with larger flakes would allow entire re-configurable circuits to be fabricated in the system through  gating. With the inclusion of annealed Pd edge contacts, such circuits could be coupled to external superconducting circuits to allow study and ultimately control of the topological superconducting states predicted in this system.

\section{Methods}

\subsection{Device fabrication}

Tungsten ditelluride crystals are either purchased from HQGraphene (Netherlands) and 2D semiconductor (USA) or provided by Sasagawa laboratory. The crystals are micromechanically exfoliated onto silicon substrates using low-adhesive dicing tape (Nitto Denko Corp.). The subsequent search for monolayer $\mathrm{WTe_{2}}$ is mediated by a homemade automatic search system based on the framework developed by Masubuchi et al. in reference \citen{Masubuchi2018}. Once monolayers are identified, they are sandwiched between thin layers of hBN and stamped onto p-type silicon substrates with $280\,$nm of thermally grown $\mathrm{SiO_{2}}$, using conventional polymer stamping techniques\cite{Purdie2018}. This process is performed in an inert argon atmosphere maintained inside a glovebox system with $\mathrm{O_{2}}$ and $\mathrm{H_{2}O}$ concentrations limited to less than $0.1\,$ppm in order to minimize the degradation of the material. Following the formation of a stack of hBN encapsulated monolayer $\mathrm{WTe_{2}}$, the device is transferred into atmosphere and contacts are processed using conventional electron beam lithography techniques. To contact the encapsulated monolayer we employ an edge contact method (Supplementary Methods 1) often used in graphene devices\cite{Wang2013,Amet2016}. The contact areas are etched in a homemade parallel plate plasma etcher in a separate chamber of the vacuum system used for the deposition of contacts to avoid any exposure to air between etching and contact deposition. We use a $\mathrm{SF_{6}}:\mathrm{Ar}$ mixture ($30:8\,$sccm) with a pressure $\sim 1\,$Pa and an rf power of $50-60\,$W. The etching rate is estimated as approximately $1\,$nm/sec. Following etching, contacts are formed through dc-magnetron sputtering of a $2\,$nm Pd layer followed by $120\,$nm of niobium titanium ($\mathrm{NbTi}$ $\mathrm{80/20\%wt}$) with $T_{\mathrm{c}}\sim 8.5\,$K. Devices that are measured immediately following contacting display relatively poor contacts but it is found that later processing improves the contacts likely through the annealing of $\mathrm{Pd}$ into the $\mathrm{WTe_{2}}$ forming $\mathrm{PdTe_{x}}$ regions (see Supplementary Note 1). We have attempted similar edge contacting methods on the monolayer using superconducting alloys such as $\mathrm{NbTi}$ and $\mathrm{MoRe}$, however in these cases we have found poor superconducting properties at the contacts (see Supplementary Note 2a). Efforts within our laboratory to stamp monolayer $\mathrm{WTe_{2}}$ onto $\mathrm{Pd}$ contacts do not result in transparent contacts upon annealing, and so we speculate that the interface between $\mathrm{Pd}$ and monolayer $\mathrm{WTe_{2}}$ produced through sputter deposition is a key factor in producing acceptable contacts. It is advantageous to form superconducting contacts to the monolayer $\mathrm{WTe_{2}}$ even if the junction being probed is formed from gating of the material as it permits the coupling of the system to an external superconducting circuit. The superconducting contacts could thus allow fabrication of SQUID geometry devices for phase control of the $\mathrm{WTe_{2}}$ junction, an important requirement for existing proposals for control of topological states\cite{ShuoPhysRevB2013}.

\subsection{Electrical measurements}

Low temperature measurements were performed in a dilution refrigerator (Oxford Instruments Kelvinox 400) at a base temperature of $\sim 15\,$mK. Transport measurements were collected using a battery powered current source and differential voltage amplifier (Delft IVVI instrumentation). Magnetic response and Shapiro measurements were performed using triangular ramps of current from the current source with voltage measurement readout from a synchronized oscilloscope (Rohde $\&$ Schwarz RTO-1022). For each trace $100-200$ individual current-voltage traces were averaged. In all measurements a correction is made for the remnant field within the superconducting magnet, the magnitude of which is dependent on the history of previous magnetic field sweeps but typically less than $3\,$mT. Shapiro measurements were performed by irradiating the device from a small antenna, made from a length of semi-rigid coaxial line with an open end, placed within a few millimeters of the sample. A continuous microwave tone was applied from a signal generator (Hewlett-Packard 83650B). All microwave powers ($P_{\mathrm{RF}}$) shown indicate the power at the output of the source and do not account for the attenuation of fridge lines.

\subsection{Extraction of critical currents}

To achieve estimates of the junction critical current from our sometimes noisy experimental $I(V)$ traces we employ a technique based on extraction of the variance of the first order moment of the voltage in a sliding window proposed by Maurand \textit{et al.} in reference \citen{Paper:VarianceMethod}, itself based on a digital filter proposed by Liu \textit{et al.} in reference \citen{LiuSignal1995}. Briefly, the first order moment of the measured voltage data $\mu(t)$ is evaluated using a rectangular window averaging filter with impulse response $h_{1}(t)=\mathrm{Rect}(t/L_{\mathrm{1}})/L_{\mathrm{1}}$, where $L_{\mathrm{1}}$ is the length and $\mathrm{Rect}$ is the normalized rectangular function. The variance of this moment is then obtained in a second rectangular moving average window, $\sigma_{\mu}^{2}(t)=<\mu(t)^{2}>-<\mu(t)>^{2}$ with impulse $h_{2}(t)=\mathrm{Rect}(t/L_{\mathrm{2}})/L_{\mathrm{2}}$. Under typical measurement conditions with sampling rate of $10\,\mathrm{kS/s}$ and current ramp rates $\sim 100-250\,\mathrm{nA/s}$, we typically employ $L_{\mathrm{1}}=L_{\mathrm{2}}=20$ and then a threshold method to detect the sharp resulting steps or peak at the supercurrent to normal branch transitions.

\subsection{BCS fitting of temperature dependent features}

An interpolation function is utilized as it is convenient for least square curve-fitting. The interpolation function for the BCS gap \cite{Gross1986} is given by 

\begin{equation}
\Delta_{\mathrm{gap}}(T)=\Delta_0\tanh\left(1.74\sqrt{\frac{T_{\mathrm{c}}}{T}-1}\right)
\end{equation}

\noindent where $\Delta_{\mathrm{0}}\approx1.76k_{\mathrm{B}}T_{\mathrm{c}}$ is the low-temperature limit of the BCS gap relation for an s-wave superconductor. Transport features are fitted with $\Delta_{\mathrm{gap}}(T)$ and a scaling prefactor $A$. Both $T_{\mathrm{c}}$ and $A$ are free fitting parameters, allowing for the evaluation of several different critical temperatures used to make judgements as to the origin of the features discussed. In figure \ref{Fig2} (d), we find features are scaled with a single $T_{\mathrm{c}}$, which we attribute to the diffused $\mathrm{Pd}$ regions within the device. It has been reported \cite{Kononov2021} that the $\mathrm{PdTe_{x}}$ formed during processing has $T_{\mathrm{c}}\approx 1.2\,$K, but this value can be lowered depending on the details of fabrication \cite{Ohtomo2022} and the geometry of the formed Josephson weak-link \cite{EndresPhysRevMaterials2022}. In figure \ref{Fig2} (e), we find that the effect of raising the gate voltage is to introduce a second feature with $T_{\mathrm{c}}\approx 1.7\,$K. Given the emergent nature of this feature with gate voltage, we attribute it to the superconducting transition of the monolayer $\mathrm{WTe_{2}}$. 

\subsection{Modeling of a 2D Josephson junction}

The effect of flux penetration in the leads of 2D superconductor junctions has been explored in a number of theoretical studies\cite{Clem2010,Kogan2001,Rosenthal1991,Clem2011,PekkerPhysRevB2005} and can be captured with a relatively simple treatment in the case where the leads are of a simple rectangular shape\cite{Clem2010}. Following reference \citen{Clem2010} we consider two leads with width $W$ and length $L/2$ forming a junction of length $l$, as shown schematically in figure \ref{SimExample2} (b). The current density flowing in the $x$ direction is $j_{\mathrm{x}}(y)=j_{\mathrm{c}}(y)\sin(\Delta\gamma(y))$, where $j_{\mathrm{c}}(y)$ describes the current density distribution across the width of junction and $\Delta\gamma(y)$ is the gauge-invariant phase difference between the two leads. Here we assume a purely sinusoidal current phase relationship. For the rectangular leads considered the phase difference can be expressed using the expansion\cite{Clem2010}

\begin{equation}
\Delta\gamma(y)=\Delta\gamma_{\mathrm{0}} + \frac{16\pi B}{\phi_{\mathrm{0}}W}\sum_{n=0}^{\infty} \frac{(-1)^{n}}{k_{n}^{3}}\tanh(k_{n}L/2)\sin(k_{n}y),
\end{equation}

\noindent where $\phi_{\mathrm{0}}=h/2e$ is the magnetic flux quantum, $B$ is the applied magnetic field and $k_{n}=2\pi(n+0.5)/W$. The maximum supercurrent through the junction is then given as 

\begin{equation}
\frac{I_{\mathrm{c}}(B)}{I_{\mathrm{c}}(0)}=\left\lvert \int_{-W/2}^{W/2} j_{\mathrm{c}}(y)\cos(\Delta\gamma(y))dy \right\rvert.
\end{equation}

\noindent The term $\Delta\gamma_{\mathrm{0}}$ includes the superconducting phase difference between the leads ($0\leq \phi\leq 2\pi$) and the phase winding due to flux penetrating the actual junction area ($2\pi lyB/\Phi_{\mathrm{0}}$). We assume the junction length is $l=100\,$nm, as defined by the center gate geometry, and find that the oscillation period of the magnetic response is mostly determined by the geometry of the gated superconducting leads. This is because magnetic flux penetrates both elements. For addition simulations with alteration of junction geometry and position  for a fixed device size see Supplementary Note 6b.

\medskip
\textbf{Supporting Information} \par 
Supporting Information is available from the Wiley Online Library or from the author.

\medskip
\textbf{Acknowledgements} \par
This work was supported by JSPS KAKENHI grant Numbers JP19H00867, JP19H05610, JP19H05790, JP20H00354, JP21H05233, JP23H02052, JP21H04652, JP21K18181 and JP21H05236. P.Z. acknowledges support from RIKENs SPDR fellowship. K.W. and T.T. acknowledge support from World Premier International Research Center Initiative (WPI), MEXT, Japan.

\medskip
\textbf{Conflict of Interest} \par
The authors declare no conflicts of interest.


\medskip
\bibliography{WTe2-RSD}

\end{document}


\title{Gate defined Josephson weak-links in monolayer WTe\textsubscript{2}\\ (Supplementary Material)}

\author{Michael D. Randle}
\affiliation{Advanced Device Laboratory, RIKEN, 2-1 Hirosawa, Wako, Saitama 351-0198, Japan}

\author{Masayuki Hosoda}
\affiliation{Fujitsu Research, Fujitsu Ltd., Atsugi, Kanagawa 243-0197, Japan}

\author{Russell S. Deacon}
\affiliation{RIKEN Center for Emergent Matter Science (CEMS), 2-1 Hirosawa, Wako, Saitama 351-0198, Japan}
\affiliation{Advanced Device Laboratory, RIKEN, 2-1 Hirosawa, Wako, Saitama 351-0198, Japan}
\email[Corresponding author: ]{russell@riken.jp}

\author{Manabu Ohtomo}
\affiliation{Fujitsu Research, Fujitsu Ltd., Atsugi, Kanagawa 243-0197, Japan}

\author{Patrick Zellekens}
\affiliation{RIKEN Center for Emergent Matter Science (CEMS), 2-1 Hirosawa, Wako, Saitama 351-0198, Japan}

\author{Kenji Watanabe}
\affiliation{Research Center for Electronic and Optical Materials, National Institute for Materials Science, 1-1 Namiki, Tsukuba, Ibaraki, 305-0044, Japan}

\author{Takashi Taniguchi}
\affiliation{Research Center for Materials Nanoarchitectonics, National Institute for Materials Science, 1-1 Namiki, Tsukuba, Ibaraki, 305-0044, Japan}

\author{Shota Okazaki}
\affiliation{Laboratory for Materials and Structures, Tokyo Institute of Technology, 4259 Nagatsuta, Midori-ku, Yokohama 226-8503, Japan}

\author{Takao Sasagawa}
\affiliation{Laboratory for Materials and Structures, Tokyo Institute of Technology, 4259 Nagatsuta, Midori-ku, Yokohama 226-8503, Japan}

\author{Kenichi Kawaguchi}
\affiliation{Fujitsu Research, Fujitsu Ltd., Atsugi, Kanagawa 243-0197, Japan}

\author{Shintaro Sato}
\affiliation{Fujitsu Research, Fujitsu Ltd., Atsugi, Kanagawa 243-0197, Japan}

\author{Koji Ishibashi}
\affiliation{RIKEN Center for Emergent Matter Science (CEMS), 2-1 Hirosawa, Wako, Saitama 351-0198, Japan}
\affiliation{Advanced Device Laboratory, RIKEN, 2-1 Hirosawa, Wako, Saitama 351-0198, Japan}

\maketitle 
\pagenumbering{arabic}

\color{black}

\renewcommand{\thesection}{S\Roman{section}}
\renewcommand{\thesubsection}{\Alph{subsection}.}
\renewcommand{\theequation}{S\arabic{equation}}
\renewcommand{\thefigure}{S\arabic{figure}}
\renewcommand{\figurename}{Figure}
\setcounter{figure}{0}
\setcounter{section}{0}

\tableofcontents
 
\justifying 

\section{Supplementary Methods 1 - Detailed fabrication procedure}

Throughout fabrication all electron beam lithography processes utilize the same steps. We spin PMMA A6 950K at $4000\,$rpm for $50\,$seconds with a ramp of $3\,$seconds then bake on a hotplate at $180^{\circ}$C. The patterns are written and then developed in an MIBK:IPA (1:2) solution at $20^{\circ}$C for $55\,$seconds then rinsed in IPA. After metal deposition, the processing resist is removed through lift-off in Acetone using gentle agitation from a pipette followed by an IPA rinse. Between processing steps, the devices are stored inside the inert argon environment of a glovebox with $<0.1\,$ppm of both $\mathrm{O_{2}}$ and $\mathrm{H_{2}O}$. In the summarized recipe below, processing steps performed in the glovebox and in-situ within a vacuum system are indicated by bracketed items.

\begin{minipage}{15cm}

\begin{tikzpicture}[remember picture, overlay]
  \node (rightenum) at (1.01\textwidth,0) {};
  \draw [decorate, decoration={brace}, ultra thick] ($({pic cs:top1} -| rightenum) + (0, 1em)$) -- ({pic cs:bot1} -| rightenum) node [midway, right] {Glovebox};
  \draw [decorate, decoration={brace}, ultra thick] ($({pic cs:top2} -| rightenum) + (0, 1em)$) -- ({pic cs:bot2} -| rightenum) node [midway, right] {Vacuum Chamber};
\end{tikzpicture}

\begin{enumerate}
\item Device chips are initially prepared through electron beam lithography and electron beam deposition of markers for later writing steps, alignment markers for flake stack positioning and bonding pads for later connection to contacts. 
\item \tikzmark{top1} $\mathrm{WTe_{2}}$ and $\mathrm{hBN}$ are exfoliated onto silicon with $280\,$nm of thermally grown oxide. 
\item An automated search is performed to locate suitable monolayer $\mathrm{WTe_{2}}$ for stacking using the methods detailed by Masubuchi \textit{et al.} in reference \citen{Masubuchi2018}. 
\item Stacking is performed using a polydimethylsiloxane (PDMS) stamp (SYLGARD 184) covered by a polycarbonate (PC) film ($6\,\%$ PC in chloroform) using the process detailed in reference \citen{Purdie2018}. Material is picked-up at temperatures between $80-100^{\circ}$C.
\item  Following pickup the stack is deposited onto the device substrate by adhering and detatching the PC film on the surface at $\sim 180^{\circ}$C. \tikzmark{bot1}
\item The device is transferred from the glovebox to atmosphere and PC is removed through soaking in chloroform overnight followed by cleaning in acetone and IPA.
\item (optional) Additional cleaning maybe performed through annealing in vacuum at $350\,\mathrm{^{\circ} C}$ for five hours. 
\item Atomic force microscopy is used to confirm the flake position, thickness and identify usable areas.
\item Contacts are defined via electron beam lithography using PMMA A6 resist and developed in $\mathrm{IPA:MIBK}$ solution ($\mathrm{1:2}$). 
\item \tikzmark{top2} Device is transferred to a UHV thin film deposition system. Contact regions are etched in-situ using home-made parallel plate rf-plasma system in a $\mathrm{SF_{6}:Ar}$ mixture ($30:8\,$sccm). Pressure is set at approximately $1\,$Pa and rf-power is $\sim 50-60\,$W depending on the specifics of the manually tuned matching.
\item In-situ the device is transferred to the deposition chamber and $2\,$nm of palladium ($\mathrm{Pd}$) is sputtered followed by sputter deposition of $120\,$nm of niobium titanium ($\mathrm{NbTi}\,80/20\%$ by weight). \tikzmark{bot2}
\item Following lift-off of the contacts in Acetone the center surface-gate is defined via electron beam lithography and $\mathrm{Ti/Au}$ ($3/80\,$nm) is deposited by electron beam deposition.
\item Atomic layer deposition (ALD) is used to grow a $20\,$nm alumina ($\mathrm{Al_{2}O_{3}}$) film that isolates the upper left and right surface-gates from the center gate and contacts. The ALD growth is performed at $150^{\circ}$C, using trimethylaluminum ($\mathrm{TMA}$) and $\mathrm{H_{2}O}$ precursors, with a total growth time of approximately one hour including the temperature stabilization and vacuum pumping wait times.
\item Final top-gates are fabricated through electron beam lithography and deposition of $\mathrm{Ti/Au}$ ($3/150\,$nm).
\item Following lift-off of the final surface-gates devices are quickly bonded and loaded into the dilution refrigerator to be measured.
\end{enumerate}

\end{minipage}


\section{Supplementary Note 1 - Effect of annealing \NoCaseChange{Pd}}

\begin{figure*}[h]
\centering
\includegraphics[width=12 cm,angle=0]{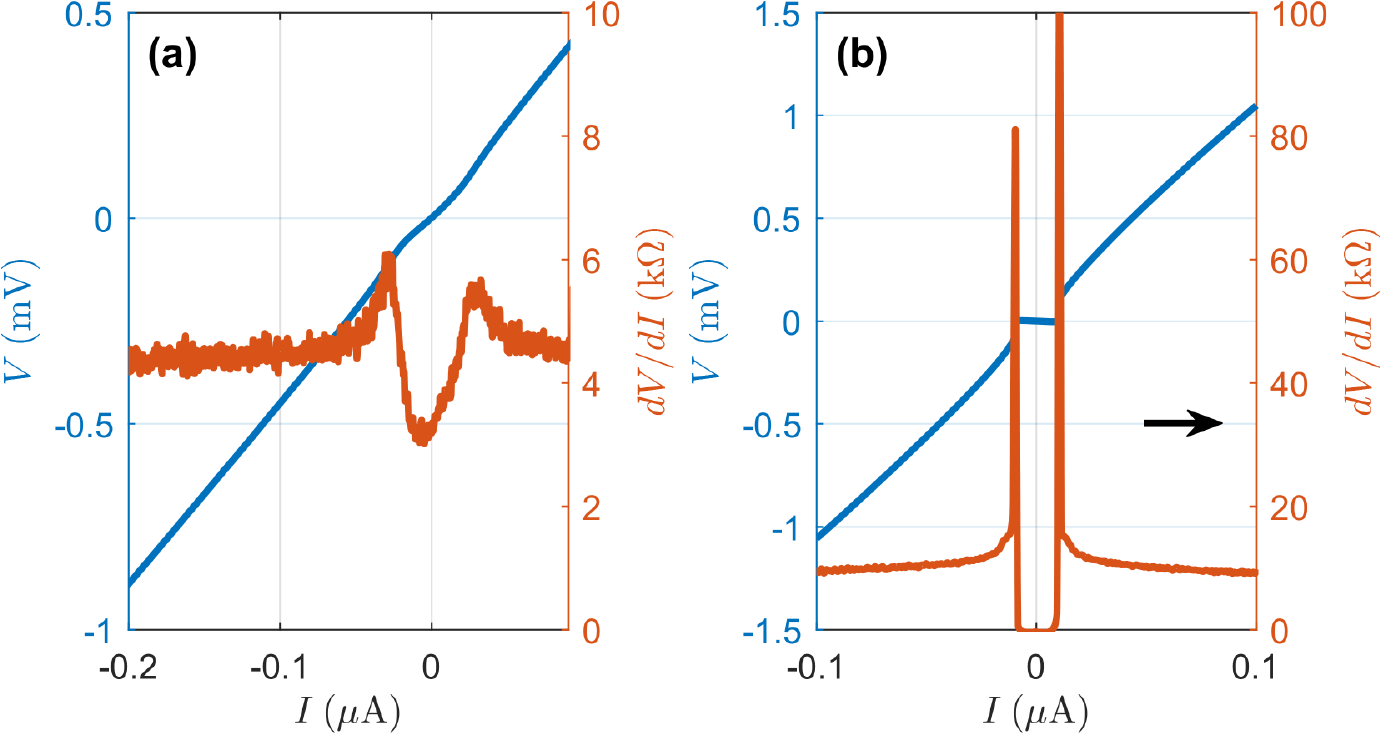}
\caption{\textbf{Effect of processing on Pd edge contacts}. Measurement of a contact pair with a nominal separation of $100\,$nm before (a) and after (b) processing of surface-gates. In both measurements $V_{\mathrm{bg}}$ and top-gates are grounded. The black arrow indicates the current sweep direction.}
\label{FigAnneal1}
\end{figure*}

During device fabrication, we form contacts using $2\,$nm of sputtered $\mathrm{Pd}$ and $120\,$nm of sputtered $\mathrm{NbTi}$ ($T_{\mathrm{c}}\sim 8.5\,$K). Initially, contacts are relatively poor, but we find that after later processing of the surface-gates that the contacts are improved (steps 9 and 10 of the detailed fabrication process in Supplementary Methods 1). In the end, this process includes two hotplate baking steps both at $180^{\circ}$C for a total of $6\,$minutes, in addition to the atomic layer deposition growth step that is performed at $150^{\circ}$C for approximately $1$ hour. We illustrate this by measurement of a contact pair with separation of $\sim 100\,$nm before and after the formation of surface-gates (figure \ref{FigAnneal1}). Prior to the additional thermal load of the surface-gate fabrication, we observe a small conductance peak about zero current (figure \ref{FigAnneal1} (a)). However, following the additional fabrication a fully formed supercurrent branch is observed (figure \ref{FigAnneal1} (b)). We conclude that the improvement is caused by the diffusion of $\mathrm{Pd}$ into the $\mathrm{WTe_{2}}$ due to the additional thermal load during processing. This leads to the formation of superconducting $\mathrm{PdTe_{x}}$ as has been recently reported in multilayer materials stamped onto $\mathrm{Pd}$ contacts\cite{Kononov2021,Ohtomo2022,EndresPhysRevMaterials2022}.

\section{Supplementary Note 2 - Additional devices}

In addition to the device discussed in the main text, we present measurements on two additional devices with different fabrication methods and geometry. The first utilizes sputtered $\mathrm{MoRe}$ edge contacts which are relatively poor as compared to the $\mathrm{Pd}$ contacts discussed earlier. While the sputtered $\mathrm{MoRe}$ does not exhibit superconducting transport at the interface with $\mathrm{WTe_{2}}$, the contacts are sufficient to use as measurement probes in a 4-terminal configuration to study a Josephson weak-link formed through gating (Supplementary Note 2a). Next, we show transport measurements in a monolayer flake of $\mathrm{MoRe}$ in which a narrow physical constriction in the flake produces quantum dot transport features with well defined coulomb blockade. When gated into the superconducting state the device displays transport features indicative of superconducting transport through a quantum dot device\cite{DefranceschiNatureNano2010,KanaiPhysRevB2010} (Supplementary Note 2b).


\subsection{Supplementary Note 2a - Second gate defined junction device}

\begin{figure*}[t]
\centering
\includegraphics[width=11 cm,angle=0]{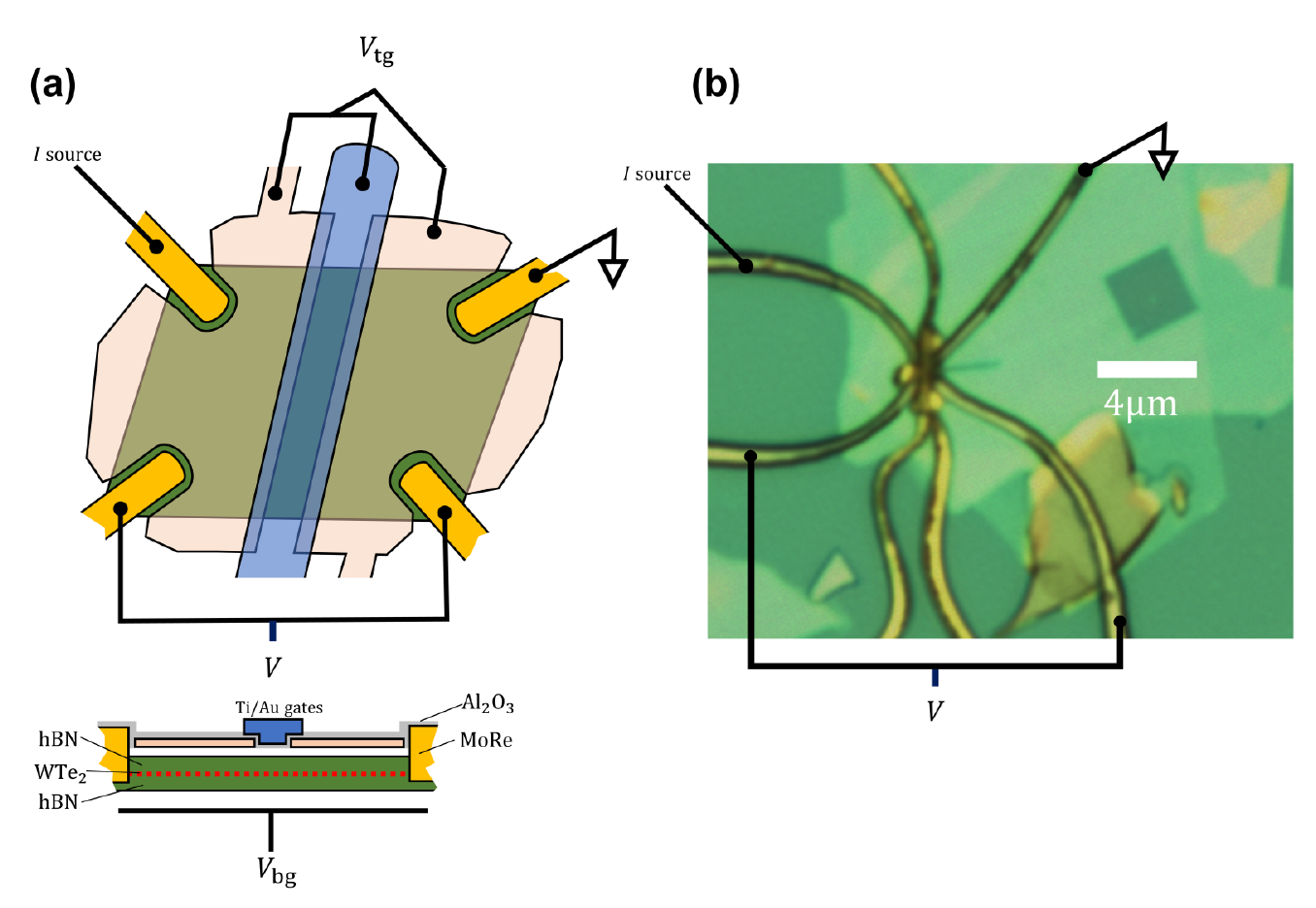}
\caption{\textbf{Schematic of second gate defined junction device}. (a) Schematic of the device design in which all gates are operated with same gate bias $V_{\mathrm{tg}}$. Edge contact are formed from $\mathrm{MoRe}$ without the $\mathrm{Pd}$ layer employed in the main text device. The center top-gate is isolated from the lower gates using a layer of ALD deposited $\mathrm{Al_{2}O_{3}}$. (b) Optical image of the device.}
\label{FigSuppdeviceBSchematic}
\end{figure*}

\begin{figure*}[h]
\centering
\includegraphics[width=11 cm,angle=0]{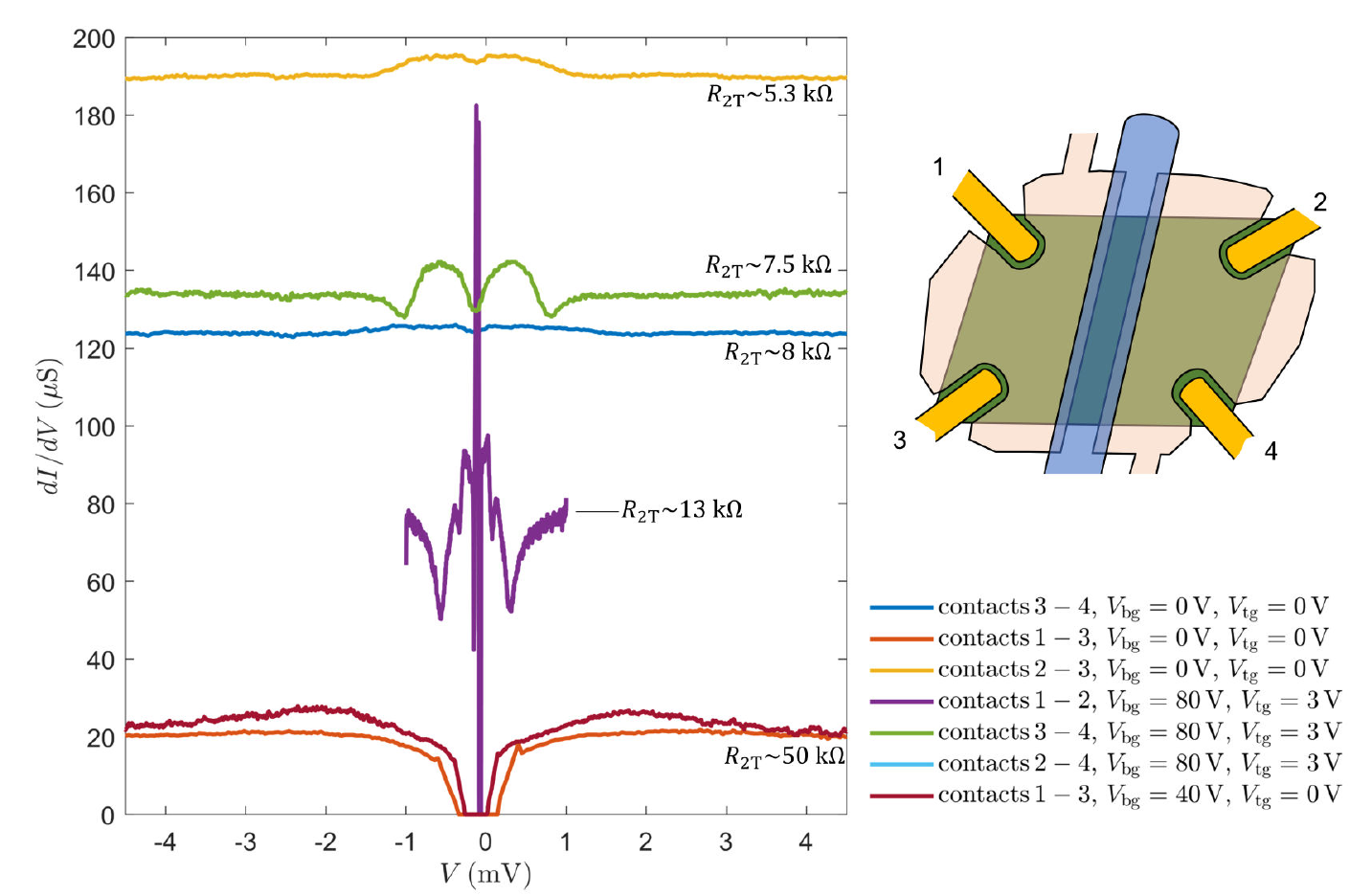}
\caption{\textbf{Two terminal conductance for contact pairs}. Plots of $dI/dV$ measured in a two terminal configuration for the indicated contact pairs and different gate conditions. Data has been corrected for the resistance of fridge wiring and filters. Contact no. 1 is found to have tunnelling behavior that prevent its use as a voltage probe.}
\label{Supp:TwoTerminalConductances}
\end{figure*}

Here we present data for a second gate defined junction device utilizing sputtered $\mathrm{MoRe}$ edge-contacts ($T_{\mathrm{c}}\sim 9.5\,$K) again processed using an in-situ etching technique but without $\mathrm{Pd}$ deposition prior to the sputtering of superconducting contacts. A schematic of the device is shown in figure \ref{FigSuppdeviceBSchematic}. During measurement, all the surface gates are common due to an unintentional short between the center top-gate and the left gate. In contrast to the device discussed in the main text, the center channel gate was fabricated after the two source/drain gates with the inclusion of an ALD grown $\mathrm{Al_{2}O_{3}}$ dielectric. The contacts formed from $\mathrm{MoRe}$ do not proximitize the $\mathrm{WTe_{2}}$ into the superconducting state and can exhibit large thresholds to conduction at low bias. To demonstrate the range of behavior of contacts we show two terminal (2T) measurements of $dI/dV$ (with corrections for fridge wiring resistances) for some combinations of contacts in figure \ref{Supp:TwoTerminalConductances}. Contact no. 1 has very poor properties with almost totally insulating behavior at low bias. The poor properties of contact no. 1 prevent its use as a voltage probe to study the superconducting transport in the device and as such it is employed as the source for current bias. Other contacts no. 2-4 have better performance. We note that the transport properties of the contacts are improved when the system is gated using the global back-gate. This is particularly evident for contact no. 1. For contacts no. 2-4 we roughly evaluate contact resistances in the range $R_{\mathrm{contact}}\sim 3-5\,\mathrm{k\Omega}$.

\begin{figure*}[p]
\centering
\includegraphics[width=12 cm,angle=0]{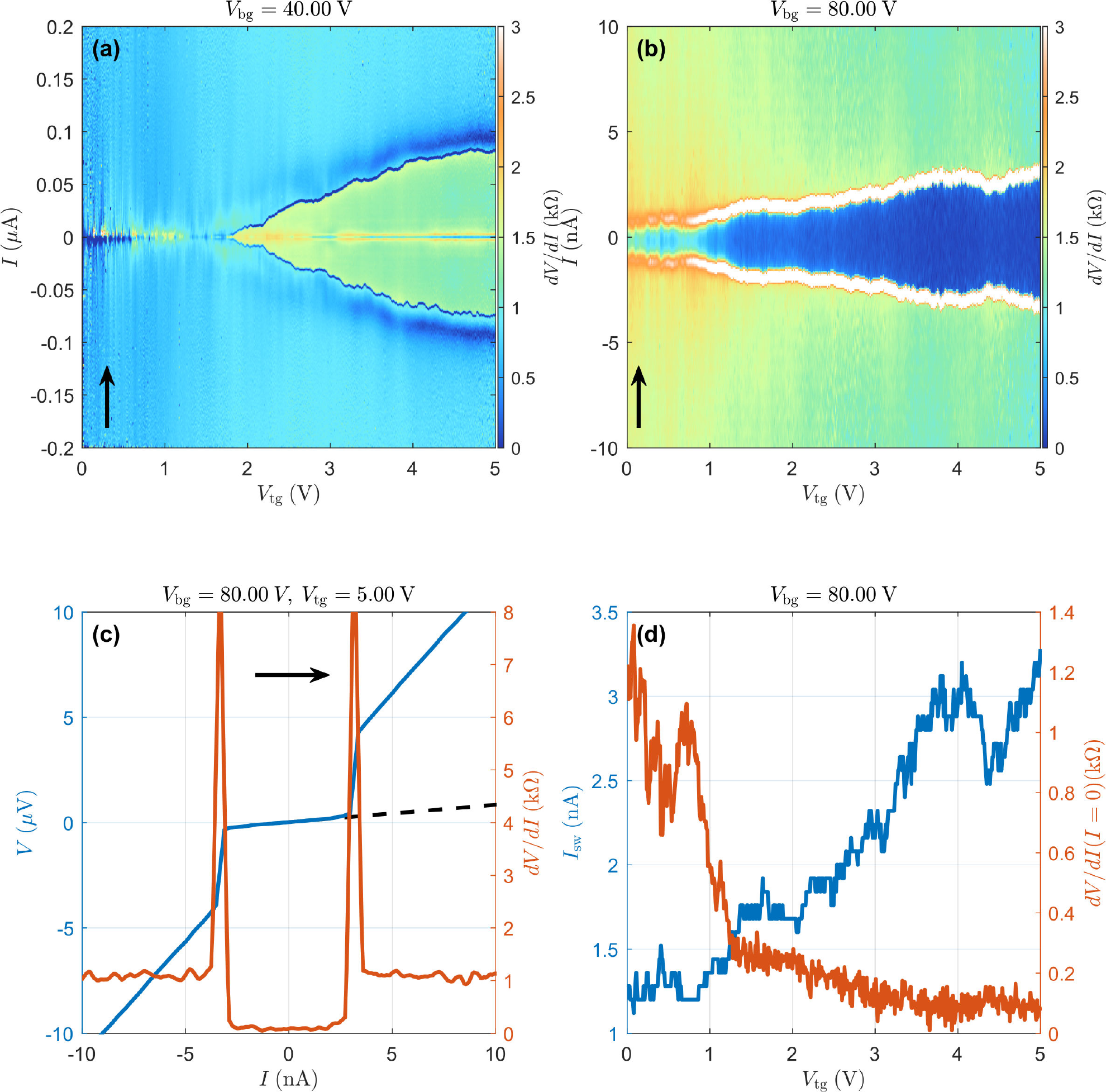}
\caption{\textbf{Gate response in a second device with $\bm{\mathrm{MoRe}}$ edge contacts}. (a) Plot of $dV/dI$ as a function of top-gate $V_{\mathrm{tg}}$ with $V_{\mathrm{bg}}=40\,$V. Above $V_{\mathrm{tg}}\sim 2\,$V the emergence of a superconducting gap can be observed. (b) Plot of $dV/dI$ as a function of top-gate $V_{\mathrm{tg}}$ with $V_{\mathrm{bg}}=80\,$V showing the gate control of the supercurrent in the junction. (c) $I-V$ trace at $V_{\mathrm{bg}}=80\,$V and $V_{\mathrm{tg}}=5\,$V. The dashed line indicates a linear fit to the supercurrent branch with resistance $82\,\mathrm{\Omega}$. (d) Evaluated $I_{\mathrm{sw}}$ and supercurrent branch resistance as a function of $V_{\mathrm{tg}}$ for $V_{\mathrm{bg}}=80\,$V. Black arrows indicate the current sweep directions used in the measurements.}
\label{figDeviceBGate1}
\end{figure*}

\begin{figure*}[p]
\centering
\includegraphics[width=12 cm,angle=0]{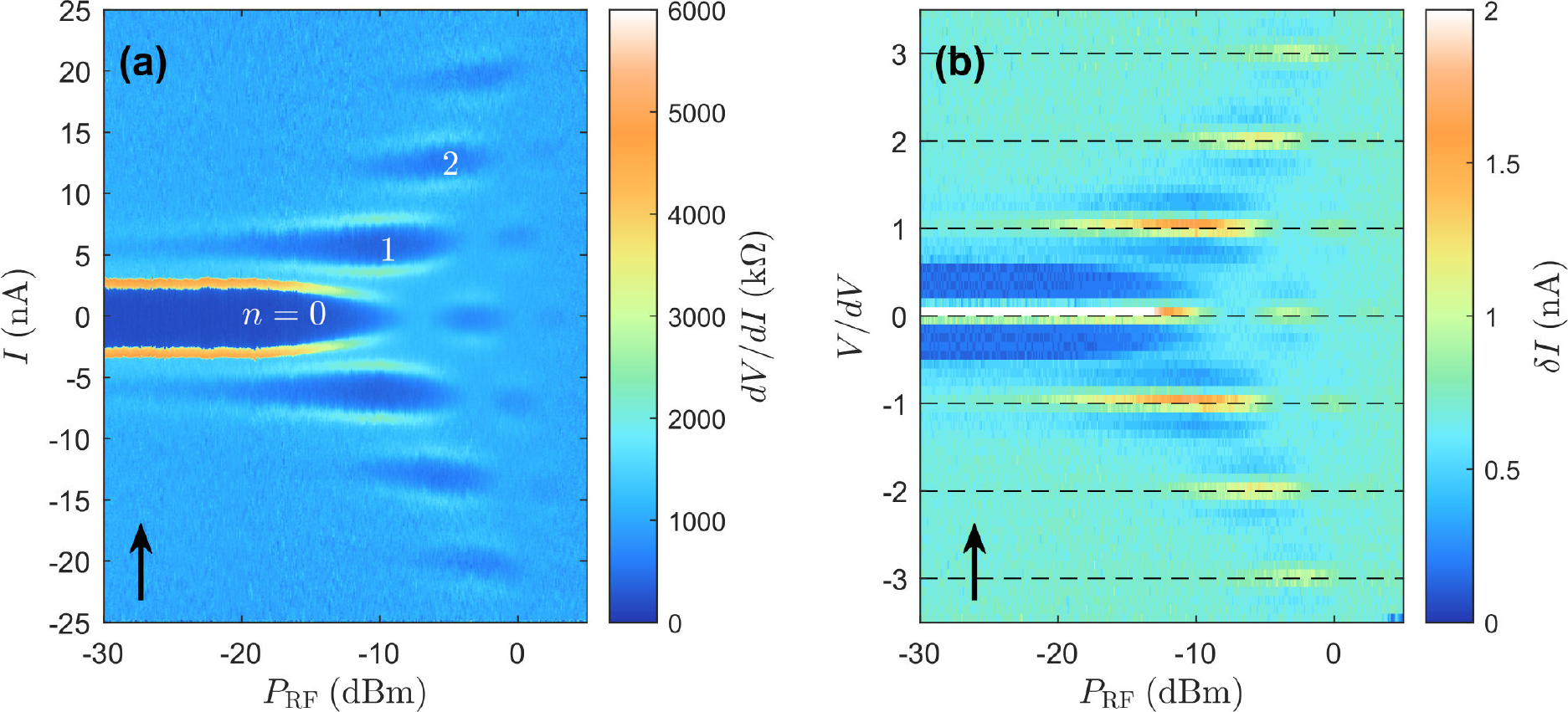}
\caption{\textbf{Shapiro response in a second gate defined junction device with external drive at $\bm{f=3.5\,}$GHz}. (a) Plot of $dV/dI$ as a function of current $I$ and microwave power $P_{\mathrm{RF}}$. The indicated microwave power is at the output from the signal generator and is not corrected for attenuation of lines on the fridge. Settings are $f=3.5\,\mathrm{GHz}$, $V_{\mathrm{tg}}=5.0\,\mathrm{V}$ and $V_{\mathrm{bg}}=80\,\mathrm{V}$. (b) Plot of histograms of binned voltage data from (a) showing the quantization of Shapiro steps. Histogram bins have a width of $\delta V=0.1dV$. Black arrows indicate the current sweep directions used in the measurements.}
\label{figDeviceBShapiro1}
\end{figure*}

To demonstrate a Josephson weak-link by gating the system superconducting using both the surface and back-gate (figure \ref{figDeviceBGate1}) we measure in the four terminal geometry indicated in figure \ref{FigSuppdeviceBSchematic} (a). Figure \ref{figDeviceBGate1} shows gate response of the system to the top-gate at fixed values of back-gate voltage. The use of the back-gate allows use of the contact no. 1 as a source contact through improvements to its performance. As with the main text device we observe the emergence of a superconducting regime induced through gating. Transport $I(V)$ traces display a supercurrent branch with clear switching to a normal branch, figure \ref{figDeviceBGate1} (c). However, the supercurrent branch exhibits a significant finite resistance. Such a resistance can arise from a parasitic series resistance in the measurement circuit which can be evaluated by considering the quantization of Shapiro steps in the microwave response of the system\cite{KamataPhysRevB2018} (figure \ref{figDeviceBShapiro1}). However, we find that the Shapiro response is well quantized, without the need to subtract resistance in the supercurrent branch and so conclude the finite resistance is due not to a series resistance but phase diffusion\cite{IngoldPRB1994,BorzenetsPRL2011,YongNanoResearch2008} possibly caused by Joule heating of the contact regions. The contacts are formed from a superconducting material but display normal transport, so we conclude that each contact region when the system is gated superconducting consists of an super-normal-super (S-N-S) connection which may be the source of hot quasi-particles within the device. The poor performance of the contacts without $\mathrm{Pd}$ may result in significant Joule heating in a thermally isolated normal region. In the device here the very high resistance of contact no. 1 is likely especially an issue.

\clearpage

\subsection{Supplementary Note 2b - Additional device with an unintentional quantum dot}

\begin{figure*}[h]
\centering
\includegraphics[width=13 cm,angle=0]{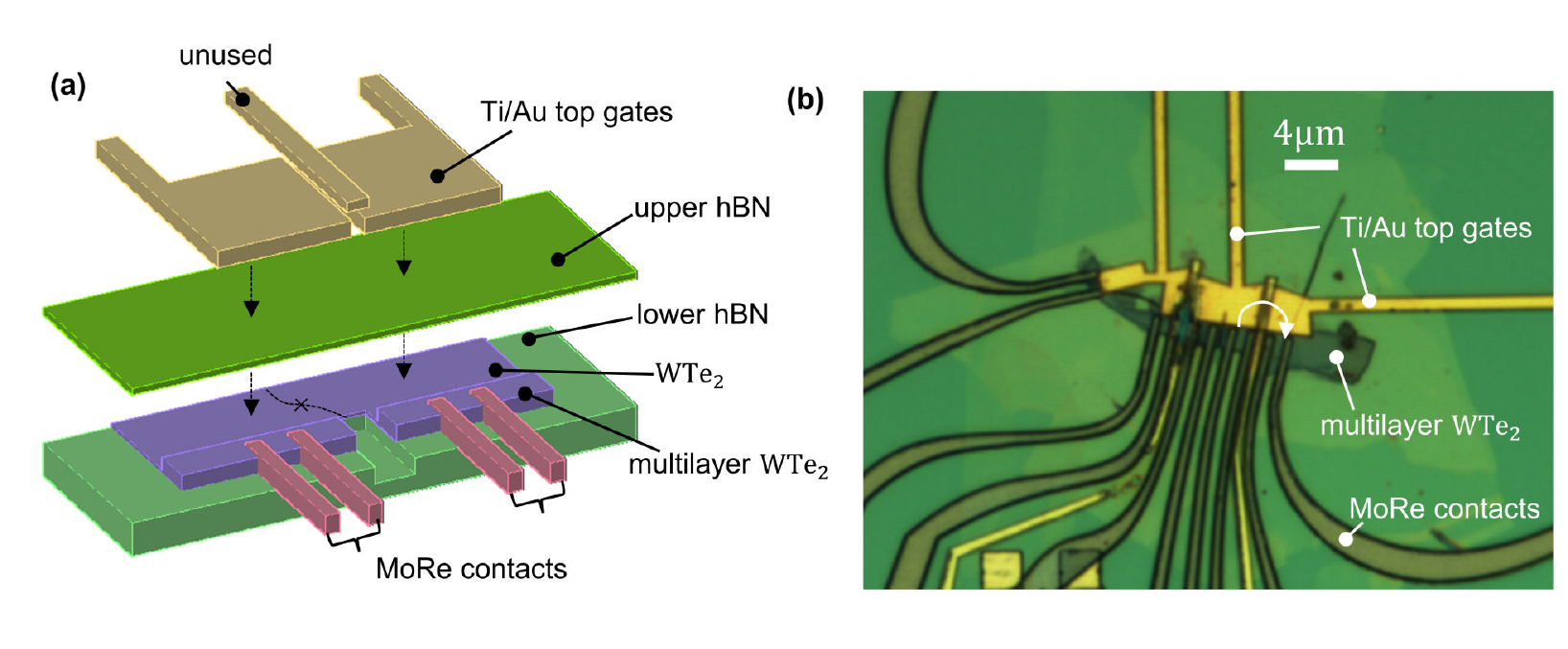}
\caption{\textbf{Schematic and image of device in which quantum dot behaviour is detected}. (a) Schematic of the device design in which contacts are formed to multilayer $\mathrm{WTe_{2}}$. The dashed line and cross indicates the speculated device structure, where a crack in the monolayer $\mathrm{WTe_{2}}$ leads to a small constriction located beneath one of the two top-gates. (b) Optical image of the actual device with transport path used indicated with a white arrow. The channel top gates were not used. Note that a second device to the left of the measured device has a broken top-gate due to electrostatic discharge.}
\label{FigSuppQD1}
\end{figure*}

\begin{figure*}[h]
\centering
\includegraphics[width=13 cm,angle=0]{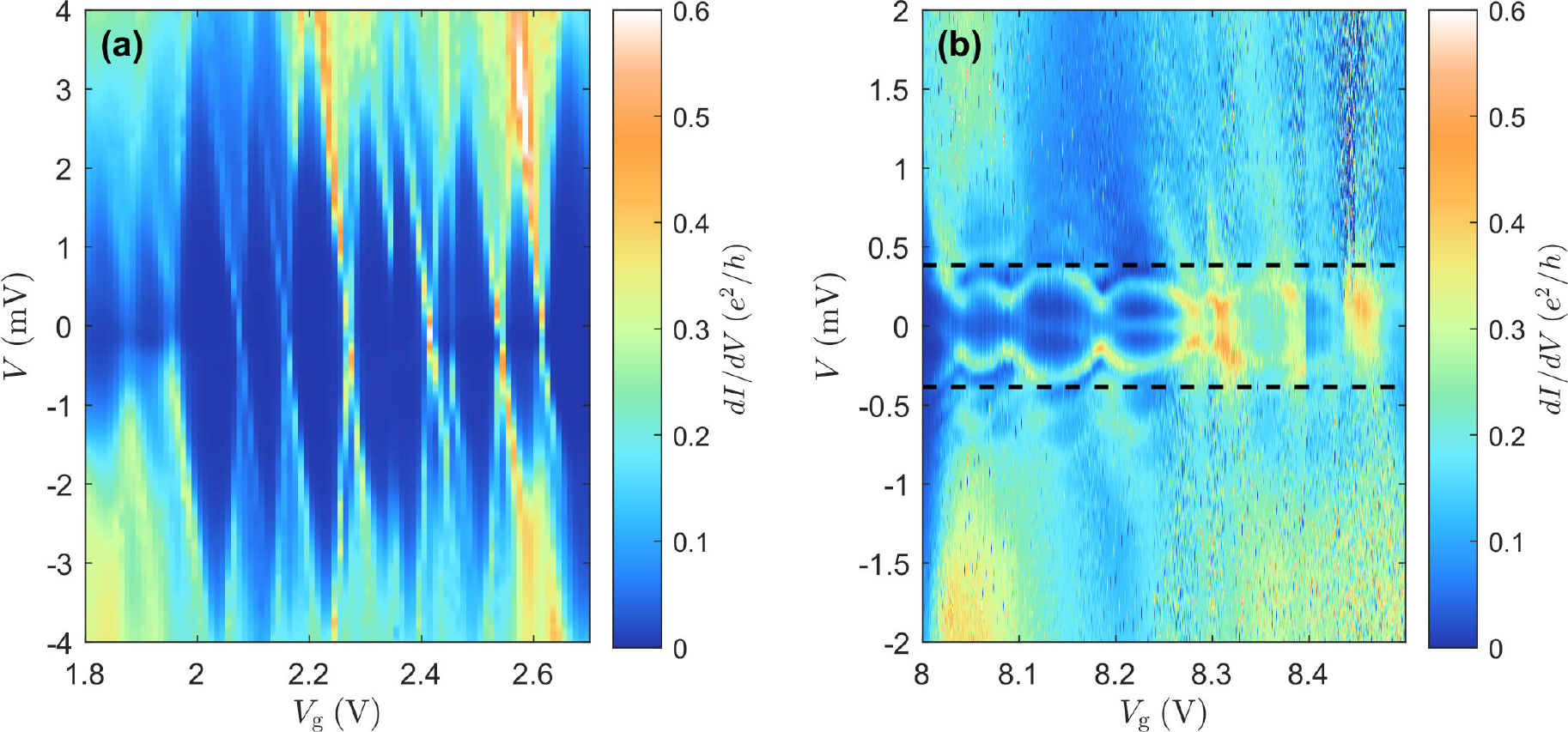}
\caption{\textbf{Measurements of a unintentional quantum dot device}. (a) Coulomb diamonds measured for a relatively low top gate voltage ($V_{\mathrm{g}}$). (b) Signatures of superconducting transport at higher gate voltages in which the $\mathrm{WTe_{2}}$ is in the superconducting state. Dashed lines indicate the gate induced superconducting gap edge at $eV=\pm 2\Delta$, where $\Delta\sim 190\,\mathrm{\mu eV}$ equivalent to a $T_{\mathrm{c}}\sim 1.3\,$K assuming a BCS relation.}
\label{FigSuppQD2}
\end{figure*}

Here we show additional data for a device in which a quantum dot is unintentionally formed within a region that is gated superconducting. The device consists of a monolayer $\mathrm{WTe_{2}}$ flake attached to a section of multilayer material. The large area of multilayer material was used to fabricate contacts and sections of the multilayer material were etched (using the same etch process as for contacting) to produce a channel in the monolayer. Finally top-gates were fabricated on a layer of ALD grown $\mathrm{Al_{2}O_{3}}$. A schematic of the device design is shown in figure \ref{FigSuppQD1}. We speculate that a crack within the monolayer $\mathrm{WTe_{2}}$ leads to a narrow constriction which dominates the transport  forming the observed transport features. The quantum dot measured is located not in the designed device channel (between the two top-gates) but beneath one of the two top-gates. From Coulomb diamonds in the low $V_{\mathrm{g}}$ range (figure \ref{FigSuppQD2} (a)) we evaluate a gate lever-arm of $\alpha=44\,\mathrm{meV/V}$. At high gate voltages we observe features consistent with superconducting transport through a quantum dot (figure \ref{FigSuppQD2}). Direct quasiparticle tunnelling features at the gate induced superconducting gap edge are observed at $eV\sim \pm 2\Delta$, indicating $\Delta\sim 190\,\mathrm{\mu eV}$. Assuming a BCS gap $\Delta=1.76k_{\mathrm{B}}T_{\mathrm{c}}$, giving $T_{\mathrm{c}}\sim 1.3\,$K.

\clearpage

\section{Supplementary Note 3 - Alternative plots of data from main article device}

\begin{figure*}[h]
\centering
\includegraphics[width=17 cm,angle=0]{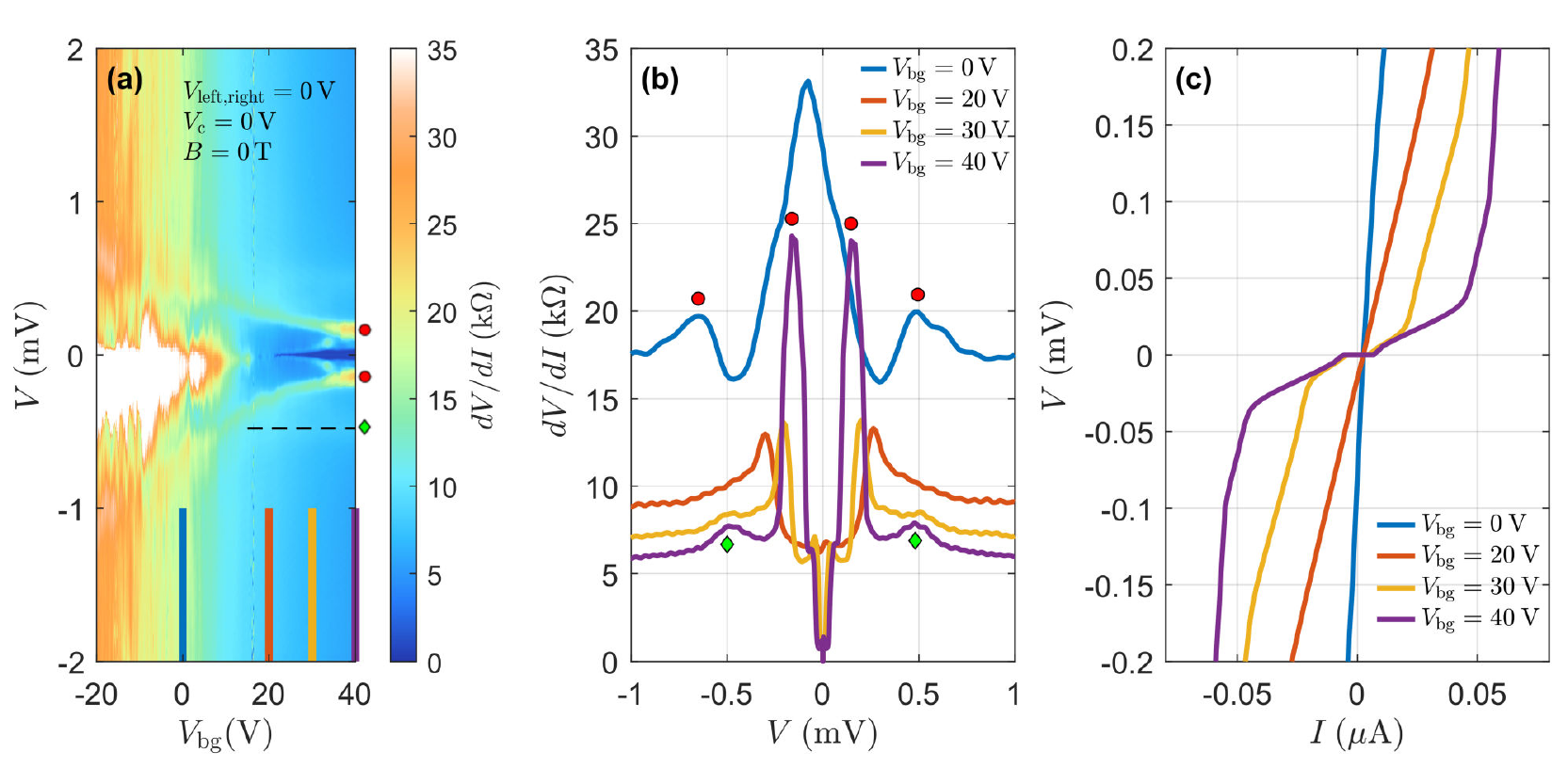}
\caption{\textbf{Influence of global back-gate $\bm{V_{\mathrm{bg}}}$.} (a) Plot of differential resistance ($dV/dI$) as a function of device bias and $V_{\mathrm{bg}}$. All other gate voltages are set to zero. (b) Differential resistance as a function of bias for four values of $V_{\mathrm{bg}}$, indicated by solid lines in (a). Symbol markers in $V_{\mathrm{bg}}=0\,\mathrm{V}$ and $V_{\mathrm{bg}}=40\,\mathrm{V}$ traces indicate the same features as in Figure 2 of the main text for consistency. (c) $I(V)$ traces for four values of $V_{\mathrm{bg}}$, indicated by solid lines in (a). Data plotted is the same as that in figure 2 of the main text only as a function of voltage as opposed to current bias. At each gate voltage a constant background voltage has been subtracted to correct for the offset of the voltage amplifier.}
\label{FigSuppBackGateVoltage}
\end{figure*}

\begin{figure*}[p]
\centering
\includegraphics[width=7 cm,angle=0]{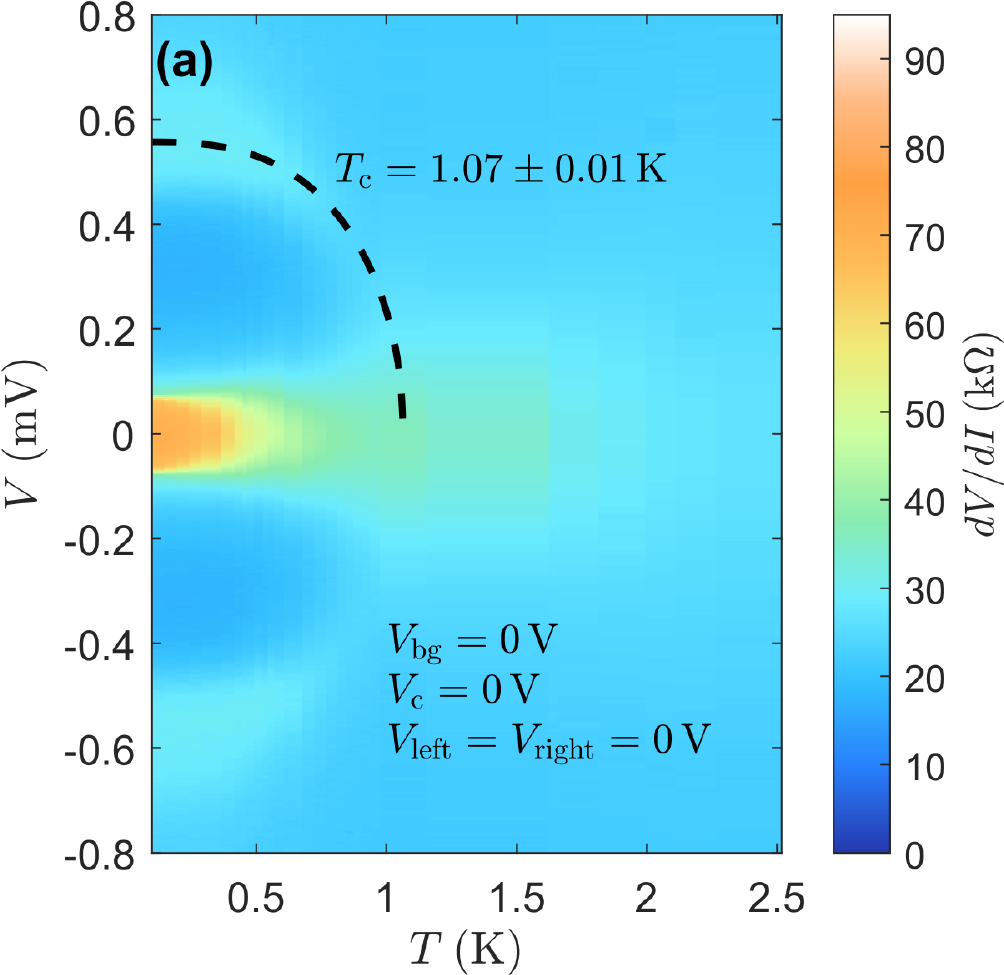}
\includegraphics[width=7 cm,angle=0]{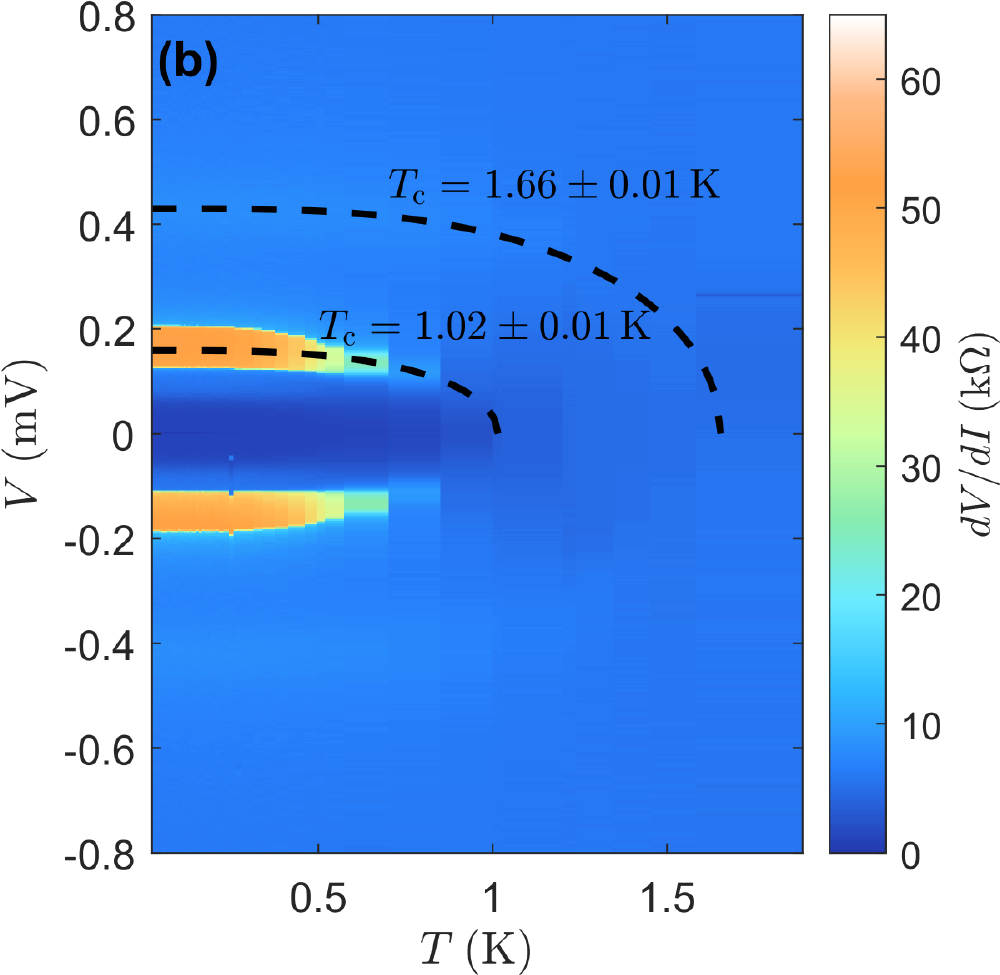}
\caption{\textbf{Temperature dependence of transport features.} Plot of $dV/dI$ as a function of temperature and voltage for $V_{\mathrm{bg}}=0\,$V in (a) and $V_{\mathrm{bg}}=50\,$V in (b). All other gate voltages are set to $0\,$V. Dashed lines indicate fits of transport features to the interpolated BCS gap relationship with indicated critical temperatures $T_{\mathrm{c}}$. Data plotted is the same as that in figure 2 (d) and (e). At each gate voltage a constant background voltage has been subtracted to correct for the offset of the voltage amplifier.}
\label{SuppTemperatureDataVoltages}
\end{figure*}

\begin{figure*}[p]
\centering
\includegraphics[width=17 cm,angle=0]{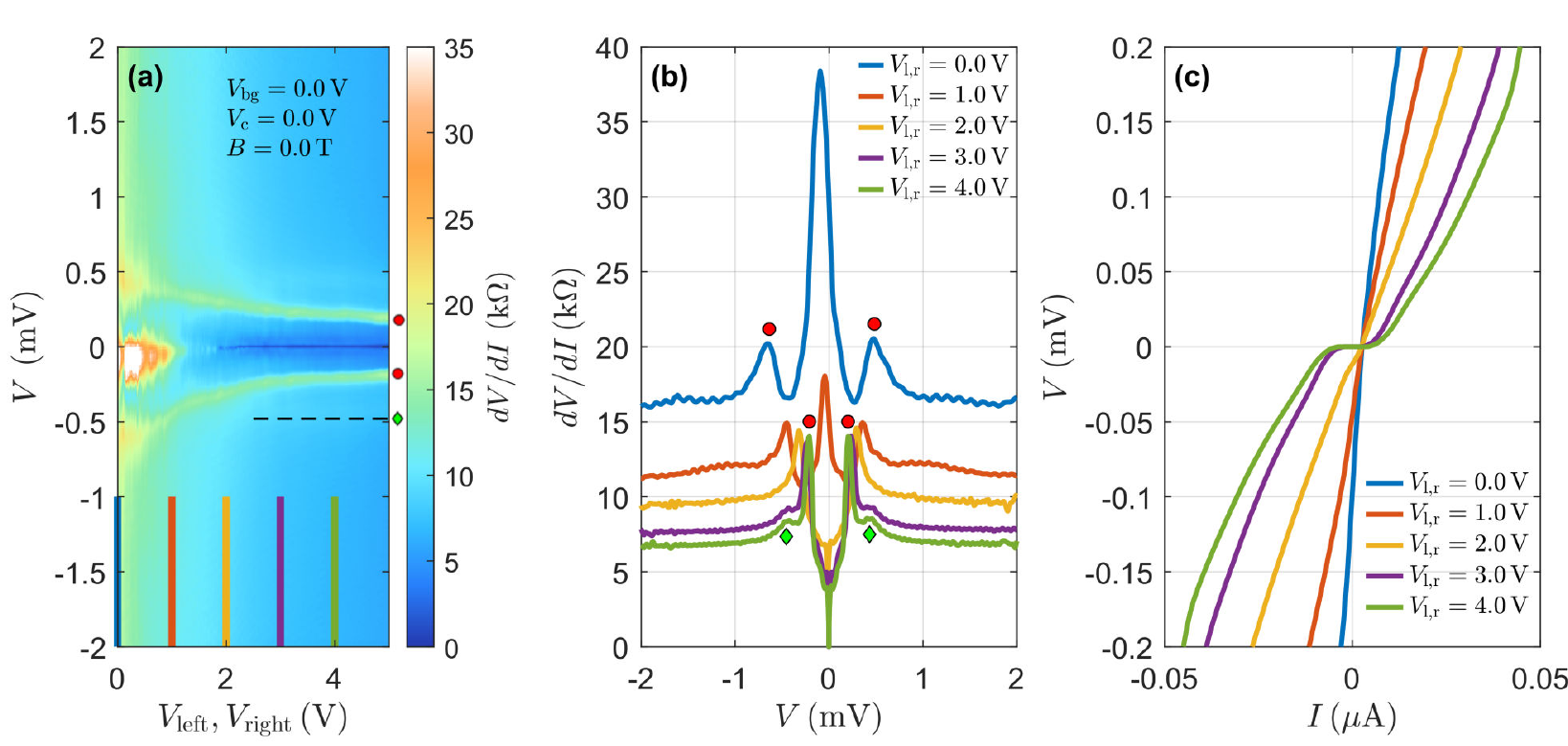}
\caption{\textbf{Influence of top gates $\bm{V_{\mathrm{left}}}$ and $\bm{V_{\mathrm{right}}}$}. (a) Plot of differential resistance ($dV/dI$) as a function of device bias and voltage applied to both $V_{\mathrm{left}}$ and $V_{\mathrm{right}}$. All other gate voltages are set to zero. (b) Differential resistance as a function of bias for five values of $V_{\mathrm{left}}=V_{\mathrm{right}}$, indicated by solid lines in (a). Data plotted is the same as that in figure 3 of the main text. At each gate voltage a constant background voltage has been subtracted to correct for the offset of the voltage amplifier.}
\label{FigSuppSideGateVoltage}
\end{figure*}

\begin{figure*}[p]
\centering
\includegraphics[width=17 cm,angle=0]{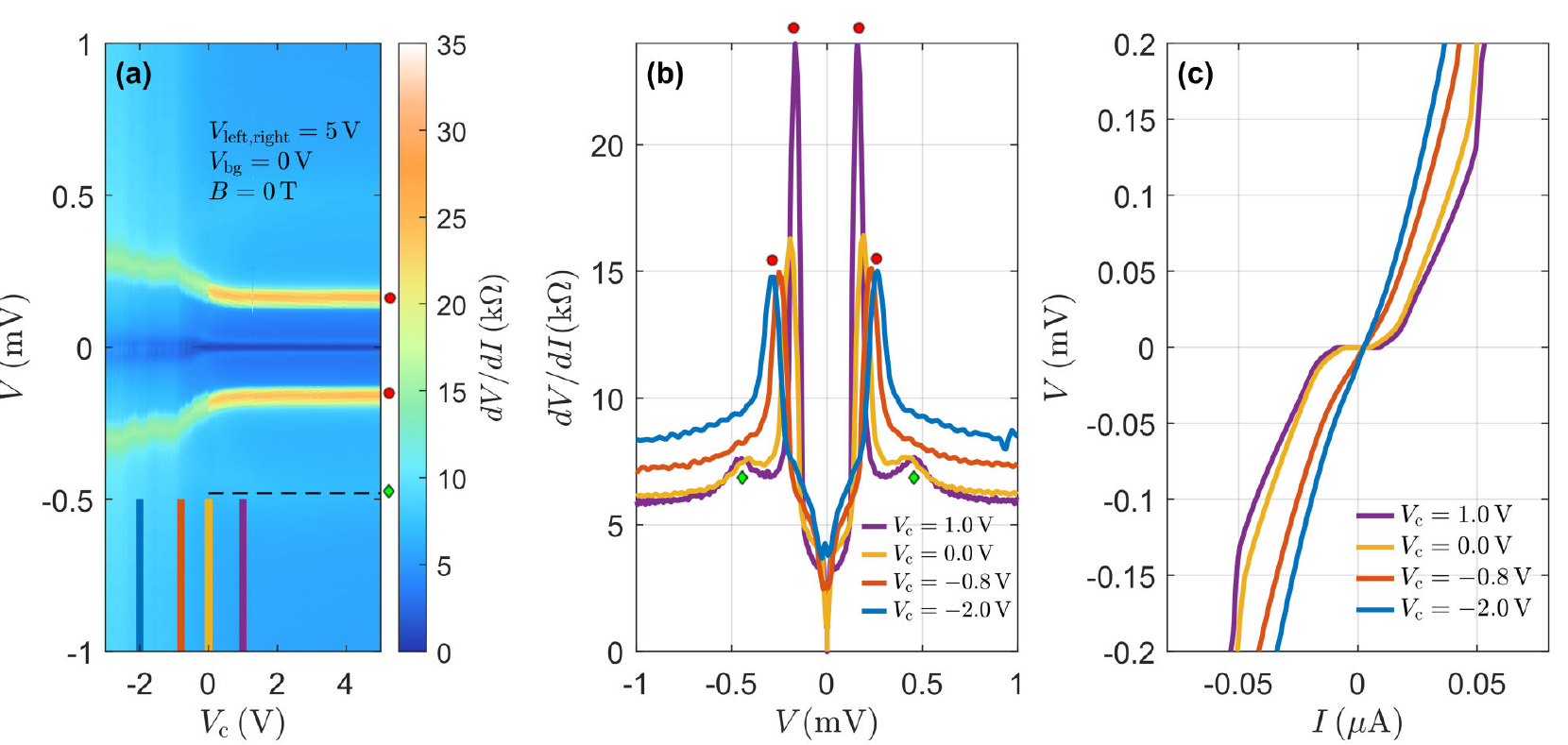}
\caption{\textbf{Control of channel supercurrent using gate $\bm{V_{\mathrm{c}}}$.} (a) Plot of differential resistance ($dV/dI$) as a function of device current and $V_{c}$. Gates $V_{\mathrm{left}}$ and $V_{\mathrm{right}}$ are set at $5.0\,$V with $V_{\mathrm{bg}}=0.0\,$V. (b) Differential resistance as a function of bias for four values of $V_{\mathrm{c}}$, indicated by solid lines in (a). Data plotted is the same as that in figure 4 of the main text. At each gate voltage a constant background voltage has been subtracted to correct for the offset of the voltage amplifier.}
\label{FigSuppTopGateVoltage}
\end{figure*}

Here we show and briefly discuss supplementary plots of data presented in the main text as a function of device source-drain bias as opposed to current. In figures \ref{FigSuppBackGateVoltage}, \ref{SuppTemperatureDataVoltages}, \ref{FigSuppSideGateVoltage} and \ref{FigSuppTopGateVoltage} we reproduce plots from main text figures 2,3 and 4 but as a function of voltage instead of current. For consistency we indicate with symbols the features identified in the main text figures. As was discussed in the main text, the $dV/dI$ peaks observed without applied gate voltages (red circles) we attribute to the critical current of contact regions with diffused $\mathrm{PdTe_{x}}$. These features are not constant in device bias and are observed to increase in current as gate voltage are applied. At high gate voltages the features are observed at currents close to $50\,$nA. Additional weaker features are observed as peaks in $dV/dI$ at higher current and voltage (green diamonds) when gates are applied to place the monolayer in the superconducting states. In contrast these features are found to be approximately constant in voltage with $V\sim \pm 480\,\mathrm{\mu V}$, and so are likely related to quasi-particle tunnelling at the gap edge. Note that such transport features are known to present as conductance minima or maxima depending on the parameters of the junction\cite{CuevasPhysRevB2006,GunelNanoLett2014}. The low supercurrent supported by the contact $\mathrm{PdTe_{x}}$ regions complicates further analysis of the dissipative transport as beyond this current, voltage is dropped over both the junction and contact regions. Such an additional series resistance will lead to an overestimation of the normal resistances and any evaluated excess currents. Consequently, we have restricted our discussion to the non-dissipative supercurrent features in the main text.

\clearpage

\section{Supplementary Note 4 - Additional discussion of magnetic field data}

Here we present additional data for the device in the main text under alternative gating conditions (figures \ref{Fig6}, \ref{Fig6b}, \ref{Fig6c}). We also explain certain nonidealities in the data such as the difficulty of determining the junction area because of flux penetration (figure \ref{SuppDeltaBDataAll}), the asymmetry present in some of the magnetic field responses due to trapped vorticies during the magnetic field sweeps (figure \ref{FigTcycle}), and finally we characterize the effect of self-screening of the junction (figure \ref{SuppSelfScreening}).

\subsection{Supplementary Note 4a - Additional data for alternative gate conditions}

\begin{figure*}[h]
\centering
\includegraphics[width=15 cm,angle=0]{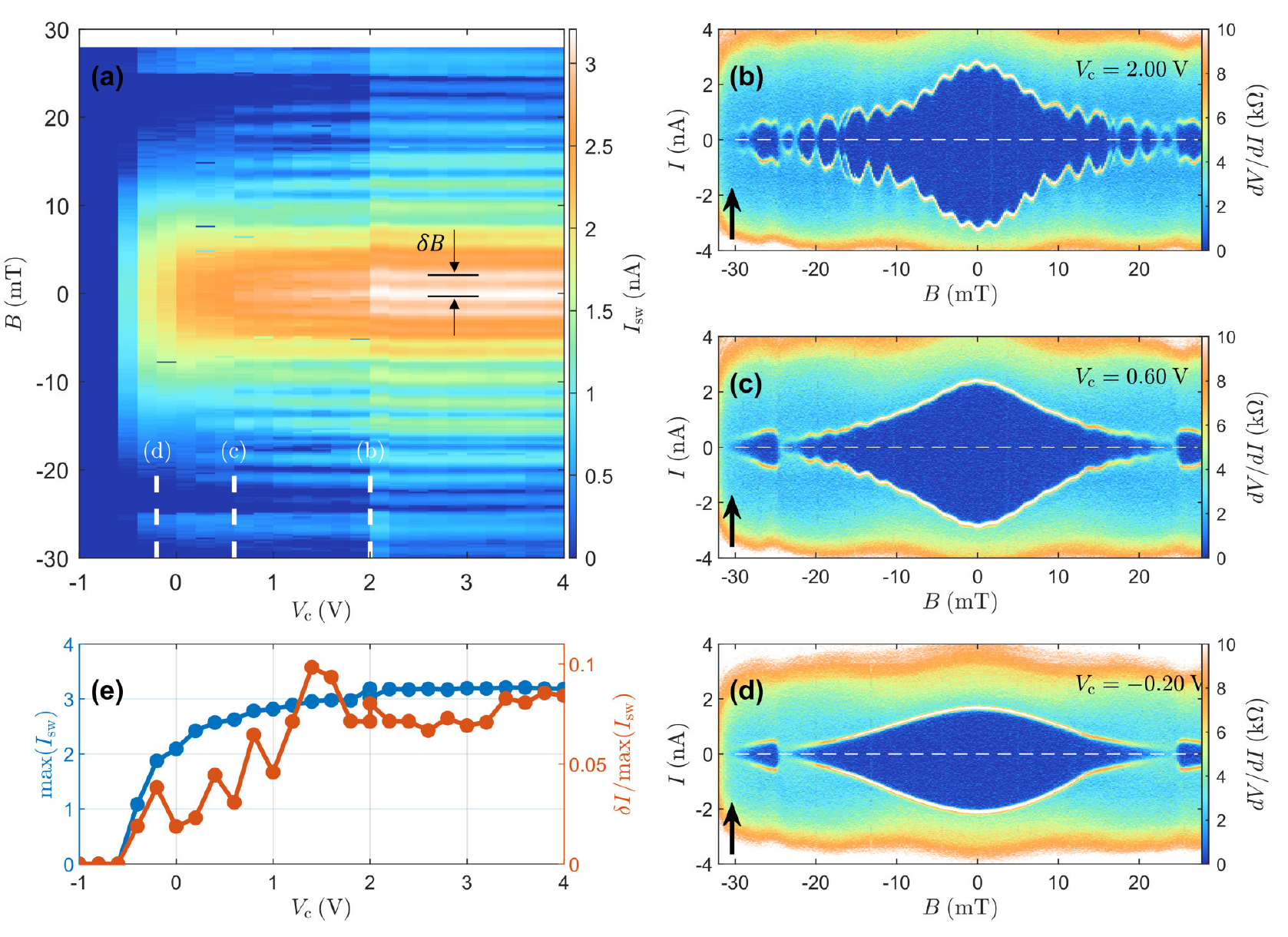}
\caption{\textbf{Influence of magnetic field on the junction transport (set A)}. (a) Plot of evaluated switching currents $I_{\mathrm{sw}}$ as a function of gate $V_{\mathrm{c}}$ with $V_{\mathrm{bg}}=0\,$V and $V_{\mathrm{left}}=V_{\mathrm{right}}=5\,$V. The SQUID-like oscillation period is $\delta B\sim 2.2\,\mathrm{mT}$. (b-d) Plots of $dV/dI$ as a function of $I$ and $B$ at values of $V_{\mathrm{c}}$ indicated in (a). (e) Plot of evaluated maximum $I_{\mathrm{sw}}$ and the visibility of SQUID-like oscillations taken as the variation between maximum and minimum $I_{\mathrm{sw}}$ for the first oscillations closest to $B=0\,$mT. In all data, a correction for a $\mathrm{2.1\,}$mT  remnant field has been subtracted to center the observed patterns. Black arrows indicate the current sweep directions used in the measurements.}
\label{Fig6}
\end{figure*}

\begin{figure*}[p]
\centering
\includegraphics[width=15 cm,angle=0]{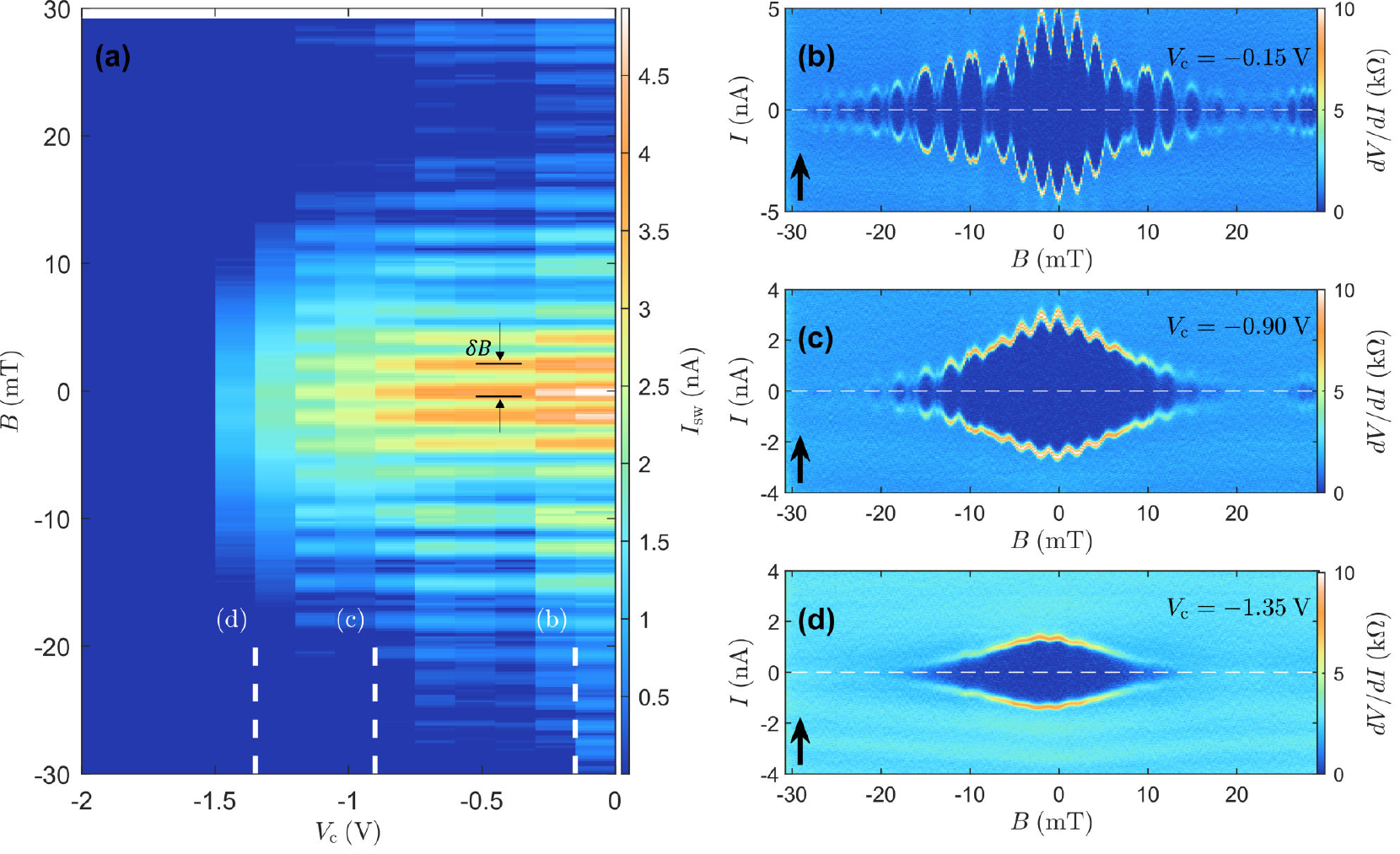}
\caption{\textbf{Influence of magnetic field on the junction transport (set B)}. (a) Plot of evaluated switching currents $I_{\mathrm{sw}}$ as a function of gate $V_{\mathrm{c}}$ with $V_{\mathrm{bg}}=40\,$V and $V_{\mathrm{left}}=V_{\mathrm{right}}=4\,$V. The SQUID-like oscillation period is $\delta B\sim 2.1\,\mathrm{mT}$. (b-d) Plots of $dV/dI$ as a function of $I$ and $B$ at values of $V_{\mathrm{c}}$ indicated in (a). In all data, a correction for a $\mathrm{0.8\,}$mT  remnant field has been subtracted to center the observed patterns. Black arrows indicate the current sweep directions used in the measurements.} 
\label{Fig6b}
\end{figure*}

\begin{figure*}[p]
\centering
\includegraphics[width=15 cm,angle=0]{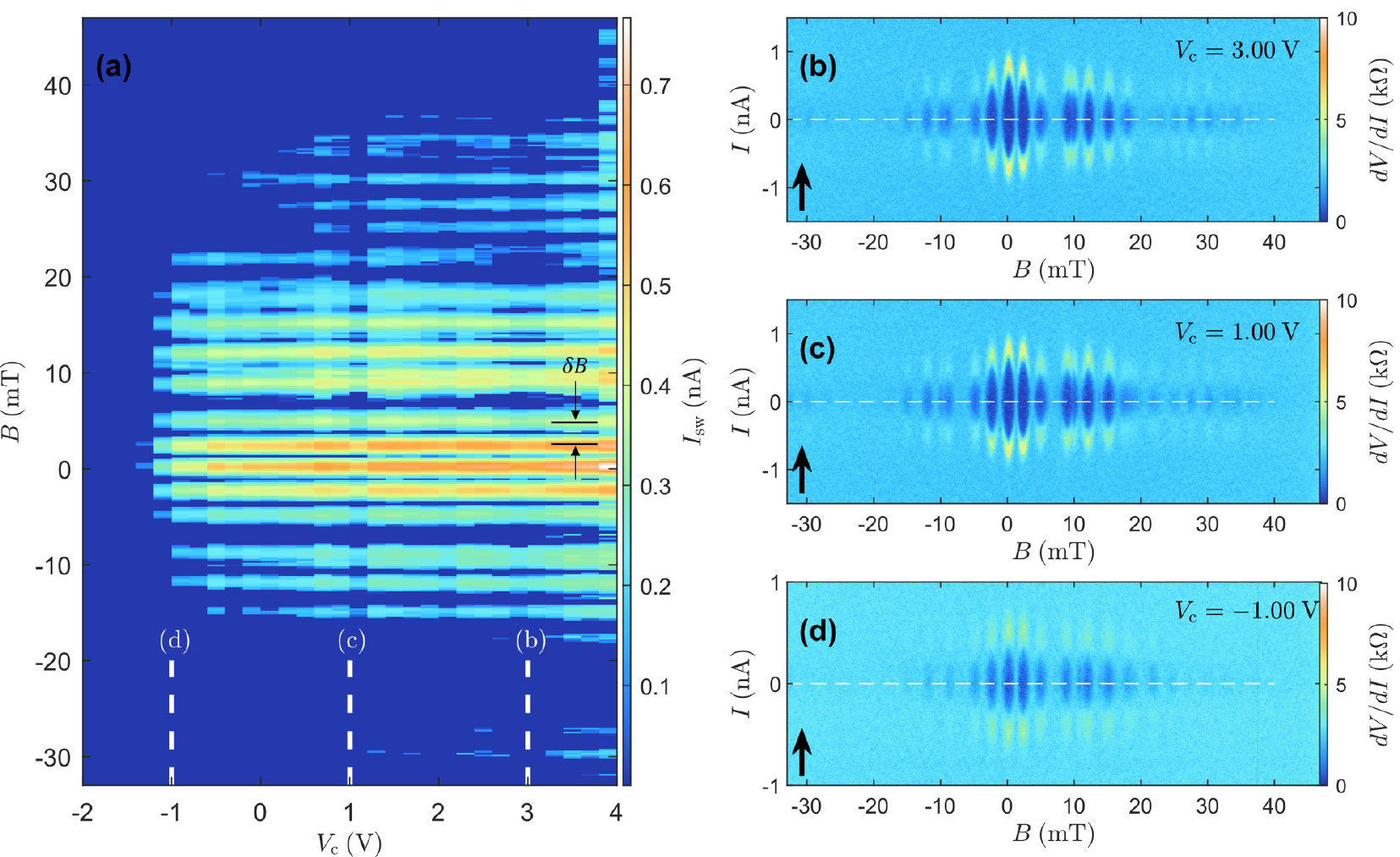}
\caption{\textbf{Influence of magnetic field on the junction transport (set C)}. (a) Plot of evaluated switching currents $I_{\mathrm{sw}}$ as a function of gate $V_{\mathrm{c}}$ with $V_{\mathrm{bg}}=50\,$V and $V_{\mathrm{left}}=V_{\mathrm{right}}=0\,$V. The SQUID-like oscillation period is $\delta B\sim 2.4\,\mathrm{mT}$. (b-d) Plots of $dV/dI$ as a function of $I$ and $B$ at values of $V_{\mathrm{c}}$ indicated in (a). In all data, a correction for a $\mathrm{5.5\,}$mT  remnant field has been subtracted to center the observed patterns. Black arrows indicate the current sweep directions used in the measurements.}
\label{Fig6c}
\end{figure*}

Here we show additional plots of the magnetic field response for several sets of gate conditions. In figure \ref{Fig6} we show data for $V_{\mathrm{bg}}=0\,$V and $V_{\mathrm{left}}=V_{\mathrm{right}}=5\,$V which reproduces data in the maintext figure 6 (a-d). In figure \ref{Fig6} (e), the maximum weak-link supercurrent for the purely top-gate induced superconducting state is plotted alongside a measure of the visibility of the SQUID-like oscillations $\delta I$, recorded as the variation between the peak and valley of the central oscillations closest to $B=0\,$mT as a fraction of the maximum $I_{\mathrm{sw}}$. In figure \ref{Fig6b} we show data for $V_{\mathrm{bg}}=40\,$V and $V_{\mathrm{left}}=V_{\mathrm{right}}=4\,$V where both back and top gates are used to induce the superconducting state resulting in a better visibility for the SQUID-like oscillations. For completeness we also show figure \ref{Fig6c} with conditions $V_{\mathrm{bg}}=50\,$V and $V_{\mathrm{left}}=V_{\mathrm{right}}=0\,$V which reproduces data in the maintext figure 6 (e-h).

\clearpage

\subsection{Supplementary Note 4b - Insensitivity of oscillation period to junction geometry}

\begin{figure*}[h]
\centering
\includegraphics[width=8 cm,angle=0]{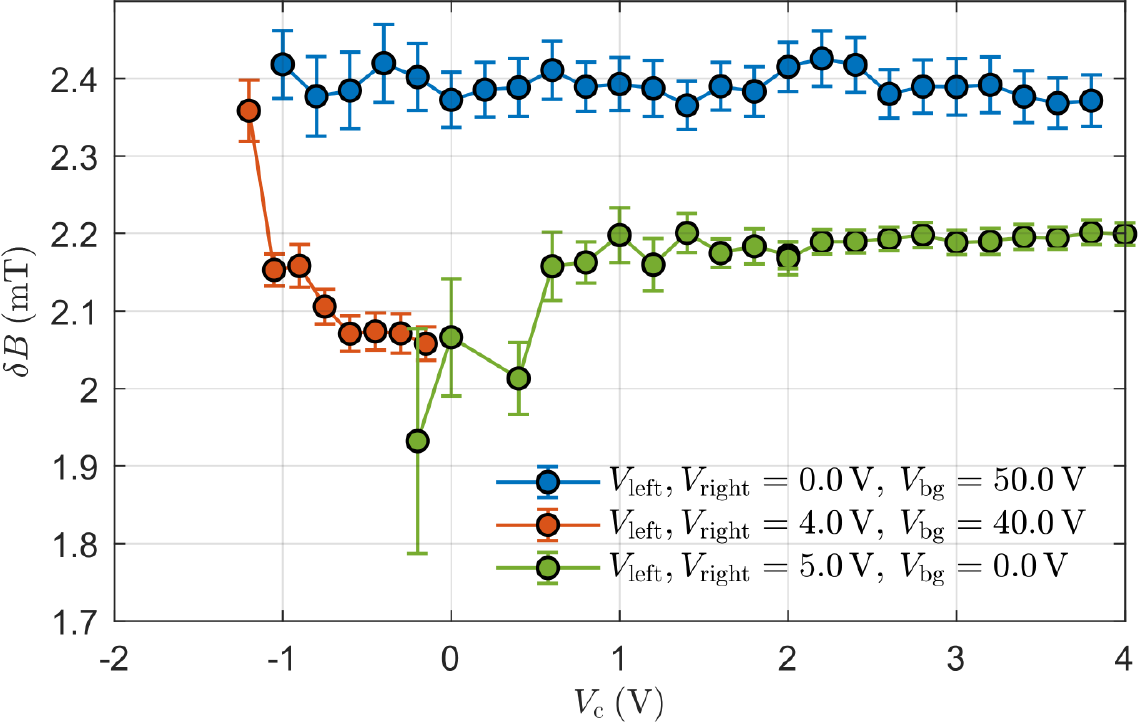}
\caption{\textbf{SQUID-like oscillation periods}. Plot of the oscillation period $\delta B$ as a function of $V_{\mathrm{c}}$ for three different sets of gate conditions. $\delta B$ is extracted from sinusoidal fits to the low magnetic field data with a background subtraction for the Fraunhofer-like envelope.}
\label{SuppDeltaBDataAll}
\end{figure*}

Figure \ref{SuppDeltaBDataAll} shows the period ($\delta B$) of the SQUID-like oscillations extracted from a sinusoidal fit to the center of the magnetic interference pattern with a background subtraction for several different sets of gate conditions. We note that the oscillation period only shows small changes with the device operated under different conditions but that this does not necessarily mean the junction size is unchanged. Due to the flux penetration in the 2D superconducting contacts we predict that alteration of junction geometry has only limited effect on the resulting interference pattern period. See also Supplementary Note 7 in which simulations are performed with variation of junction position and junction length for a fixed size system.

\subsection{Supplementary Note 4c - Origin of asymmetry in the magnetic field response}

\begin{figure*}[h]
\centering
\includegraphics[width=6 cm,angle=0]{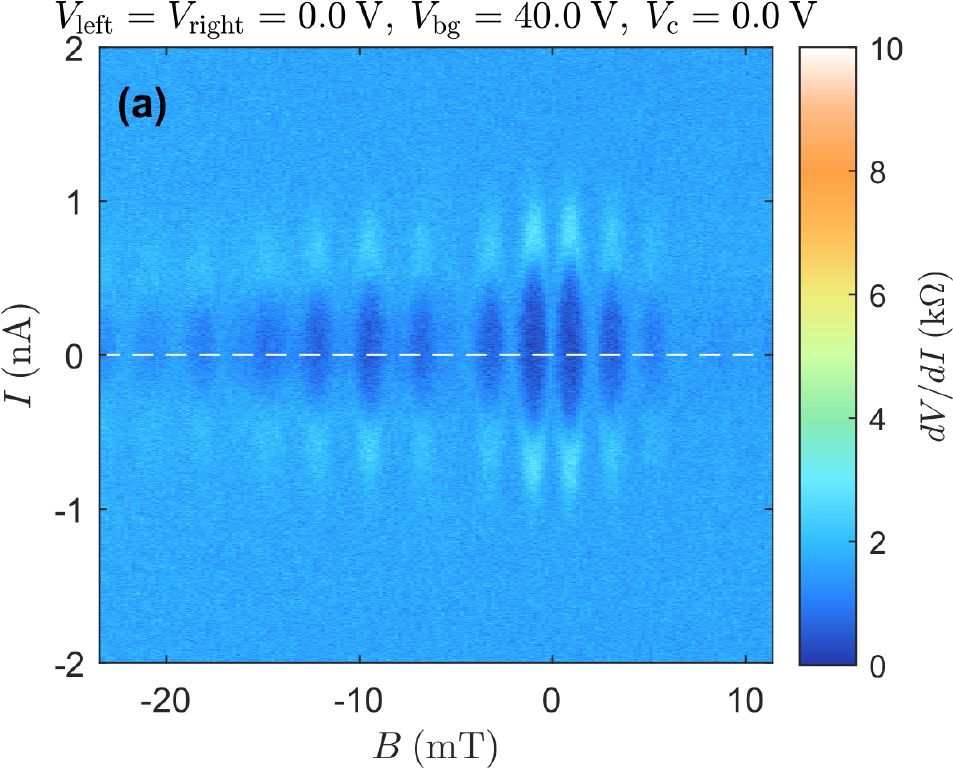}
\includegraphics[width=6 cm,angle=0]{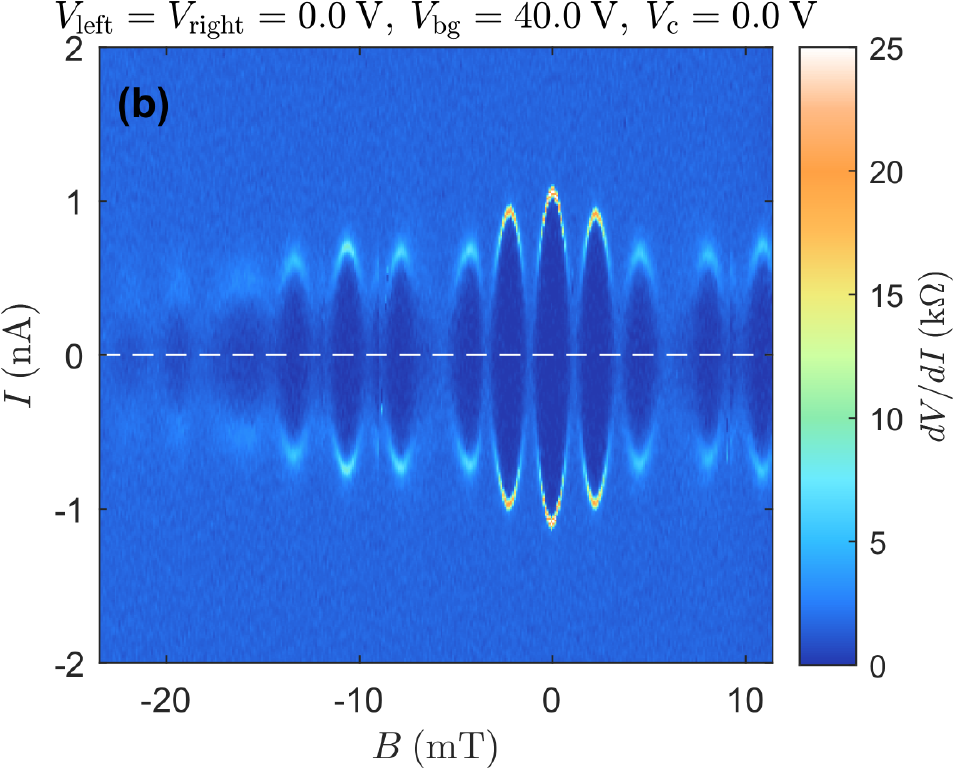}
\caption{\textbf{Recovery of asymmetric magnetic response with thermal cycling}. (a) Plot of $dV/dI$ as a function of current and $B$ showing an asymmetric pattern of peaks. (b) A measurement under the same conditions taken after a cycle of the fridge temperature restoring the symmetry in the pattern. Black arrows indicate both the current and magnetic field sweep directions used in the measurements.}
\label{FigTcycle}
\end{figure*}

In figure \ref{FigTcycle} we demonstrate that cycling of temperature can restore the symmetry of the magnetic response patterns with respect to the applied field polarity. Here two measurements have been conducted under the same gate conditions but before and after a thermal cycle of the fridge. Discrepancy in the quality of data is a consequence of modification of the measurements setup between the two measurements. Regardless the two experiments demonstrate that the symmetry of the magnetic response pattern can be restored with a thermal cycle and that the cause is likely vortex states trapped in surrounding superconducting leads or the junction itself if external field is driven too high during the experiment. Such trapped vortices have been considered in a number of report such as that in reference \citen{StolyarovNanoLetter2022} and references therein.

\clearpage

\subsection{Supplementary Note 4d - Self-screening currents}

\begin{figure*}[h]
\centering
\includegraphics[width=6 cm,angle=0]{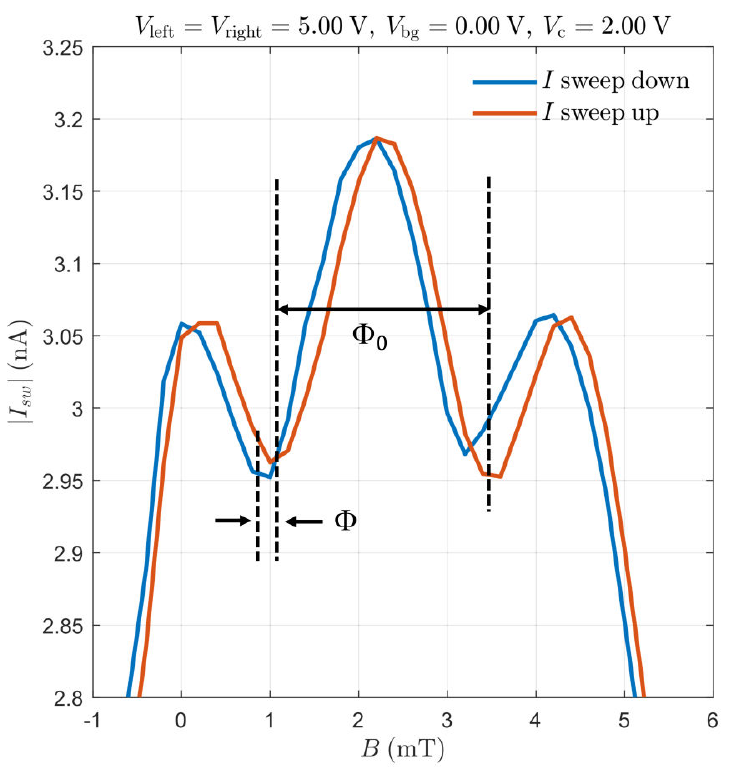}
\caption{\textbf{Evaluation of screening parameter.} Here the offset of the $B$-axis due to a remnant field in the superconducting coil has not been corrected.}
\label{SuppSelfScreening}
\end{figure*}

Self-screening effects can become important if the junction has a screening parameter $\beta_{\mathrm{L}}=(2\pi/\Phi_{\mathrm{0}})LI_{\mathrm{c}}>1$ where $L$ is the junction inductance which includes both geometric and kinetic inductance. We can estimate the screening parameter by considering the asymmetry in magnetic field between peaks in the absolute switching current in forward and reverse current directions\cite{BorisBook2004}. We consider the case of a simple SQUID with two junctions with currents $I_{\mathrm{1}}$ and $I_{\mathrm{2}}$. The SQUID has two arms with inductance $L_{\mathrm{1}}$ and $L_{\mathrm{2}}$. The shift of forward and reverse supercurrent maxima in flux is given as $\Phi=\beta_{\mathrm{L}}(\alpha_{\mathrm{L}}+\alpha_{\mathrm{I}})/\Phi_{\mathrm{0}}$, where $\alpha_{\mathrm{L}}$ is the asymmetry in inductance $\alpha_{\mathrm{L}}=(L_{\mathrm{2}}-L_{\mathrm{1}})/(L_{\mathrm{2}}+L_{\mathrm{1}})$ and $\alpha_{\mathrm{I}}=1-(I_{\mathrm{1}}/I_{\mathrm{0}})$, where $I_{\mathrm{0}}=(I_{\mathrm{1}}+I_{\mathrm{2}})/2$. We consider the device to have a symmetric geometry such that $L_{\mathrm{1}}=L_{\mathrm{2}}$ and $\alpha_{\mathrm{L}}=0$. From experiment (figure \ref{SuppSelfScreening}) we observe $\Phi=0.02\Phi_{\mathrm{0}}$ for a case with $\alpha_{\mathrm{I}}=0.9375$ giving a estimate of $\beta_{\mathrm{L}}=0.0213$. Note also that a large screening parameter $\beta_{\mathrm{L}}>1$ can also cause noise to switch the junction resulting in asymmetric sinusoidal features in the magnetic response\cite{EnglishPRB2016}.

\section{Supplementary Note 5 - Minimization fitting of the magnetic response data}

\begin{figure*}[h]
\centering
\includegraphics[width=5 cm,angle=0]{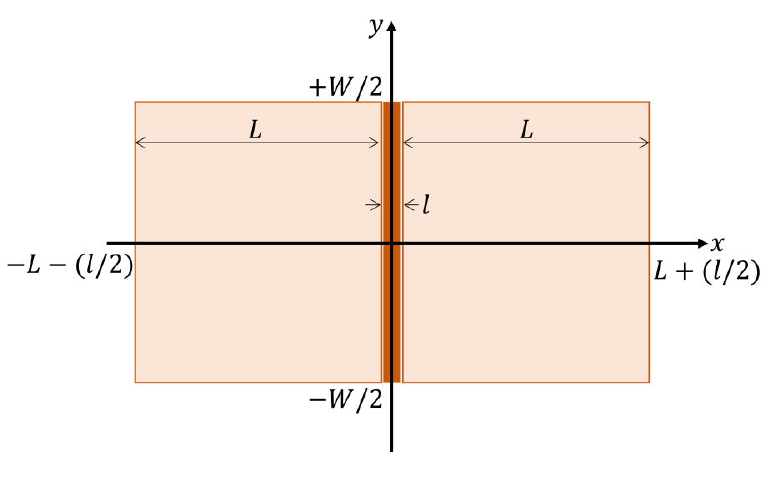}
\caption{\textbf{Schematic of the simulation system geometry.} We assume $W=1.6\,\mu$m, $L=0.9\,\mu$m and $l=100\,$nm.}
\label{FigSuppSimSchematic}
\end{figure*}

We perform analysis of the distribution of current across the junction width using the minimization method described by Hui et al. in reference \citen{HuiPRB2014}. We assume a sinusoidal current phase relationship and that the leads consist of regions of two-dimensional superconductor which does not screen the applied magnetic field. The additional phase shift due to the lack of screening in the leads is accounted for using the model of reference \citen{Clem2010}. Elements of the model are described in the main text and methods, but are also reproduced here for completeness. The current density flowing in the $x$ direction is $j_{\mathrm{x}}(y)=j_{\mathrm{c}}(y)\sin(\Delta\gamma(y))$, where $j_{\mathrm{c}}(y)$ describes the current density distribution across the width of junction and $\Delta\gamma(y)$ is the gauge-invariant phase difference between the two leads. We assume the device geometry as shown in figure \ref{FigSuppSimSchematic}. For the rectangular leads considered the phase difference can be expressed using the expansion\cite{Clem2010}

\begin{equation}
\Delta\gamma(y)=\Delta\gamma_{0} + \frac{16\pi B}{\phi_{\mathrm{0}}W}\sum_{n=0}^{\infty} \frac{(-1)^{n}}{k_{n}^{3}}\tanh(k_{n}L/2)\sin(k_{n}y),
\end{equation}

\noindent where $\phi_{\mathrm{0}}=h/2e$ is the magnetic flux quantum, $B$ is the applied magnetic field and $k_{n}=2\pi(n+0.5)/W$. The maximum supercurrent through the junction is then given as 

\begin{equation}
\frac{I_{\mathrm{c}}(B)}{I_{\mathrm{c}}(0)}=\left\lvert \int_{-W/2}^{W/2} j_{\mathrm{c}}(y)\cos(\Delta\gamma(y))dy \right\rvert.
\label{equationsimIc}
\end{equation}

\noindent The term $\Delta\gamma_{\mathrm{0}}$ includes the superconducting phase difference between the leads ($0\geq \phi\leq 2\pi$) and the phase winding due to flux penetrating the actual junction area $2\pi lyB/\Phi_{\mathrm{0}}$.

\begin{figure*}[h]
\centering
\includegraphics[width=13 cm,angle=0]{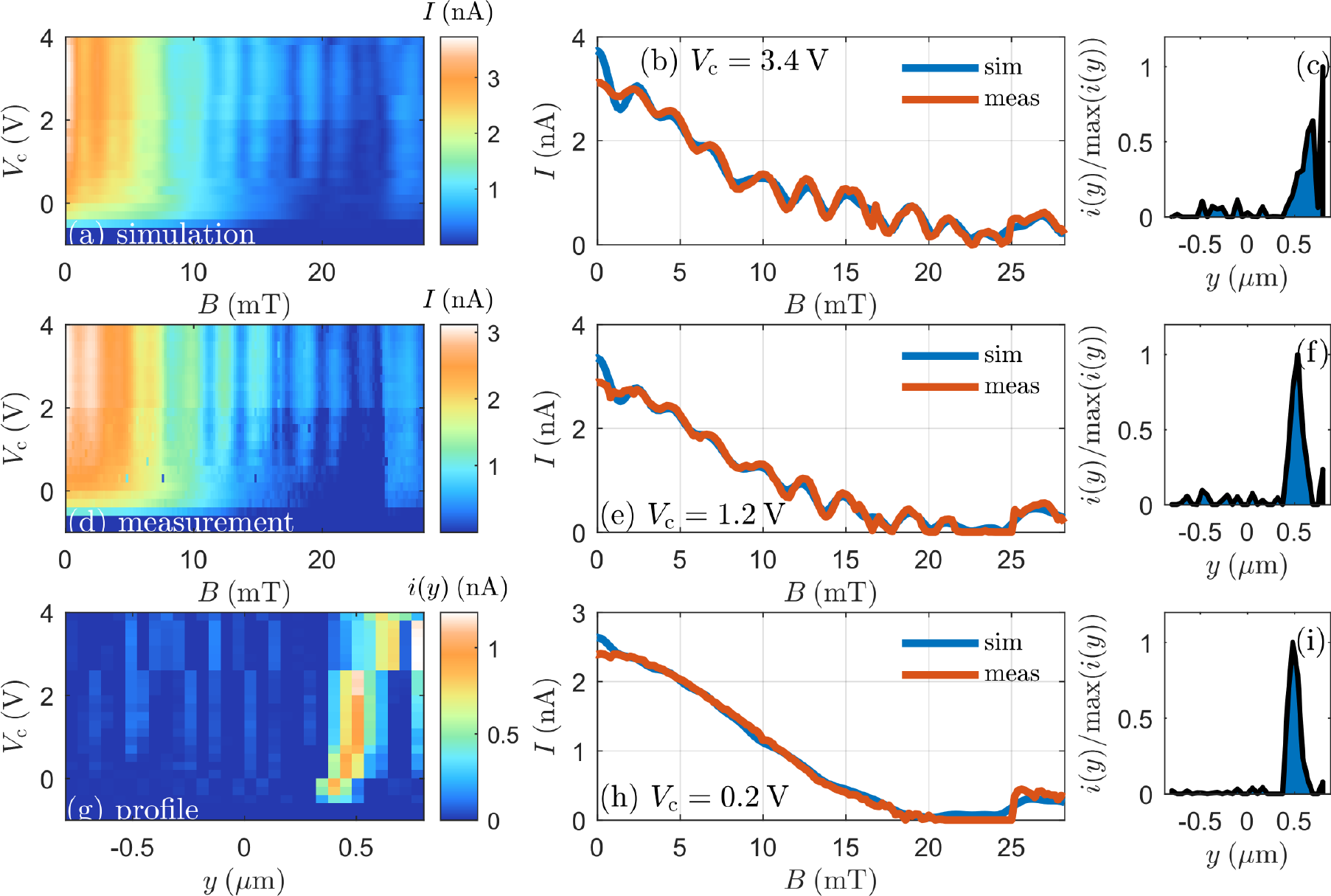}
\caption{\textbf{Minimization fitting of junction current profile (set A).} Device settings are $V_{\mathrm{bg}}=0\,$V, $V_{\mathrm{left}}=V_{\mathrm{right}}=5\,$V. (a) Shows the simulated magnetic field response of the junctions. (d) shows the experimental data. (g) shows the fitted current profiles. (b,e,h) Show minimization fits and experimental data at indicated values of $V_{\mathrm{c}}$. (c,d,i) show the corresponding current profiles.}
\label{MinimizationFittingSet2}
\end{figure*}

\begin{figure*}[p]
\centering
\includegraphics[width=13 cm,angle=0]{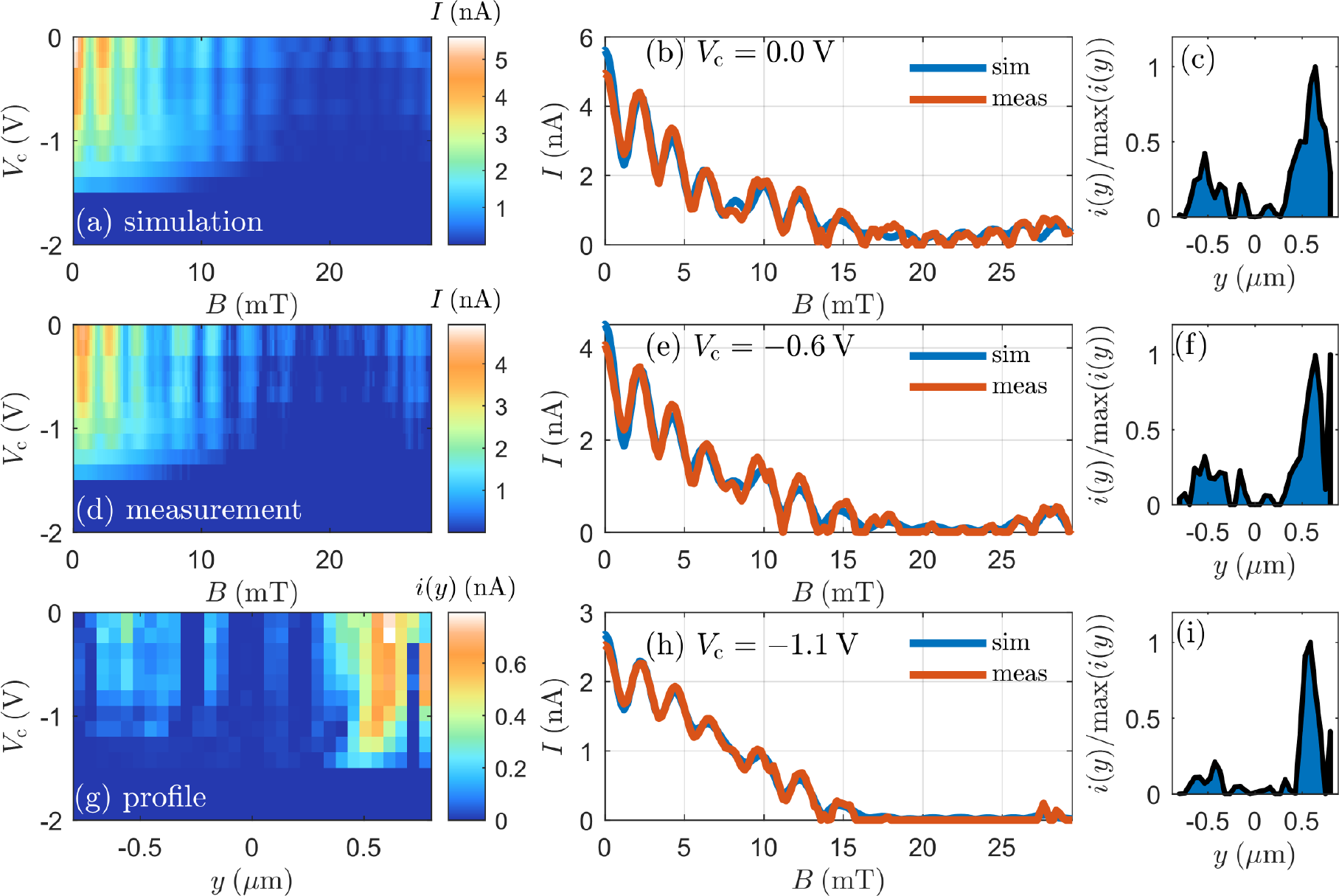}
\caption{\textbf{Minimization fitting of junction current profile (set B).} Device settings are $V_{\mathrm{bg}}=40\,$V, $V_{\mathrm{left}}=V_{\mathrm{right}}=4\,$V. (a) Shows the simulated magnetic field response of the junctions. (d) shows the experimental data. (g) shows the fitted current profiles. (b,e,h) Show minimization fits and experimental data at indicated values of $V_{\mathrm{c}}$. (c,d,i) show the corresponding current profiles.}
\label{MinimizationFittingSet3}
\end{figure*}

\begin{figure*}[p]
\centering
\includegraphics[width=13 cm,angle=0]{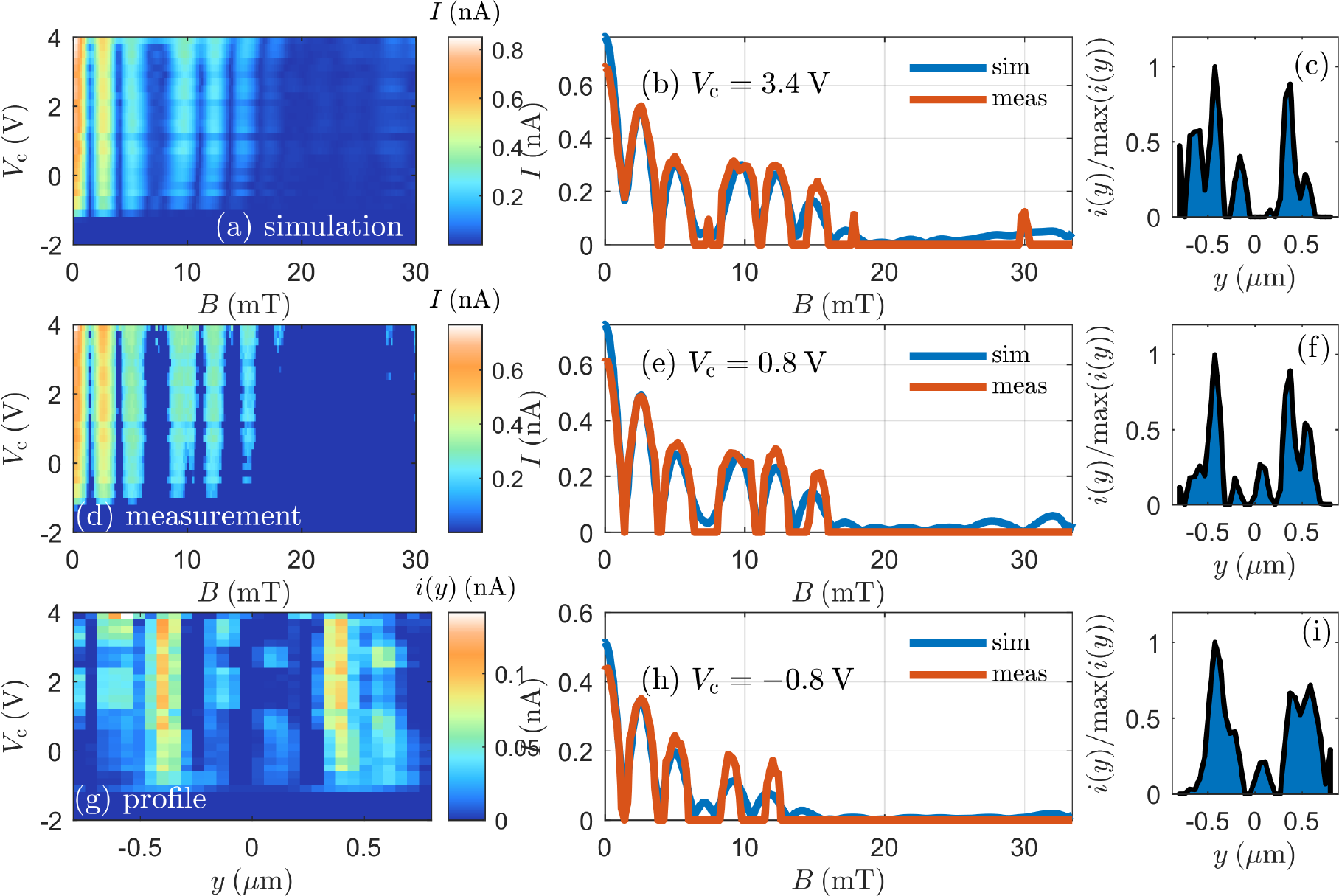}
\caption{\textbf{Minimization fitting of junction current profile (set C).} Device settings $V_{\mathrm{bg}}=50\,$V, $V_{\mathrm{left}}=V_{\mathrm{right}}=0\,$V. (a) Shows the simulated magnetic field response of the junctions. (d) shows the experimental data. (g) shows the fitted current profiles. (b,e,h) Show minimization fits and experimental data at indicated values of $V_{\mathrm{c}}$. (c,d,i) show the corresponding current profiles.}
\label{MinimizationFittingSet4}
\end{figure*}

Following the method of reference \citen{HuiPRB2014} we assign a cost function given as 

\begin{equation}
C[j_{\mathrm{c}}(y)]=\frac{1}{2B_{\mathrm{max}}}\int_{-B_{\mathrm{max}}}^{B_{\mathrm{max}}}dB \left[\frac{I_{\mathrm{c}}(B)-I_{\mathrm{meas}}(B)}{I_{\mathrm{0}}(0)}\right]^{2},
\end{equation}

\noindent where $I_{\mathrm{0}}(B)$ is the experimental data and $I_{\mathrm{c}}(B)$ is given by equation \ref{equationsimIc}. We use the "active-set" constrained non-linear optimization algorithm in Matlab to locate the optimal current profile $j_{\mathrm{c}}(y)$. We fit the data using $31$ $y$ sites evenly spaced in the range $\pm (W/2)$ fixing the maximum junction width. We perform an initial fitting using randomised initial profiles of $j_{\mathrm{c}}$ for the highest value of $V_{\mathrm{c}}$ considered. We perform the minimization only for positive magnetic field data as the method cannot account for asymmetries between positive and negative fields. Following this initial fitting (for the highest $V_{\mathrm{c}}$ data set) we perform fitting for following sets of $V_{\mathrm{c}}$ data using the last found minimized solution as a starting point to monitor how the profile develops as a function of gating strength. Example minimization results are shown in figure \ref{MinimizationFittingSet2} for conditions $V_{\mathrm{bg}}=0\,$V, $V_{\mathrm{left}}=V_{\mathrm{right}}=5\,$V and a range of $V_{\mathrm{c}}$. Additional minimization fitting results are shown in figure \ref{MinimizationFittingSet3} for conditions $V_{\mathrm{bg}}=40\,$V, $V_{\mathrm{left}}=V_{\mathrm{right}}=4\,$V and a range of $V_{\mathrm{c}}$. Also figure \ref{MinimizationFittingSet4} for $V_{\mathrm{bg}}=0\,$V, $V_{\mathrm{left}}=V_{\mathrm{right}}=5\,$V and again a range of $V_{\mathrm{c}}$. It should be noted that as discussed in detail by Hui et al.\cite{HuiPRB2014} obtaining detailed quantitative information on the Josephson current distribution from analysis of the magnetic response alone is challenging. The fitting here can be useful to allow discussion of a possible solution but the uniqueness of the resulting profiles cannot be confirmed. As can be seen from the extracted current profiles we find that inducing superconductivity using the top gates results in a current peak at one edge of the device while use of the back gate results in two peaks located at the device edges.

\clearpage

\section{Supplementary Note 6 - Additional simulation plots}

Here we present some additional simulation plots to compliment those in the main text. In figure \ref{TransparencyFig}, we show that consideration of a non-sinusoidal current phase relationship due to high transparancy in the junction is able to lift the zeros of the magnetic field response, but is insufficient to explain the results presented in the main text. In figures \ref{SimJJLength} and \ref{SimJJPosition}, we demonstrate the insensitivity of the magnetic response period to junction position and size as a consequence of 2D flux penetration in the leads.

\subsection{Supplementary Note 6a - Effect of transparency on SQUID junctions}

\begin{figure*}[h]
\centering
\includegraphics[width=7 cm,angle=0]{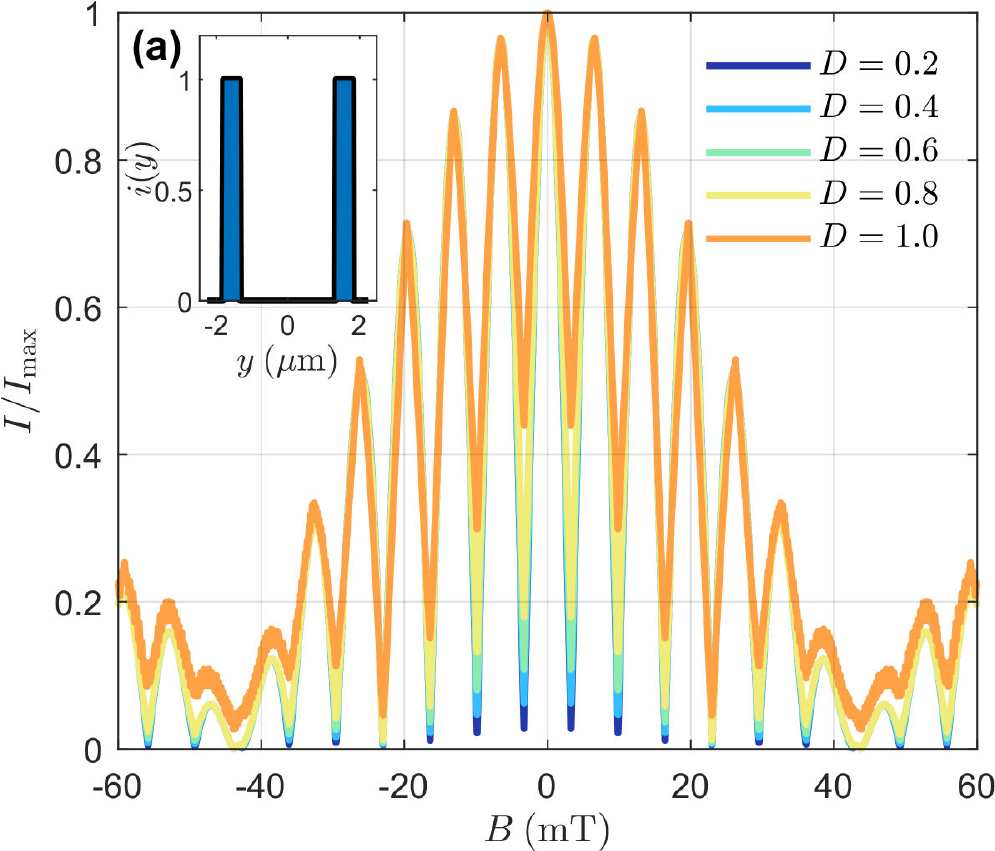}
\includegraphics[width=7 cm,angle=0]{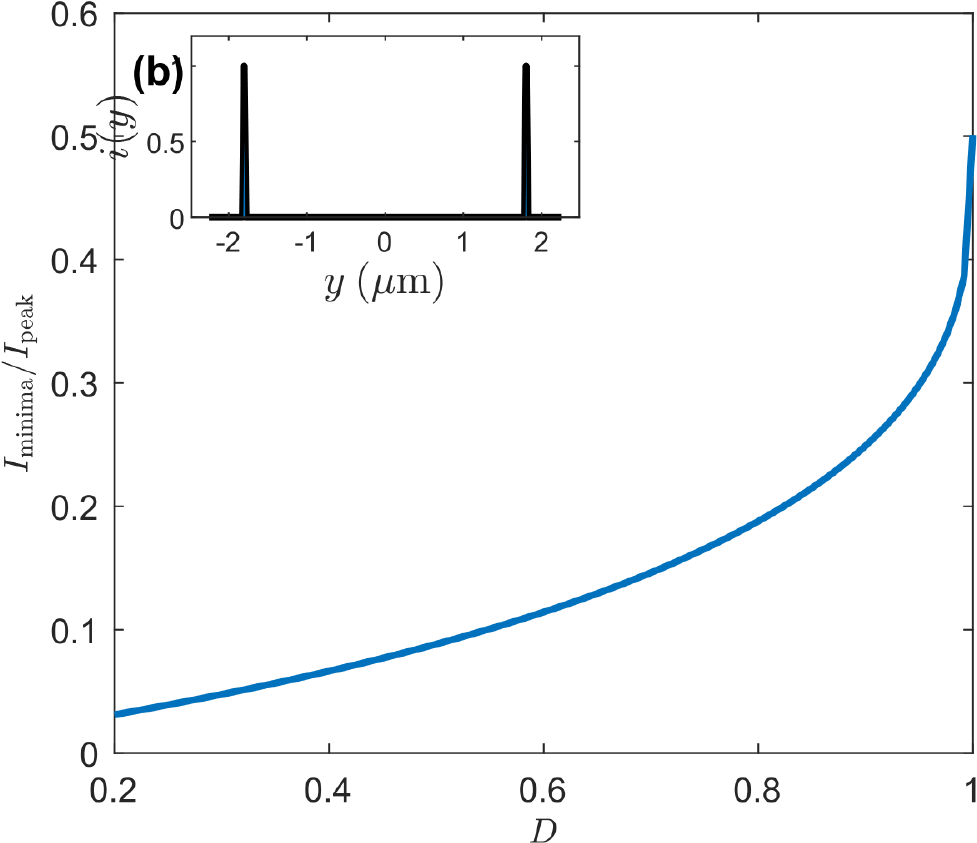}
\caption{\textbf{Effect of transparency $D$ on magnetic response.} (a) Simulation modifying transparency for a SQUID geometry device with screening bulk superconducting leads. The transition to high transparency will lift SQUID minima to roughly half visibility. (b) Simulation modifying transparency for a SQUID geometry device (with two zero width junctions) with screening bulk superconducting leads. For a SQUID device the effect of unity transparency will lift the zeros to exactly $0.5\times$ the sum of junction currents.}
\label{TransparencyFig}
\end{figure*}

In principle, the simulation of the magnetic response data does not require the system to be modeled using a sinusoidal current phase relationship. Following the work in reference \citen{HuiPRB2014}, we consider the non-sinusoidal current phase relationship\cite{Haberkorn1978,GolubovRevModPhys2004} created when an arbitrary transparency $D$ is considered for the junction

\begin{equation}
I(\varphi)=\frac{\pi\Delta}{2eR_{\mathrm{N}}}\frac{\sin(\varphi)}{\sqrt{1-D\sin^{2}(\varphi)}}\tanh\left[ \frac{\Delta}{2T}\sqrt{1-D\sin^{2}(\varphi)}\right],
\end{equation}

\noindent where $\Delta$ is the superconducting gap, $R_{\mathrm{N}}$ is the normal resistance and $T$ is the temperature. Parameter $D$ accounts for the transmission of the junction with perfect transparency for $D=1$. Figure \ref{TransparencyFig} (a) shows the effects of modifying the transparency and assuming conventional, screening leads. The simulation is performed using a junction length of $l=100\,\mathrm{nm}$ and the London penetration depth is set as zero. As the junction transparency is increased, the minima of the interference pattern are lifted from zero. Figure \ref{TransparencyFig} (b) shows the ratio of peak minima and maxima ($I_{\mathrm{minima}}/I_{\mathrm{peak}}$) for a SQUID containing two zero width junctions and again $l=100\,\mathrm{nm}$ showing that as $D$ approaches unity the ratio approaches $0.5$.

\clearpage

\subsection{Supplementary Note 6b - Junction size and position in a 2D superconductor junction}

\begin{figure*}[h]
\centering
\includegraphics[width=12 cm,angle=0]{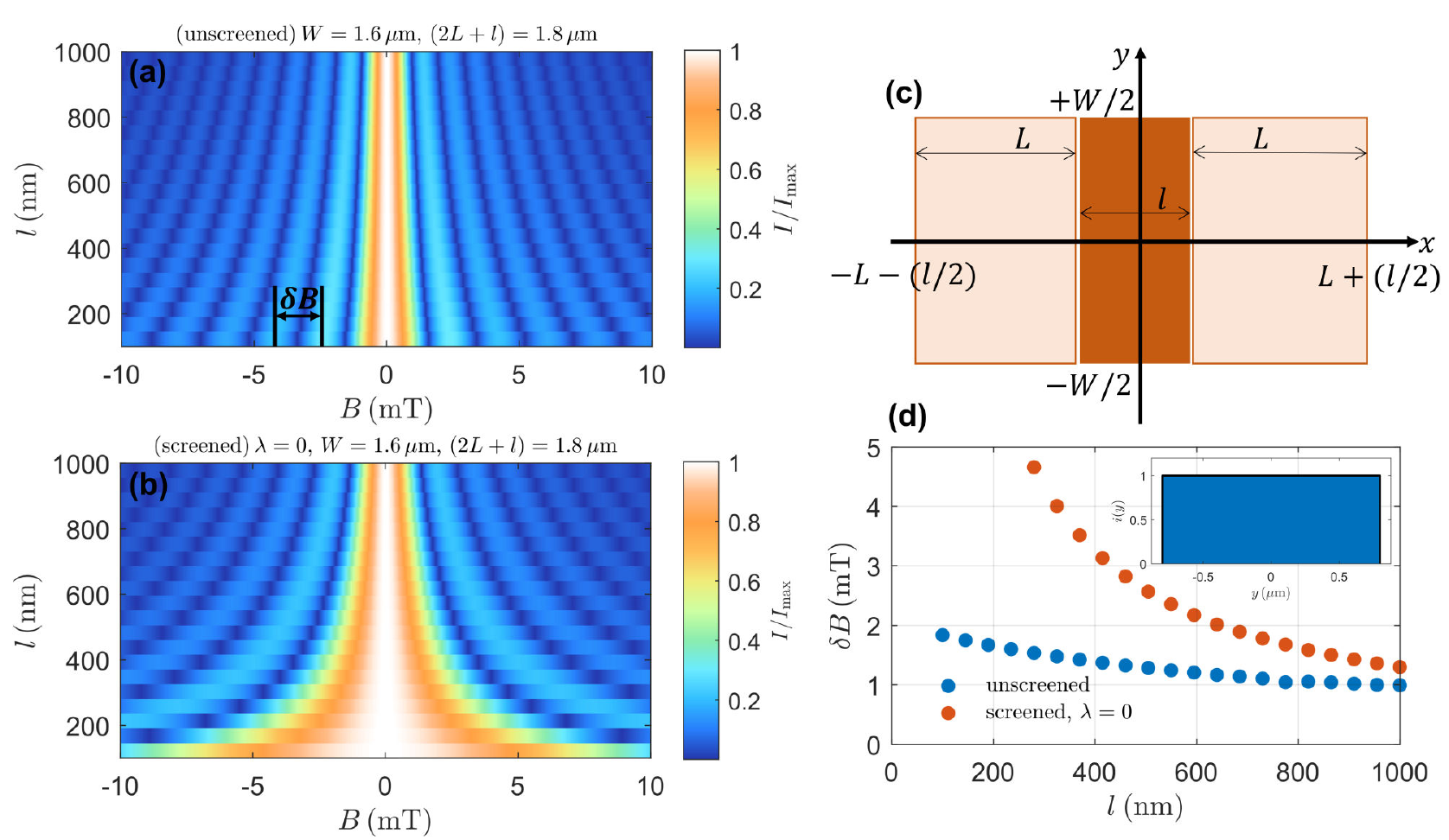}
\caption{\textbf{Simulation results with changing junction length $l$}. (a) Plot of simulation results showing influence of junction length $l$ using the device geometry shown in (c) for the case of the unscreened junction. (b) Simulation as in (a) but for the screened junction case with $\lambda=0$. (c) Schematic of the simulation device geometry. (d) Extracted oscillation period $\delta B$ as indicated in (a) for both screened and unscreened cases. Simulation parameters include $W=1.6\,\mu\mathrm{m}$, $2L+1=1.8\,\mathrm{\mu m}$ and $l=100\,$nm. A uniform current profile across the junction width is used as shown inset in (d).}
\label{SimJJLength}
\end{figure*}

We consider the effects of changing the junction size and position on the periodicity of oscillations in the magnetic response assuming a uniform current distribution across the junction width. The effect of changing junction length ($l$) is shown in figure \ref{SimJJLength}. We find that due to the penetration of flux in both the junction and 2D leads that the modification of junction length only causes small modifications to the oscillation period as compared to that expected in the screened case.

\begin{figure*}[h]
\centering
\includegraphics[width=12 cm,angle=0]{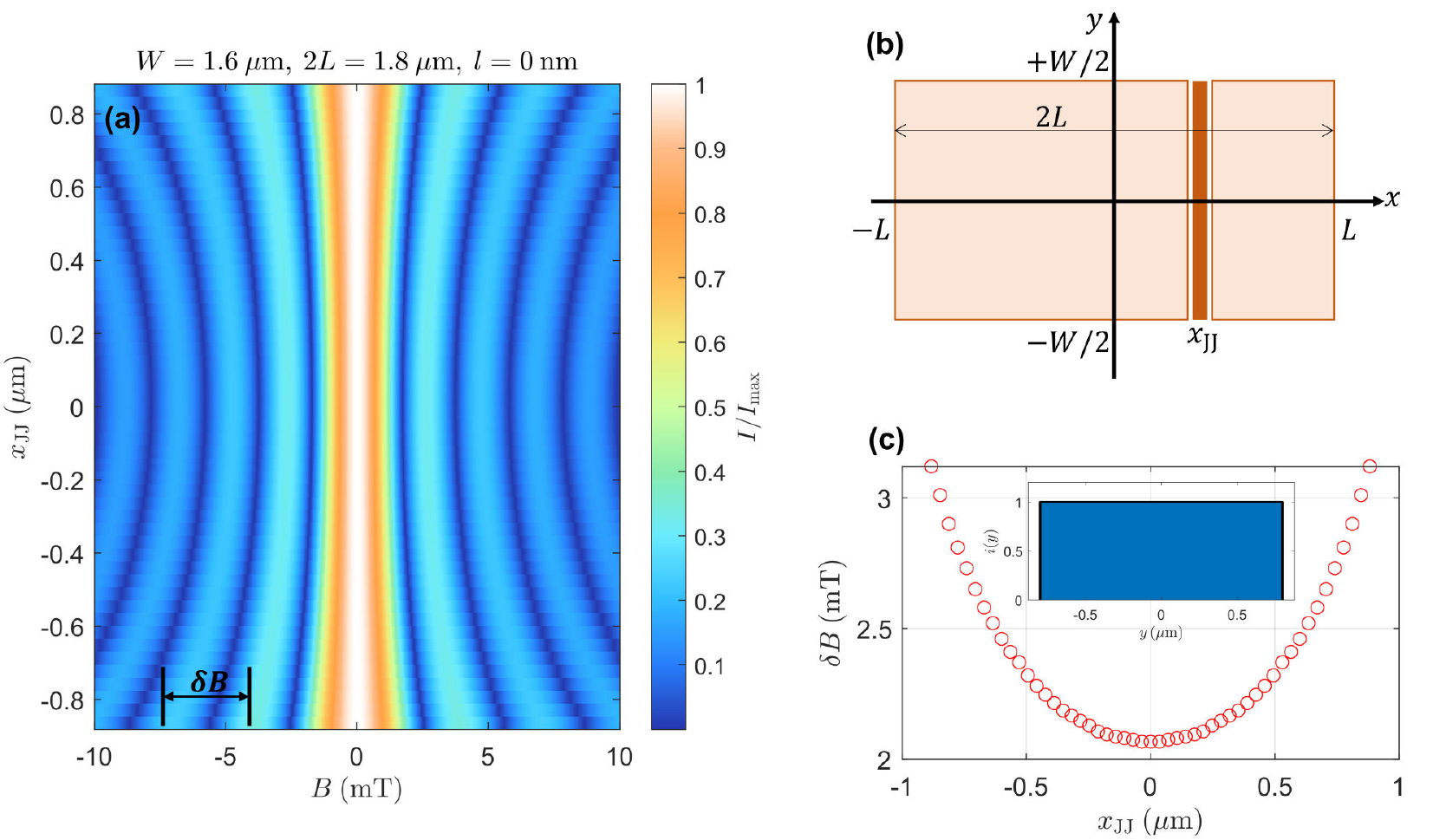}
\caption{\textbf{Simulation results with changing junction position $x_{\mathrm{JJ}}$}. (a) Plot of simulation results showing influence of junction position $x_{\mathrm{JJ}}$ in an unscreened junction using the device geometry shown in (b). (c) Extracted oscillation period $\delta B$ as indicated in (a). Simulation parameters include $W=1.6\,\mu\mathrm{m}$, $2L=1.8\,\mathrm{\mu m}$. A uniform current profile across the junction width is used as shown inset in (c).}
\label{SimJJPosition}
\end{figure*}

The effect of changing junction position ($x_{\mathrm{JJ}}$) is shown in figure \ref{SimJJPosition}. Here the junction length $l$ is taken to be zero and the phase accumulation in each lead is calculated following the work of Pekker et al. in reference \citen{PekkerPhysRevB2005}. As with the junction length, the junction position has a relatively small influence on the observed oscillation period.

\clearpage

\section{Supplementary Note 7 - Additional Shapiro response data}

\begin{figure*}[h]
\centering
\includegraphics[width=14 cm,angle=0]{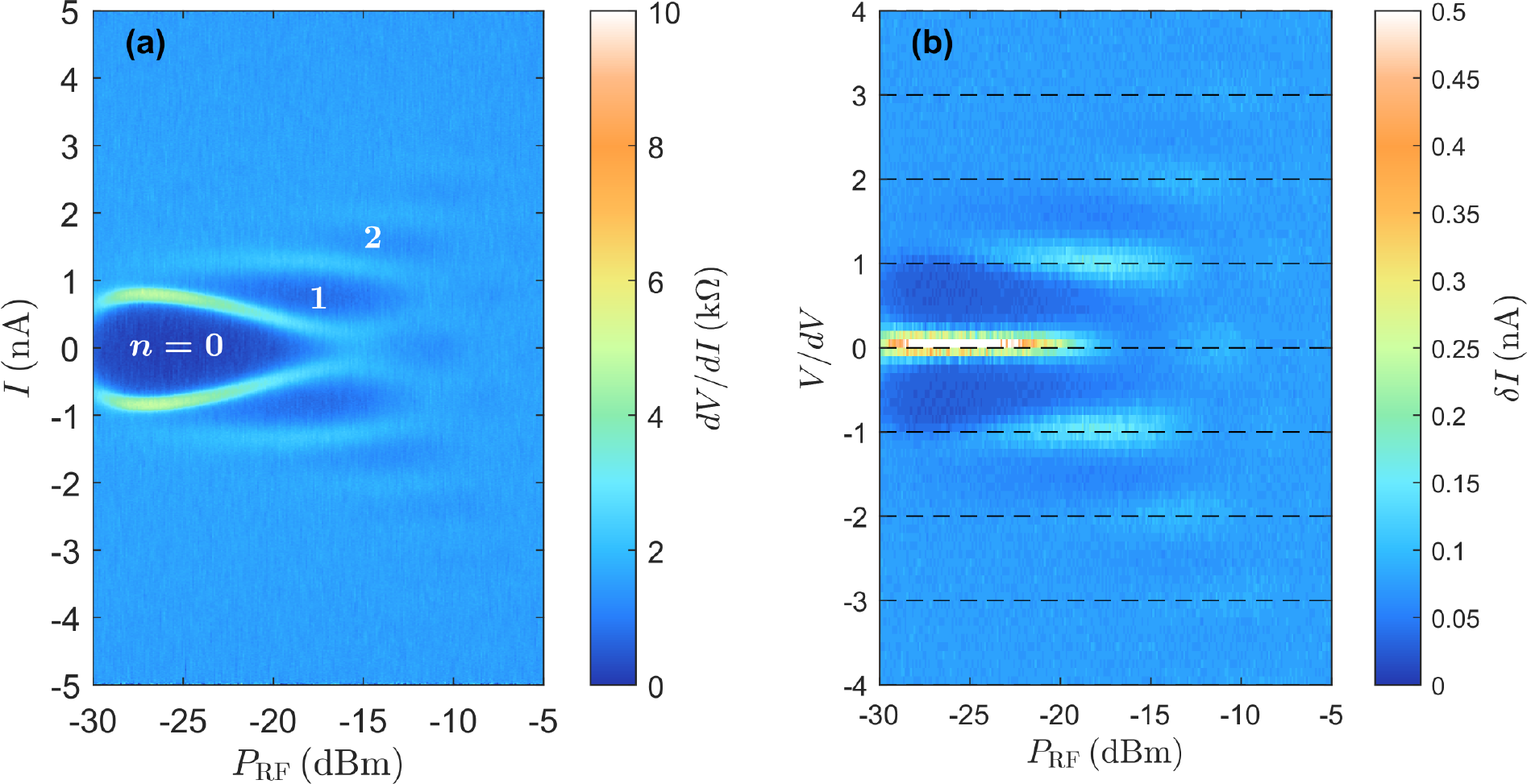}
\caption{\textbf{Shapiro response with external drive at $\bm{f=0.6\,}$GHz and $\bm{V_{\mathrm{bg}}=40\,}$V} (a) Differential resistance ($dV/dI$) as a function of device current ($I$) and microwave power ($P_\mathrm{RF}$). The indicated microwave power is at the output from the signal generator and is not corrected for attenuation of lines on the fridge. Gate settings are $V_{\mathrm{c}}=0.0\,$V, $V_{\mathrm{left}}=V_{\mathrm{right}}=0.0\,$V and $V_{\mathrm{bg}}=40.0\,$V. (c) Voltage histograms of $I(V)$ traces revealing the quantization of the Shapiro steps $dV=hf/2e=1.2424\,\mathrm{\mu V}$. Histogram bins have a width of $\delta V=0.1dV$.}
\label{fig:ShapiroVbg}
\end{figure*}

\begin{figure*}[p]
\centering

\includegraphics[width=8.5 cm,angle=0]{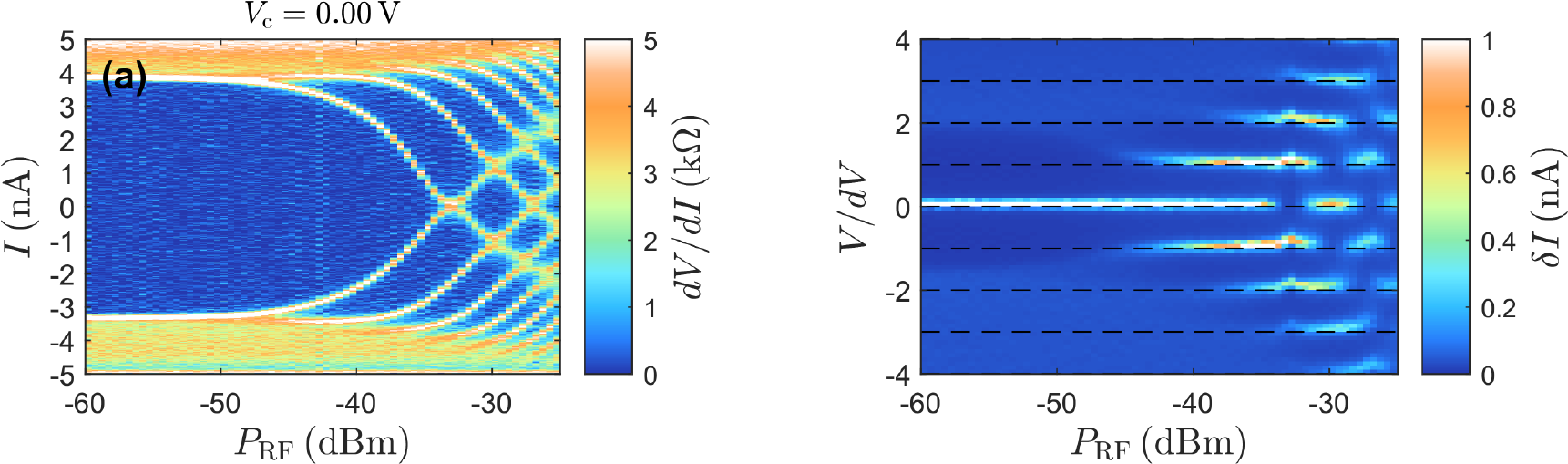} \includegraphics[width=8.5 cm,angle=0]{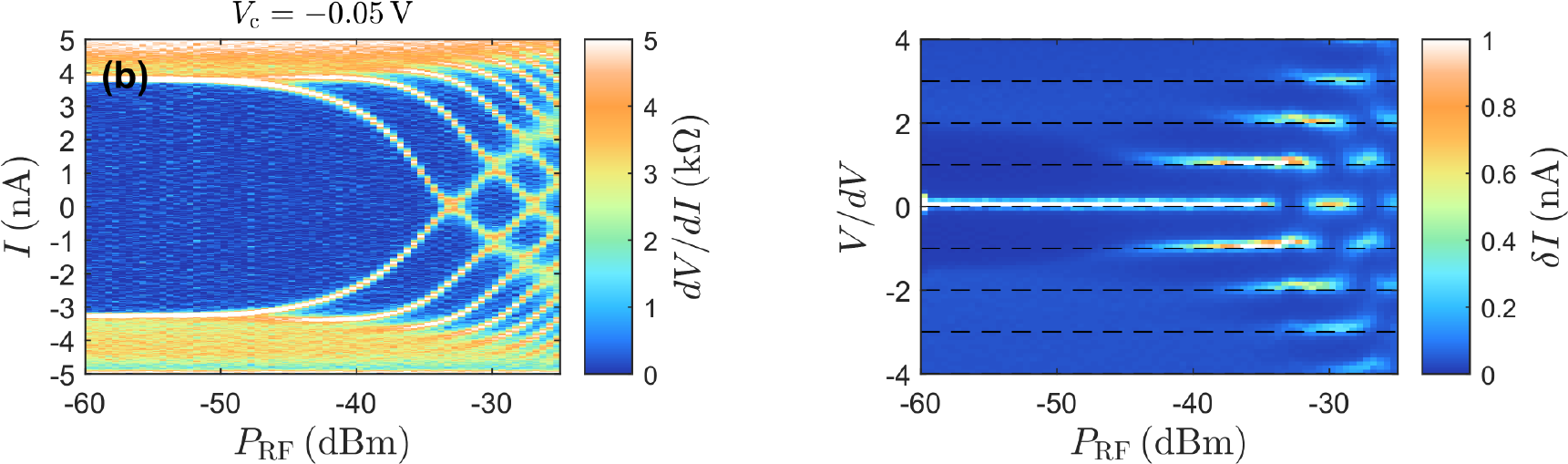}
\includegraphics[width=8.5 cm,angle=0]{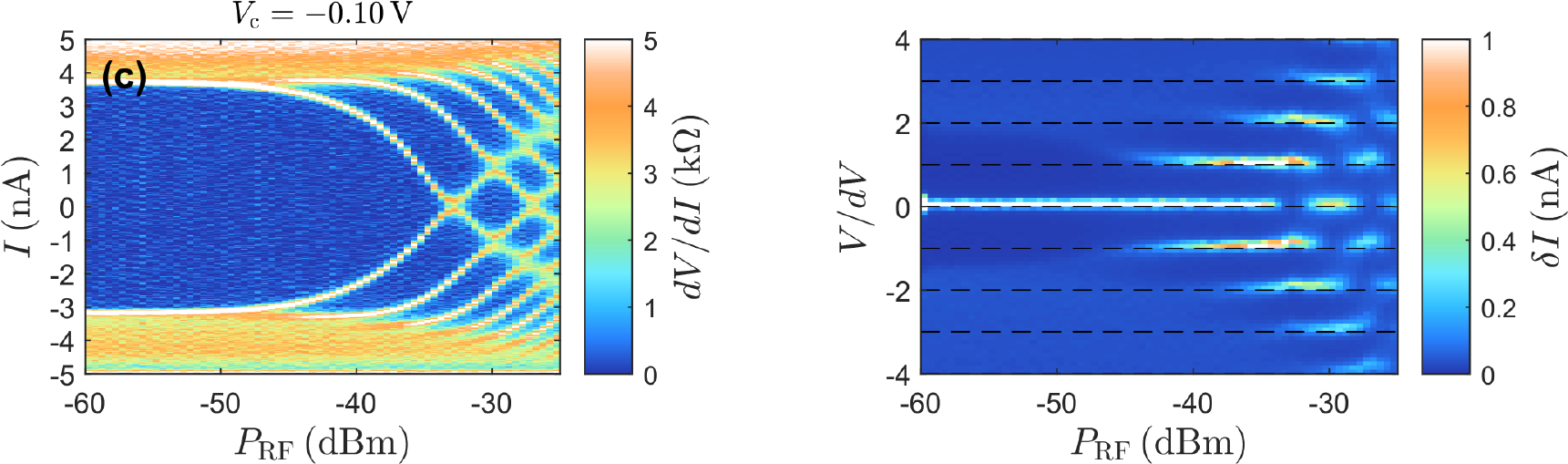} \includegraphics[width=8.5 cm,angle=0]{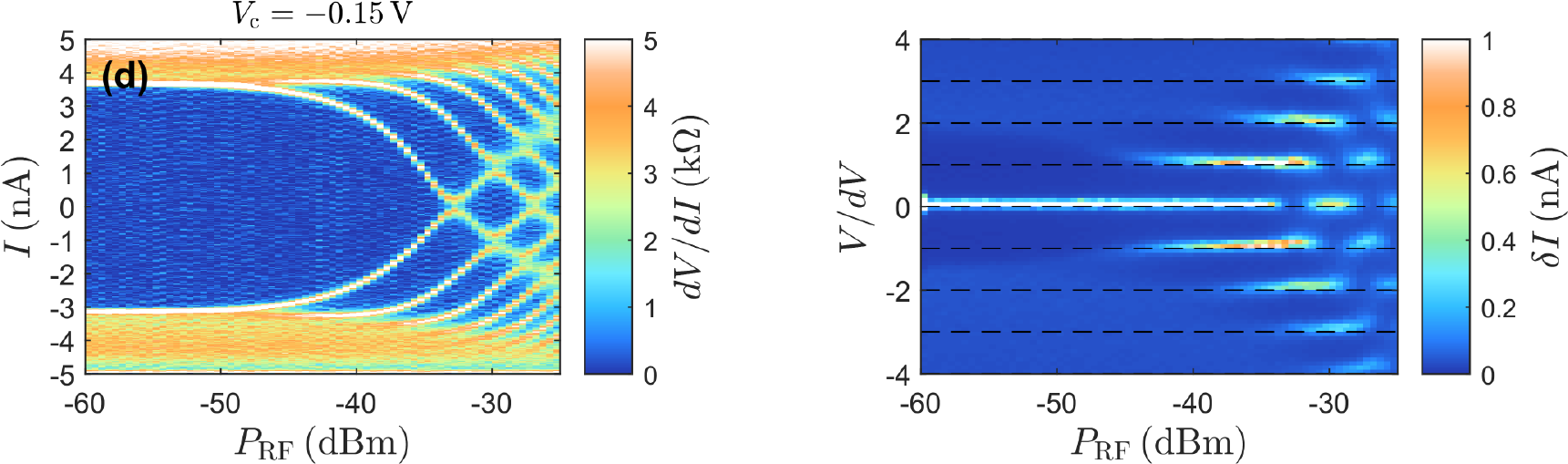}
\includegraphics[width=8.5 cm,angle=0]{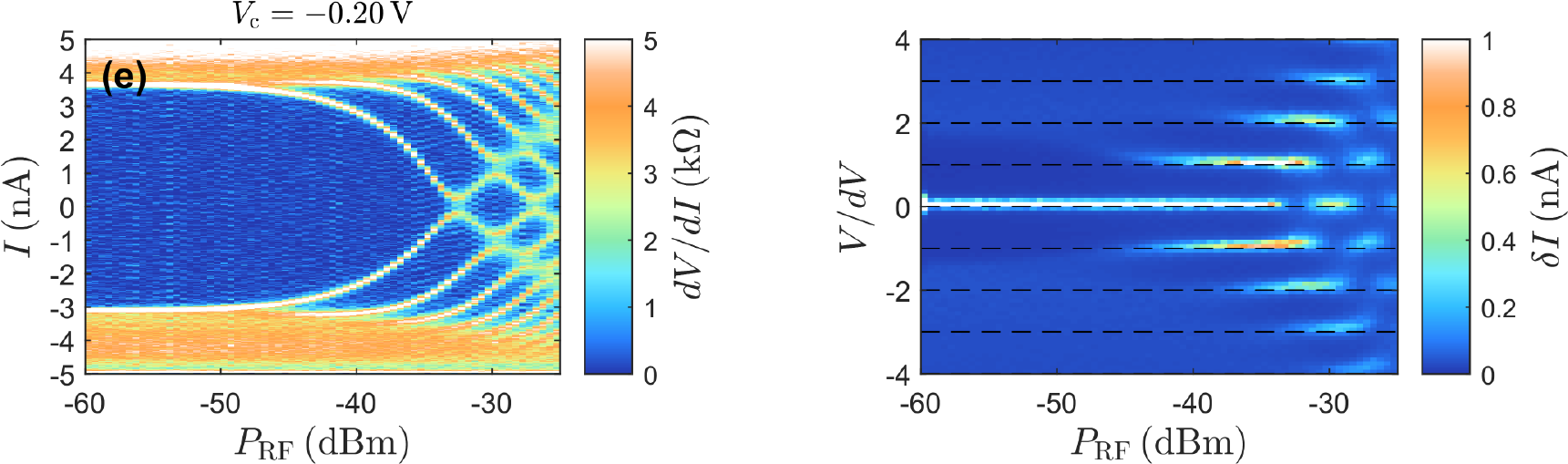} \includegraphics[width=8.5 cm,angle=0]{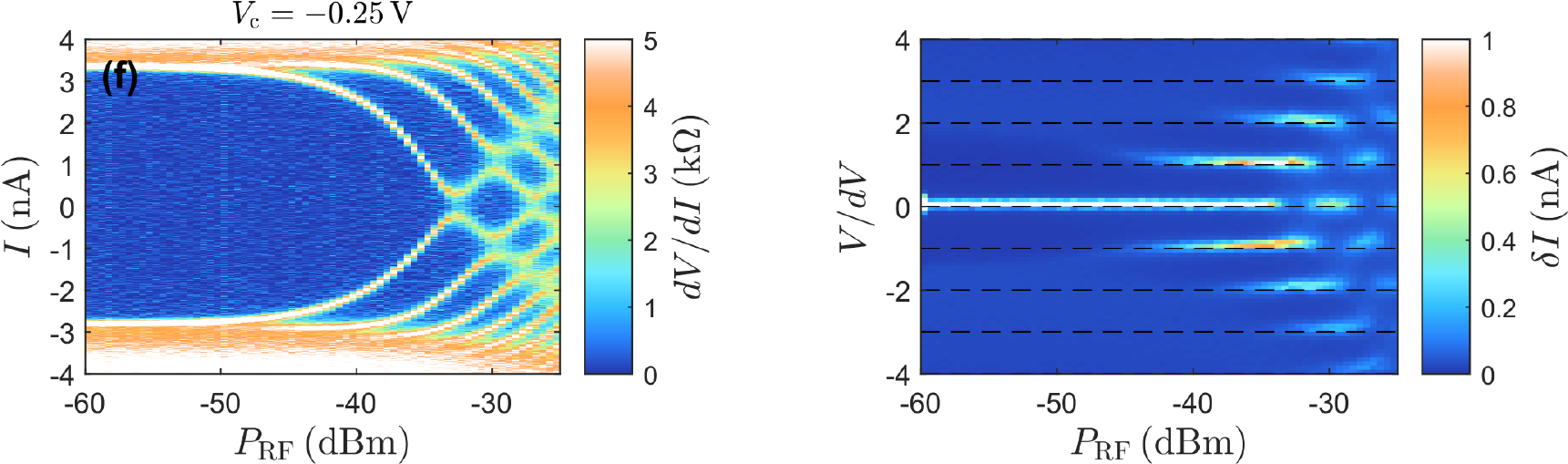}
\includegraphics[width=8.5 cm,angle=0]{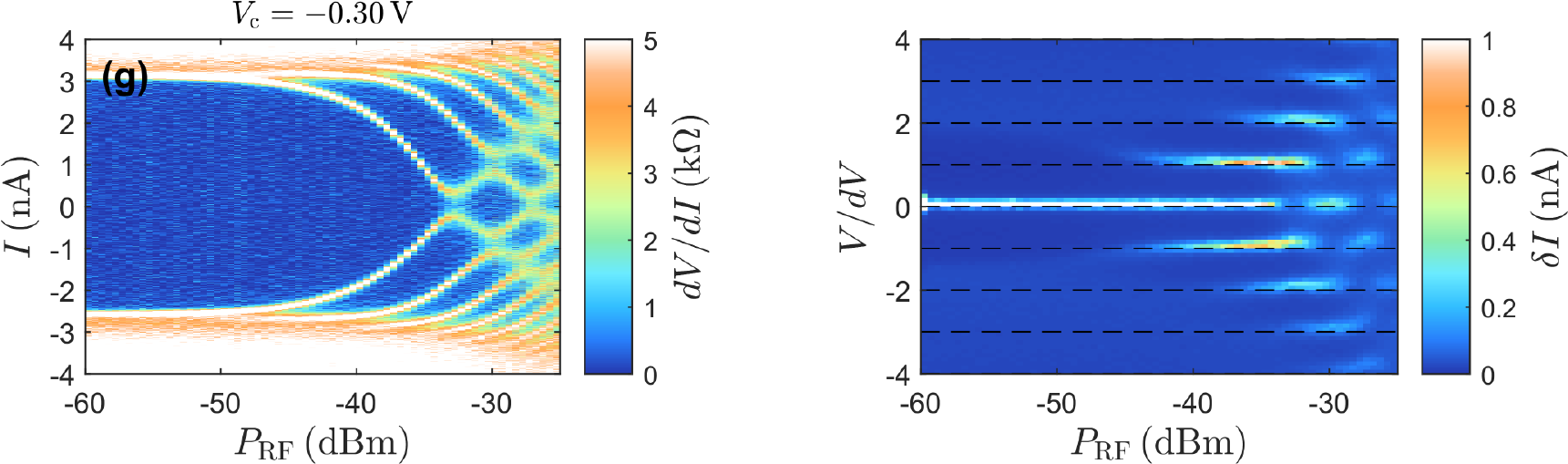} \includegraphics[width=8.5 cm,angle=0]{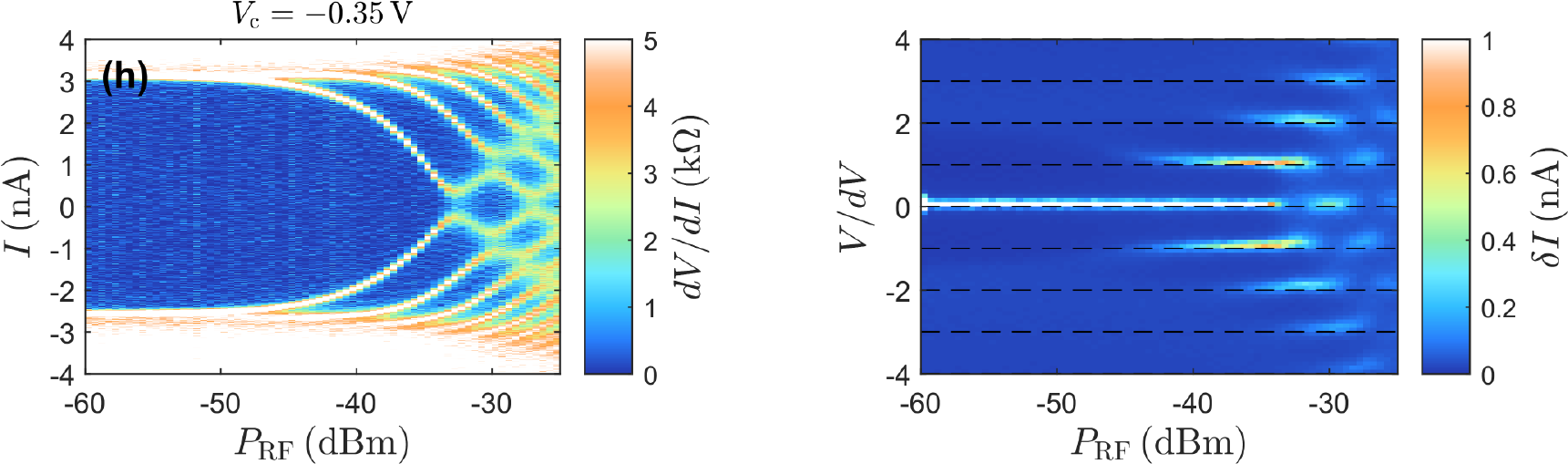}
\includegraphics[width=8.5 cm,angle=0]{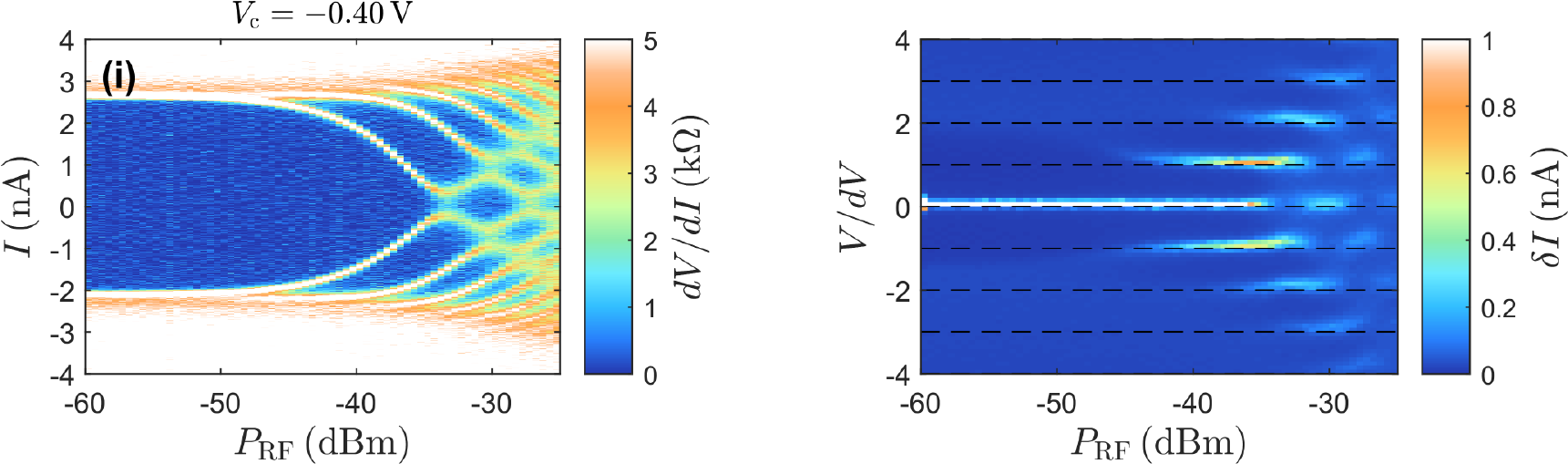} \includegraphics[width=8.5 cm,angle=0]{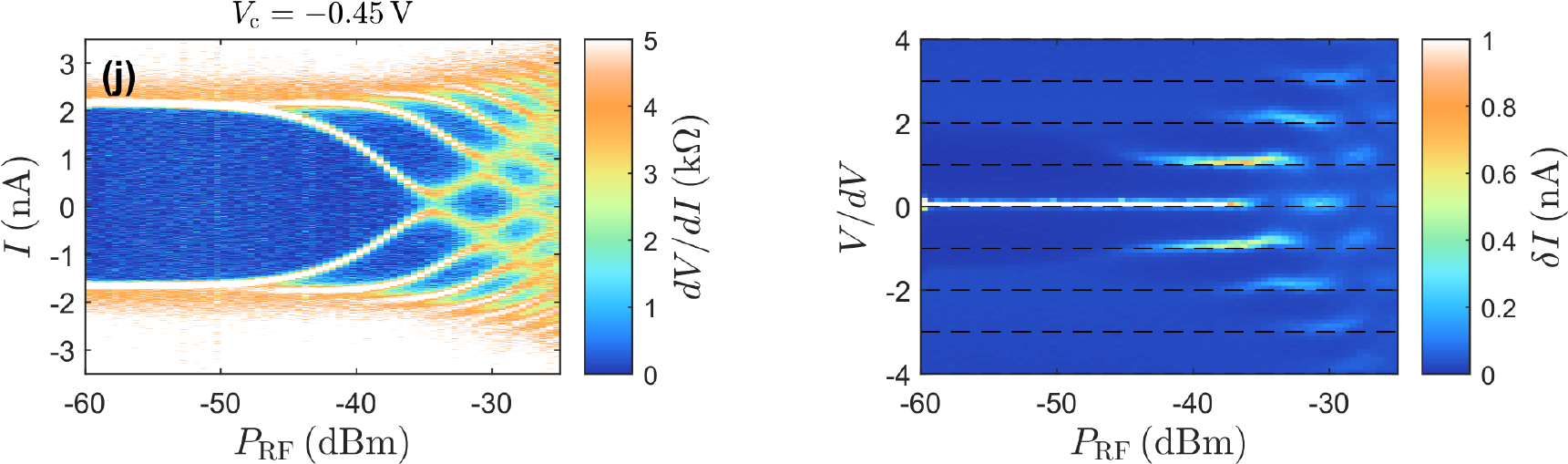}
\includegraphics[width=8.5 cm,angle=0]{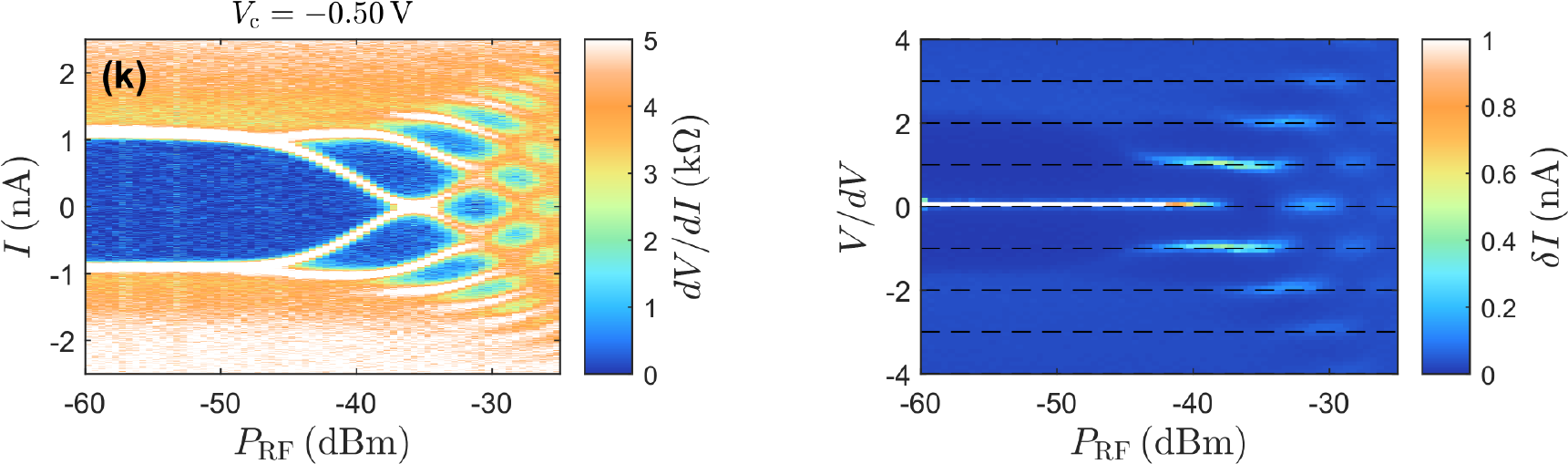} \includegraphics[width=8.5 cm,angle=0]{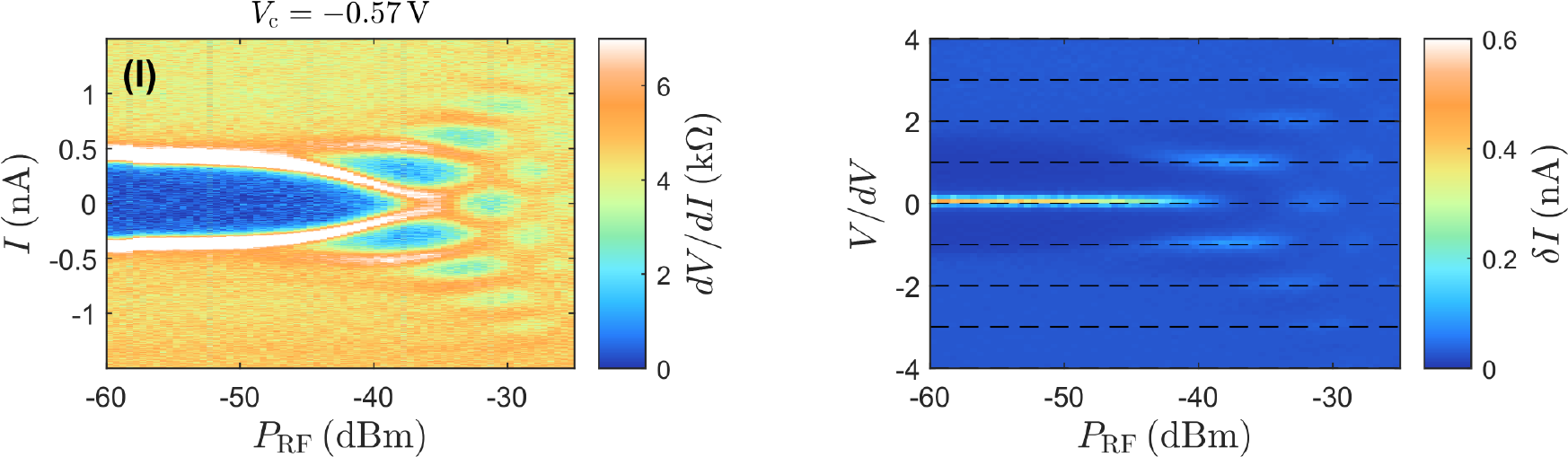}
\includegraphics[width=8.5 cm,angle=0]{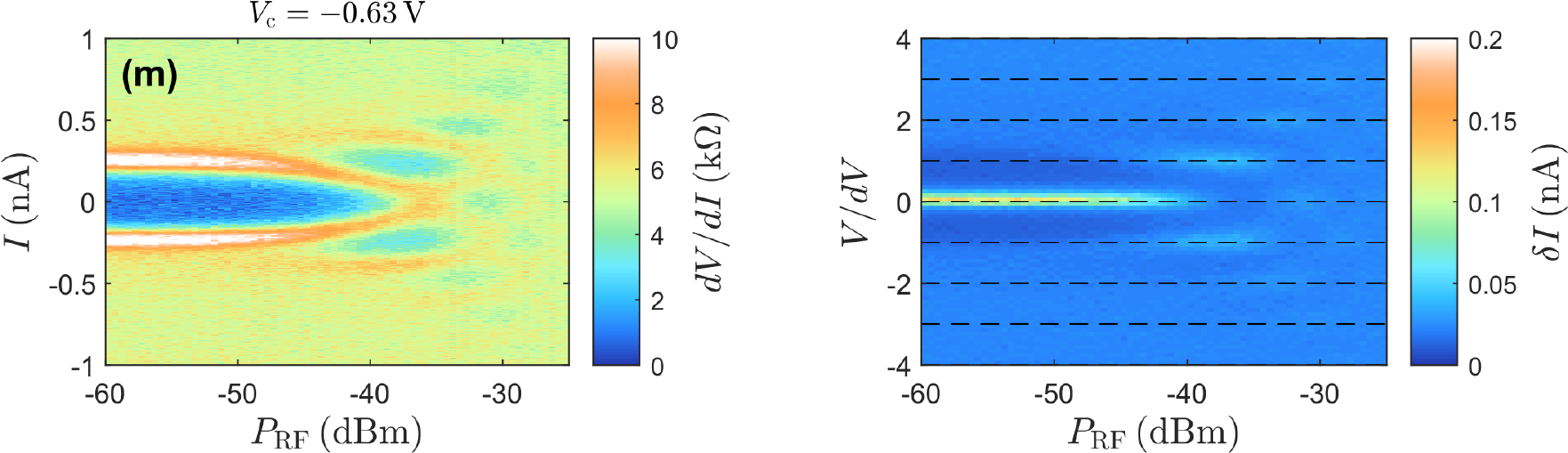} \includegraphics[width=8.5 cm,angle=0]{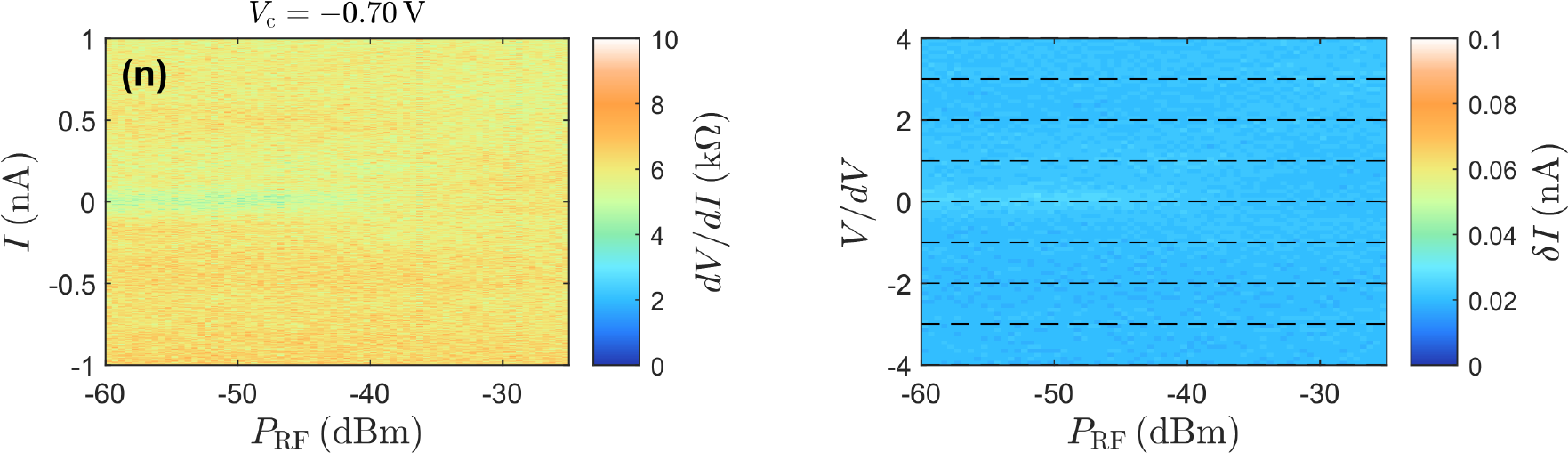}
\caption{\textbf{Shapiro response with external drive at $\bm{f=0.6\,}$GHz}. Plots of differential resistance ($dV/dI$) as a function of device current ($I$) and microwave power ($P_\mathrm{RF}$) with $V_{\mathrm{left}}=3.4\,$V, $V_{\mathrm{right}}=4.0\,$V and $V_{\mathrm{bg}}=0.0\,$V. Each plot includes voltage histograms of $I(V)$ traces revealing the quantization of the Shapiro steps. Histogram bins have a width of $\delta V=0.1dV$. The indicated microwave power is at the output from the signal generator and is not corrected for attenuation of lines on the fridge. Plots are presented for a range of $V_{\mathrm{c}}$ as follows. (a) $V_{\mathrm{c}}=0.00\,\mathrm{V}$, (b) $V_{\mathrm{c}}=-0.05\,\mathrm{V}$, (c) $V_{\mathrm{c}}=-0.10\,\mathrm{V}$, (d) $V_{\mathrm{c}}=-0.15\,\mathrm{V}$, (e) $V_{\mathrm{c}}=-0.20\,\mathrm{V}$, (f) $V_{\mathrm{c}}=-0.25\,\mathrm{V}$, (g) $V_{\mathrm{c}}=-0.30\,\mathrm{V}$, (h) $V_{\mathrm{c}}=-0.35\,\mathrm{V}$, (i) $V_{\mathrm{c}}=-0.40\,\mathrm{V}$, (j) $V_{\mathrm{c}}=-0.45\,\mathrm{V}$, (k) $V_{\mathrm{c}}=-0.50\,\mathrm{V}$, (l) $V_{\mathrm{c}}=-0.57\,\mathrm{V}$, (m) $V_{\mathrm{c}}=-0.63\,\mathrm{V}$, and (n) $V_{\mathrm{c}}=-0.70\,\mathrm{V}$.}
\label{FigExtraShapiro}
\end{figure*}

Here we plot additional Shapiro response data all for a low drive frequency of $f=0.6\,\mathrm{GHz}$. Data is shown for the condition in which only a back-gate voltage of $V_{\mathrm{bg}}=40\,\mathrm{V}$ is applied in figure \ref{fig:ShapiroVbg}. In addition, data is shown for $V_{\mathrm{left}}=3.4\,$V, $V_{\mathrm{right}}=4.0\,$V, $V_{\mathrm{bg}}=0.0\,$V and a wide range of $V_{\mathrm{c}}$ in figure \ref{FigExtraShapiro}. Collectively, these data are an additional confirmation of the inverse ac Josephson effect for both of the gating configurations presented in the main text.

\clearpage

\medskip
\bibliography{WTe2-RSD}